\documentclass[a4paper,11pt]{article}
\usepackage[utf8]{inputenc}

\PassOptionsToPackage{unicode, psdextra}{hyperref}

\usepackage{jcappub}
\usepackage{aas_macros}
\usepackage{graphicx}
\usepackage{amsmath}
\usepackage{amssymb}
\usepackage{mathrsfs}
\usepackage{xfrac}
\usepackage{lmodern} 
\usepackage{anyfontsize} 

\usepackage{fabien}
\newcommand{\vv}{\mathbf{v}}

\def\ds{\displaystyle}

\def\simlt{\lower.5ex\hbox{$\; \buildrel < \over \sim \;$}}
\def\simgt{\lower.5ex\hbox{$\; \buildrel > \over \sim \;$}}

\makeatletter
\gdef\@fpheader{}
\g@addto@macro\bfseries{\boldmath}
\makeatother

\title{2+1 is not 3: Angular bispectrum and Super-Sample Covariance on the light cone}

\author[a,b]{Fabien Lacasa,}
\emailAdd{fabien.lacasa@ulb.be}

\author[b]{Julien Grain}
\emailAdd{julien.grain@ias.u-psud.fr}

\affiliation[a]{
Service de Physique Th\'{e}orique \& Brussels Laboratory of the Universe BLU-ULB, CP 225, Universit\'{e} Libre de Bruxelles, Boulevard du Triomphe, B-1050 Brussels, Belgium
}

\affiliation[b]{
Université Paris-Saclay, CNRS, Institut d'astrophysique spatiale, 91405, Orsay, France.
}

\date{today}

\begin{document}

\abstract{
Observations of the Large Scale Structure sit on the light cone. Cosmological statistics must then be projected from 3D to angular space. For the bispectrum and higher orders, the polyspectra crucially depend on the angles between Fourier modes in addition to their moduli. This leads to delicate cancellations in the squeezed limit and make Limber's approximation fail. This is problematic for predicting the Super-Sample Covariance (SSC), with no known full-sky computation of the angle-dependent terms.\\
Here we develop a new analytical approach to the angular bispectrum, writing Fourier vectors as derivatives that we project on the sphere and tackle with spin-weighted spherical harmonics.\\
We find bispectrum equations free of cancellations and amenable to beyond-Limber numerical methods. Limber's approximation becomes well-behaved and works in the squeezed limit and any other configuration.\\
We then derive the consequence for the covariance between 1-point (e.g. cluster or peak counts) and 2-point statistics (e.g. galaxy clustering, weak lensing or 21cm), which relates to the squeezed bispectrum. The results split into intra-survey covariance and SSC, where we find significant differences with past literature. Our results are in line with the 2+1 geometry of the light cone, having e.g. temporal derivatives in addition to spatial ones. We also find that the tidal term is degenerate with the simpler growth-only term. The total SSC contains a wealth of new terms including wide-angle effects and responses to the background velocity. However we show numerically that it is well approximated by a single term with a density-only response. This ultimately makes our results simpler than past literature.\\
This full-sky result should be more representative than the past flat-sky limit for large survey areas, thus being the one relevant for coming observations with e.g. Euclid, LSST and Roman.
}



\maketitle

\par\noindent\rule{\textwidth}{0.4pt}

\newpage
\section{Introduction}

With the advent of stage IV galaxy surveys, DESI \cite{DESI2023} Euclid \cite{Mellier2024} Rubin/LSST \cite{Ivezic2019} and Roman \cite{Roman}, comes the necessity of a precise statistical characterisation of the Large Scale Structure (LSS) at the high resolution of these coming datasets. On these small scales, the challenge lies in the non-linear evolution of the considered fields, as we can no longer consider that $\delta\ll1$ with $\delta\equiv \delta\rho/\rho$ the matter density contrast. In particular, even for Gaussian initial conditions, the non-linear growth of the LSS generates non-Gaussianity in the statistical distributions of tracer fields.

In this article, we are interested in the lowest-order statistical indicator of non-Gaussianity: the bispectrum, i.e. the 3-point correlation function in harmonic space. Specifically, we will study the theoretical projection of the 3D bispectrum onto the angular bispectrum. The latter is the 3-point function in the space of spherical harmonics, which is a natural and popular basis since observables are projected on the celestial sphere. There are two goals to this article: (i) find projection equations adequate for numerically fast and precise implementation, and (ii) make the link with previous literature on non-Gaussian covariance terms and, in particular, Super-Sample Covariance \citep[SSC][]{Takada2013}.

Regarding the first goal, the context is that the raw projection equations involve integrals of several Bessel functions, whose oscillations are costly to sample in a brute force numerical integration. This has been well studied for the power spectrum, with a variety of approaches available to solve the problem. The most popular approach is the Limber approximation \cite{Limber1953,Kaiser1992,Kaiser1998,LoVerde2008}, which is the numerically fastest approach as it collapses the multidimensional integral into a one-dimensional one with a smooth integrand. Because the Limber approximation is sometimes inaccurate, other methods have been developed which treat the integrals exactly (to numerical precision) \cite[e.g.][]{Assassi2017,Grasshorn2017,Fang2019,LSST2022,Feldbrugge2023,Raccanelli2023,Chiarenza2024}; we call them beyond-Limber methods hereafter. Both the Limber approximation and the beyond-Limber methods are applicable to terms of the bispectrum (or higher-order polyspectra) which only depend on the moduli of the Fourier wave-vectors. The issue is that cosmological perturbation theory (PT) and the bias expansion predict terms that are sensitive not only to these moduli but also to the angles between the Fourier wave-vectors.
Hence, our goal is to devise another approach for these angled terms that puts them in a form amenable to a safe use of the Limber approximation (as well as beyond-Limber methods).

Regarding the second goal, non-Gaussian covariance terms, the context is that the seminal SSC equations \cite{Takada2013} were derived in 3D and projected using the flat sky approximation. This approximation is inadequate for coming stage IV surveys which cover a large fraction of the sky, sometimes distributed over the two celestial hemispheres \citep[e.g. Euclid,][]{Mellier2024}. And indeed, when \cite{Lacasa2016} derived the functional form for SSC in full sky, they found a form substantially different from that given by the flat sky approximation. The problem with \cite{Lacasa2016} is that, due to the complexity of the exact spherical projection, they had to explicitly neglect the angled terms. As a result, they did not recover all SSC terms found by the 3D literature, missing the so-called dilation \cite{Li2014} and tidal \cite{Akitsu2017,Li2018} terms. To date, there still exist no exact derivation of these SSC terms when accounting for the sky curvature, i.e. with spherical harmonics. This article intends to fill this gap. Furthermore, in Sect.~\ref{Sect:context-SSC-bisp-proj-issues} we will show that these dilation and tidal terms in fact raise questions with regards to the self-coherence of the SSC equations. For instance, the 3D-derived equations contain a spatial derivative, but they contain no corresponding time derivative that one would expect since observables sit on the light cone. 
Our goal is thus to have an exact (curved sky) derivation of the covariance (both SSC and intra-survey covariance) that should answer these questions and show whether we find new covariance terms missed in past literature. Spoiler alert: we will.

To reach these goals, we need to perform lengthy calculations. We considered it worthwhile to present the general skeleton of the derivations, leaving the details to numerous appendices. Indeed, the ideas could prove useful to other analyses, and in particular they will need to be generalised to derive the projection of the trispectrum, which is necessary for the covariance of the standard power spectrum. For the readers who are more interested in the results than the derivation, we advise to skip to Table \ref{Table:bisp-summary-noLimber} and \ref{Table:bisp-summary-Limber} for the bispectrum equations respectively in the no-Limber (exact) projection and with Limber approximation, to section \ref{Sect:SSC-total} for the results on super sample covariance, or to section \ref{Sect:ISC} for intra-survey covariance (i.e. other covariance terms).

The outline of this article is the following:
in section \ref{Sect:motivations-notations} we lay out the context with the power spectrum and bispectrum, and we show in greater detail the issue that arise with the projection of the bispectrum in the squeezed configuration.
In section \ref{Sect:way-spin-spherical-harmonics} we develop the first part of the calculation: turn the angled terms into new equations without angles but instead Bessel derivatives.
In section \ref{Sect:tractable-forms} we develop the second part of the calculation: remove the Bessel derivatives to get equations amenable to beyond-Limber methods or Limber's approximation. For readers interested in the bispectrum, these will be the main results.
In section \ref{Sect:application-covariance} we apply the previous findings to the covariance, where we need a mixed projection case (exact on the first multipole and Limber on the two other). We then derive the corresponding SSC and intra-survey covariance terms. For readers interested in covariances, these will be the main results.
Finally, we discuss our results in section \ref{Sect:discussion}.

\section{Motivations and notations}\label{Sect:motivations-notations}

In this section, we introduce our notations and the problem we want to solve in this article. To do so we first recall some useful elements of cosmological perturbation theory, then we introduce the 3D power spectrum and bispectrum of LSS tracers and their projection to angular observables, before showing the link with the covariance and the issue of angled terms for Super-Sample Covariance.

\subsection{Cosmological fields and useful results from perturbation theory}\label{Sect:perturbation-theory}

We work in a Friedmann-Lemaître background and note comoving position as $\xx$. For a point on the light cone we have $\xx= r(z)\hn$ with $r(z)$ the comoving distance from the observer to redshift $z$ and $\hn$ the direction of observation. Conformal time is noted with $\tau$, but on the light cone we may choose instead to denote it by redshift or comoving distance through some mild abuse of terminology, e.g. the scale factor is $a(\tau)=a(z)=a(r)$. The Fourier wavevector is noted $\kk$ with modulus $k=|\kk|$ and unit vector $\hk=\kk/k$. To shorten equations, repetition of indices denotes either multiplication, e.g. $k_{12}\equiv k_1 \, k_2$, or repetition of arguments in a function, e.g. $f(r_{12})\equiv f(r_1,r_2)$.

The three fundamental cosmological fields that describe the LSS are the gravitational potential $\Phi$, the peculiar velocity field $\vv$ and the (total) matter density $\rho$. As standard, we define the mean/cosmic density $\bar{\rho}(z) = \lbra \rho(\xx,z)\rbra$ and the density contrast $\delta(\xx,z)=\frac{\rho(\xx,z)-\bar{\rho}(z)}{\bar{\rho}(z)}$. The nonlinear equations of motion for the three fields are \cite[e.g.][]{Bernardeau2002}:
\ba
\Delta\Phi =\frac{3}{2} \Omega_m \mathcal{H}^2 \delta \label{Eq:Poisson} \\
\frac{\partial \delta}{\partial \tau} + \nabla \cdot \left[(1+\delta)\vv\right] = 0 \label{Eq:Continuity} \\
\frac{\partial \vv}{\partial \tau} + \mathcal{H} \vv + (\vv\cdot\nabla)\vv = - \nabla\Phi \label{Eq:Navier-Stokes}.
\ea
where $\tau$ is the conformal time, $\mathcal{H}=\left.\frac{\dd a}{\dd\tau}\middle/a\right.$, and all spatial derivatives are in comoving coordinates. 
In the following we consider this effectively as fluid dynamics (on an expanding background) and call standardly these three equations as respectively Poisson (Eq.~\ref{Eq:Poisson}), continuity (Eq.~\ref{Eq:Continuity}) and Navier-Stokes (Eq.~\ref{Eq:Navier-Stokes}). The Poisson equation is straightforwardly solved in Fourier space as $k^2\Phi(\kk) \propto\delta(\kk)$. We call standardly $\theta=\nabla\cdot\vv$ the divergence of the velocity field.

Cosmological perturbation theory solves these equations perturbatively by assuming that all fields are small. The first order is linear theory and the fields have a solution separable into time-dependent and space-dependent functions \cite[see e.g.][]{Bernardeau2002}
\ba
\delta(\xx,\tau) = G(\tau) \ \delta_{(1)}(\xx) + \mathcal{O}\left(\delta_{(1)}^2\right) \qquad \theta(\xx,\tau) &= G'(\tau) \ \theta_{(1)}(\xx) + \mathcal{O}\left(\delta_{(1)}^2\right)
\ea
where the density contrast relates to the velocity divergence as
$\theta_{(1)} = - \delta_{(1)}$
and $G$ is the linear growth factor, found by solving the differential equation
$\frac{\dd^2 G}{\dd \tau^2} + \mathcal{H}\frac{\dd G}{\dd \tau} = \frac{3}{2} \Omega_m \mathcal{H}^2 \ G$ \cite{Bernardeau2002}.
The velocity field is best written in Fourier space as
\ba\label{Eq:first-order-velocity}
\vv_{(1)}(\kk) = - \delta_{(1)}(\kk) \, \frac{\kk}{k^2}.
\ea

Beyond first order, on an Einstein-de Sitter background the solution is still separable and can be found recursively. Perturbation theory thus allows to relate all the space-dependent functions $\delta_{(n)}$ and $\theta_{(n)}$ to the first order $\delta_{(1)}$. In particular at second order the density contrast writes \cite{Bernardeau2002}
\ba\label{Eq:delta-expansion}
\delta(\xx,\tau) &= G(\tau) \ \delta_{(1)}(\xx) + G(\tau)^2 \ \delta_{(2)}(\xx) +\mathcal{O}\left(\delta_{(1)}^3\right)
\ea
where $\delta_{(2)}$ is given in Fourier space by
\ba
\delta_{(2)}(\kk) = \int \dd^3\kk_1 \dd^3\kk_2 \ \delta_D(\kk-\kk_1-\kk_2) \ F_2(\kk_1,\kk_2) \, \delta_{(1)}(\kk_1) \, \delta_{(1)}(\kk_2)
\ea
with $\delta_D$ a Dirac distribution and $F_2$ the (symmetrised, EdS) second-order kernel \cite{Fry1984}
\ba
F_2(\kk_1,\kk_2) &= \frac{5}{7} + \frac{1}{2} \left(\frac{1}{k_1^2}+\frac{1}{k_2^2}\right) \left(\kk_1\cdot\kk_2\right) + \frac{2}{7} \left(\frac{\kk_1\cdot\kk_2}{k_1 \, k_2}\right)^2 \label{Eq:F_2_physform} \\
&= \frac{5}{7} + \frac{1}{2} \left(\frac{k_1}{k_2}+\frac{k_2}{k_1}\right) \frac{\kk_1\cdot\kk_2}{k_1 \, k_2} + \frac{2}{7} \left(\frac{\kk_1\cdot\kk_2}{k_1 \, k_2}\right)^2. \label{Eq:F_2_mathsform}
\ea
The first form is more suitable for physical interpretation ; the second form is more suitable for calculations as it expands as a series over dot products $\left(\hk_1\cdot\hk_2\right)^n$ with terms for $n=0$, 1 and 2. \\
This form for the $F_2$ kernel is stricto sensu only valid in an Einstein-de Sitter cosmology. However, \cite{Scoccimarro1998} showed that the equations of motion in $\Lambda$CDM can be approximated to lead to the same kernels as the Einstein-de Sitter case to any order. The cosmology dependence is then encapsulated in the linear growth factor $G(z)$. We use this approximation in this article.

These three terms in the $F_2$ kernel can be interpreted physically, which will prove useful to interpret our results throughout the article. Indeed, the second order density contrast $\delta_{(2)}$ rises from three contributions:
\begin{itemize}
    \item $(\delta_{(1)})^2$. This is the $n=0$ term and is called the "growth-only" contribution in the literature.
    \item $\nabla \delta_{(1)}\cdot \vv_{(1)}$. This is the $n=1$ term. It can be traced back to two terms: (i) $\nabla\cdot (\delta \mathbf{v})$ in the  continuity equation, and (ii) a part of $\left( \mathbf{v} \cdot \nabla \right) \mathbf{v}$ in Navier-Stokes
    . We call this the transport contribution.
    \item $\partial_i v_j^{(1)} \times \partial_j v_i^{(1)}$. This is the $n=2$ term. It can be traced back to (the other part of) the term $\left( \mathbf{v} \cdot \nabla \right) \mathbf{v}$ in Navier-Stokes. We call this the advection contribution. 
\end{itemize}

We generally cannot observe matter directly and instead characterize the statistic of a LSS tracer. The spatial distribution of the density contrast $\delta_\mr{t} \equiv \frac{n_\mr{t}-\overline{n}_\mr{t}}{\overline{n}_\mr{t}} $ of such tracer is related to the matter field (and other physical fields beyond linear order) by the bias expansion \cite{Desjacques2018}. To second order, this expansion reads:
\be
\delta_\mr{t} = b_1^\mr{t} \, \delta + b_2^\mr{t} \, \left(\delta^2 - \lbra\delta^2\rbra\right) + b_{s^2}^t \, s^2 \label{Eq:bias-expansion}
\ee
where the biases $b_i$ are implicitly functions of time and scale: $b_i^{\mr{t}} = b_i^{\mr{t}}(k,r)$. Here $s^2= s_{ij} \, s^{ij}$ with $s_{ij}=\partial_i\partial_j\Phi-\delta_{ij}\Delta\Phi/3$ the tidal tensor. In Fourier space, $s^2$ is linked to $\delta$ via the tidal kernel \cite{Baldauf2012}
\ba
S_2(\kk_1,\kk_2) = & \left(\frac{\kk_1\cdot\kk_2}{k_1 \, k_2}\right)^2 -\frac{1}{3}.
\ea
The statistics of $\delta_\mr{t}$ can then be computed order by order using this double expansion, Eqs~\ref{Eq:delta-expansion} and \ref{Eq:bias-expansion}.

\subsection{Power spectrum of LSS tracers}\label{Sect:context-power-spectrum}

\subsubsection*{3D spectrum}
We note $P(k|z)$ the 3D matter power spectrum at a redshift $z$. Moreover, in this article we need the cross-spectrum between different redshifts/times, and it proves more useful to indicate redshift/time with the corresponding comoving distance $r(z)$. Hence we define the cross-spectrum $P(k|r_1,r_2)$:
\be
\lbra \delta(\kk|z_1) \ \delta(\kk'|z_2) \rbra = (2\pi)^3 \, \delta_D(\kk+\kk') \ P(k|r_1,r_2) 
\ee
where $\delta_D$ is a Dirac distribution and $r_i=r(z_i)$. In the following we further shorten $P(k|r_{12})\equiv P(k|r_1,r_2)$.\\
Throughout this article we use the tree-level matter power spectrum which stems from first order perturbation theory
\be
P(k|r_{12}) = G(r_1) \, G(r_2) \ P(k)
\ee
where $P(k)$ is the matter power spectrum today ($r=0$ i.e. $z=0$).

Using the bias expansion to first order, we find that the tree-level power spectrum of tracers' density contrasts is
\be
P_{\delta_{t_1} \delta_{t_2}}(k|r_{12}) = b_1^\mr{t_1} b_1^\mr{t_2} \times P(k|r_{12}) = \left( b_1^\mr{t_1} G(r_1)\right) \left( b_1^\mr{t_2} G(r_2)\right) \ P(k).
\ee
In this article, we find it simpler to work directly with the tracer density $n_\mr{t}$ instead of its density contrast  $\delta_\mr{t}$. In that case we have the density power spectrum
\be
P_{n_{t_1} n_{t_2}}(k|r_{12}) = n_\mr{t_1} \, n_\mr{t_2} \ P_{\delta_{t_1} \delta_{t_2}}(k|r_{12}) = \left( n_\mr{t_1} b_1^\mr{t_1} G(r_1) \right) \left( n_\mr{t_2} b_1^\mr{t_2} G(r_2) \right)  P(k)
\ee
which we further shorten as $P_{n_{t_{12}}} \equiv P_{n_{t_1} n_{t_2}}$

\subsubsection*{Angular projection}
3D quantities are not directly observable, instead we observe quantities projected on the sky. Thus the projected tracer density is an integral over the light cone: $N_\mr{t}(\hn) = \int r^2\dd r \ n_\mr{t}(r\hn)$, where $\hn$ is a unit vector on the sphere and we assume that redshift selection functions were absorbed into $n_\mr{t}$. From this projected density one can standardly define the harmonic coefficients $a_{\ell,m}^\mr{t} = \int \dd^2\hn \, N_\mr{t}(\hn) \, Y_{\ell,m}^*(\hn)$ and the angular power spectrum through $\lbra a_{\ell,m}^\mr{t_1} \, (a_{\ell',m'}^\mr{t_2})^*\rbra = C_\ell^{t_1 t_2} \, \delta^K_{\ell,\ell'} \,  \delta^K_{m,m'}$ with $\delta^K$ the Kronecker symbol. Then, straightforward algebra shows the relation between the 3D power spectrum and the angular one as
\ba
C_\ell^{t_{12}} &= \int r_{12}^2 \, \dd r_{12} \ c_\ell^{t_{12}}(r_{12}) \label{Eq:Cl-from-cl}\\
c_\ell^{t_{12}}(r_{12}) &= \frac{2}{\pi} \int k^2 \dd k \ P_{n_{t_{12}}}(k|r_{12}) \ j_{\ell}(k r_1) \, j_{\ell}(k r_2). \label{Eq:cl}
\ea
As mentioned in the introduction, numerical methods are available to tackle these integrals \cite[e.g.][]{Assassi2017,Grasshorn2017,Fang2019,LSST2022,Feldbrugge2023,Raccanelli2023,Chiarenza2024}. A popular approach however is to use the numerically faster Limber's approximation \cite{Limber1953}, more precisely its extension in Fourier space \cite{Kaiser1992,Kaiser1998,LoVerde2008}. Indeed, this approximation allows to simplify some integrals with Bessel functions:
\ba\label{Eq:Limber}
\int k^2 \dd k \ f(k) \ j_{\ell}(k r_1) \, j_{\ell}(k r_2) \approx f(k_\ell) \ \frac{\pi}{2 r_1^2} \, \delta_D(r_1-r_2)
\ea
with $k_\ell=(\ell+\sfrac{1}{2})/r_1$ the peak of the Bessel function. We note here that the approximation is exact when $f(k)$ is a constant\footnote{This shows that the approximation assumes that $f(k)$ varies slowly with k with respect to the oscillations of the Bessel functions $\Delta k = 1/r$. In turns this explains why the approximation typically performs better at high $\ell$, meaning high $k$, since this yields smaller relative variations $\Delta k/k$.} due to the closure relation of spherical Bessel functions
\ba\label{Eq:closure-Bessel}
\int k^2 \dd k \ j_{\ell}(k r_1) \, j_{\ell}(k r_2) = \frac{\pi}{2 r_1^2} \ \delta_D(r_1-r_2)
\ea
In the case of the power spectrum, Limber's approximation gives
\ba
c_\ell^{t_{12}}(r_{12}) &= P_{n_{t_{12}}}(k_\ell|r_{12}) \ \frac{\delta_D(r_1-r_2)}{r_1^2} \label{Eq:cl-with-Limber} \\
C_\ell^{t_{12}} &= \int r^2 \dd r \ P_{n_{t_{12}}}(k_\ell|r) \label{Eq:Cl-with-Limber}.
\ea

\subsection{Bispectrum of LSS tracers}\label{Sect:context-bispectrum}

\subsubsection*{3D bispectrum}
Having introduced the power spectrum and its projection, we now turn to the more complex case of the bispectrum.
On large scales, this is derived using the bias expansion Eq.~\ref{Eq:bias-expansion} and perturbation theory, both to second order. We find the tree-level bispectrum of (the dimensionless density contrast of) three tracers $\mr{t_1,t_2,t_3}$ as
\ba
\nonumber B_{\delta_\mr{t_{123}}}(\kk_{123}|r_{123}) =& \ b_1^\mr{t_1}(k_1,r_1) \, b_1^\mr{t_2}(k_2,r_2) \, b_1^\mr{t_3}(k_3,r_3) \ B_\mr{2PT}(\kk_{123}|r_{123}) \\
\nonumber +& \ b_1^\mr{t_1}(k_1,r_1) \, b_1^\mr{t_2}(k_2,r_2) \, b_2^\mr{t_3}(k_3,r_3) \ B_{b_2}(\kk_{123}|r_{123}) \\
+& \ b_1^\mr{t_1}(k_1,r_1) \, b_1^\mr{t_2}(k_2,r_2) \, b_{s^2}^\mr{t_3}(k_3,r_3) \ B_{s^2}(\kk_{123}|r_{123}) \label{Eq:Bispectrum-3D-original}
\ea
where we used the shorthand notation $B_{\delta_\mr{t_{123}}}(k_{123}|r_{123}) \equiv B_{\delta_\mr{t_1},\delta_\mr{t_2},\delta_\mr{t_3}}(k_1,k_2,k_3|r_1,r_2,r_3)$.
In the equation above, we have
\ba
B_\mr{2PT}(\kk_{123}|r_{123}) &= 2 \ F_2(\kk_1,\kk_2) \ P(k_1|r_{13}) \ P(k_2|r_{23}) + 2 \ \mr{perm.} \\
B_{b_2}(\kk_{123}|r_{123}) &= 2 \qquad \frac{1}{2} \qquad \ P(k_1|r_{13}) \ P(k_2|r_{23}) + 2 \ \mr{perm.} \\
B_{s^2}(\kk_{123}|r_{123}) &= 2 \ S_2(\kk_1,\kk_2) \ P(k_1|r_{13}) \ P(k_2|r_{23}) + 2 \ \mr{perm.}
\ea
In the following we drop the dependence of the bias on $(k,r)$ when it can be understood from the tracer superscript, e.g. $b_1^\mr{t_1} \equiv b_1^\mr{t_1}(k_1,r_1)$. Then equation~\ref{Eq:Bispectrum-3D-original} can be simplified into
\ba
B_{\delta_\mr{t_{123}}}(\kk_{123}|r_{123}) =& \ 2 \ b_1^\mr{t_1} \; b_1^\mr{t_2} \; b_{\Sigma_2}^\mr{t_3}(\kk_{123},r_3) \ P(k_1|r_{13}) \, P(k_2|r_{23}) + 2 \ \mr{perm.}
\ea
where $\Sigma_2$ stands for the sum of all second-order contributions and we have
\ba
b_{\Sigma_2}^\mr{t_3}(\kk_{123},r_3) = F_2(\kk_1,\kk_2) \, b_1^\mr{t_3} + \frac{1}{2} \, b_2^\mr{t_3} + S_2(\kk_1,\kk_2) \, b_{s^2}^\mr{t_3} 
\ea
We expand the dependence of $b^\mr{t_3}_{\Sigma_2}$ over angles as
\ba
b_{\Sigma_2}^\mr{t_3}(\kk_{123},r_3) = \sum_{n=0}^2  b_{(n)}^\mr{t_3}(k_{3},r_3) \ q_{(n)}(k_1,k_2) \ \left(\hat{k}_1 \cdot \hat{k}_2\right)^n \label{Eq:implicit-def-bias_(n)}  
\ea
where 
\ba
b^\mr{t_3}_{(0)} &= \frac{5}{7} b^\mr{t_3}_1 + \frac{1}{2} b^\mr{t_3}_2 - \frac{1}{3} b^\mr{t_3}_{s^2} \qquad & q_{(0)}(k_1,k_2) &= 1 \label{Eq:def-bias_(0)} \\ 
b^\mr{t_3}_{(1)} &= b^\mr{t_3}_1  \qquad & q_{(1)}(k_1,k_2) &= \frac{1}{2} \left(\frac{k_1}{k_2}+\frac{k_2}{k_1}\right) \label{Eq:def-bias_(1)} \\
b^\mr{t_3}_{(2)} &= \frac{2}{7} b^\mr{t_3}_1 + b^\mr{t_3}_{s^2}  \qquad & q_{(2)}(k_1,k_2) &= 1 \label{Eq:def-bias_(2)} 
\ea
As in the power spectrum case, in this article we find more useful to look at the bispectrum of the number density of the tracers, $n_\mr{t}$, instead of their density contrast, $\delta_\mr{t}$. This just adds a prefactor:
\be
B_\mr{n_{t_{123}}}(\kk_{123}|r_{123}) = n_\mr{t_1} \, n_\mr{t_2} \, n_\mr{t_3} \ B_{\delta_{t_{123}}}(\kk_{123}|r_{123}).
\ee
In this case, a summary of the previous equations is that we have
\be
B_\mr{n_{t_{123}}}(\kk_{123}|r_{123}) = 2 \sum_{n=0}^2 \tilde{B}_n \ \frac{\left(\kk_1\cdot\kk_2\right)^n}{k_1^n \, k_2^n} + 2 \ \mr{perm.}
\ee
where we defined $\tilde{B}_n$ (with a tilde to avoid confusion with the full bispectrum) as
\be
\tilde{B}_n = n_\mr{t_1}(r_1) \, n_\mr{t_2}(r_2) \, n_\mr{t_3}(r_3)\ P(k_1|r_{13}) P(k_2|r_{23}) \ b^\mr{t_1}_1(k_1,r_1) \, b^\mr{t_2}_1(k_2,r_2) \, b^\mr{t_3}_{(n)}(k_{3},r_3) \, q_{(n)}(k_1,k_2) 
\ee
or more compactly with implicit dependences
\be
\tilde{B}_n = n_\mr{t_1} \, n_\mr{t_2} \, n_\mr{t_3}\ P(k_1|r_{13}) P(k_2|r_{23}) \ b^\mr{t_1}_1 \, b^\mr{t_2}_1 \, b^\mr{t_3}_{(n)} \, q_{(n)} \label{Eq:def-B_n} 
\ee
The interest of this $\tilde{B}_n$ is that it has the same dimension regardless of $n$: 1/Volume.

\subsubsection*{Angular projection}
As in the power spectrum case, the 3D bispectrum is not directly observable. The true observable is the angular bispectrum defined from the harmonic coefficients of the projected counts: $\lbra a_{1}^\mr{t_1} \, a_{2}^\mr{t_2} \, a_{3}^\mr{t_3} \,\rbra = b_{\ell_{123}}^{n_\mr{t_{123}}} \ G_{123}$ where we shortened the harmonic coefficient as $a_i \equiv a_{\boldell_i} \equiv a_{\ell_i,m_i}$, introduced the harmonic doublet $\boldell \equiv (\ell,m)$, and used the Gaunt coefficient
\be
G_{123} \equiv G_{m_1,m_2,m_3}^{\ell_1,\ell_2,\ell_3} = \int \dd^2\hn \ Y_{\boldell_1}(\hn) \, Y_{\boldell_2}(\hn) \, Y_{\boldell_3}(\hn) .
\ee
Some straightforward algebra shows that the relation between the 3D and angular bispectra is:
\ba
\nonumber b_{\ell_{123}}^{n_\mr{t_{123}}} \ G_{123} =& \  i^{\lu+\ld+\lt} \, (4\pi)^3 \!\int r^2_{123} \, \dd r_{123}  \ \frac{\dd^3\kk_{123}}{(2\pi)^9} \ j_{\lu}(k_1 r_1) \, j_{\ld}(k_2 r_2) \, j_{\lt}(k_3 r_3) \\
& \times Y^*_{\boldell_1}(\hk_1) \, Y^*_{\boldell_2}(\hk_2) \, Y^*_{\boldell_3}(\hk_3) \ (2\pi)^3 \, \delta_D(\kk_1+\kk_2+\kk_3) \ B_\mr{n_{t_{123}}}(\kk_{123}|r_{123}) \label{Eq:bispectrum-3D-to-angular}
\ea
where $\hk = \kk/k$.\\
In the general case, Eq.~\ref{Eq:bispectrum-3D-to-angular} does not reduce further in a simple manner. One case where it simplifies nicely is when the bispectrum only depends on the moduli of the Fourier modes: $B(\kk_{123})=B(k_{123})$. That is the case for the $n=0$ term of the bispectrum, and in this case we find
\ba
b_{\ell_{123}}^{n_\mr{t_{123}}}  =& \  (-1)^{\lu+\ld+\lt} \, \left(\frac{2}{\pi}\right)^3 \!\int x^2 \, \dd x \left[ \prod_{i=1,2,3} r^2_{i} \, \dd r_{i}  \ k_{i}^2 \, \dd k_{i} \ j_{\ell_i}(k_i r_i) \, j_{\ell_i}(k_i x) \right] B_\mr{n_{t_{123}}}(k_{123}|r_{123}). \label{Eq:bispectrum-3D-to-angular-n=0}
\ea
This form is amenable to numerical treatment with the fore-cited beyond-Limber methods developed for the angular power spectrum. A more popular and numerically faster approach is to use Limber's approximation which gives
\ba
b_{\ell_{123}}^{n_\mr{t_{123}}}  =& \  (-1)^{\lu+\ld+\lt} \!\int x^2 \, \dd x \ B_\mr{n_{t_{123}}}(k_{\ell_{123}}|x). \label{Eq:bispectrum-3D-to-angular-n=0-Limber}
\ea
where $k_{\ell_i}=(\ell_i+\sfrac{1}{2})/x$.\\
Equations~\ref{Eq:bispectrum-3D-to-angular-n=0} and \ref{Eq:bispectrum-3D-to-angular-n=0-Limber} are restricted to the case where the bispectrum does not depends on the angles between the Fourier modes, which we call the angle independent case from now on. We now turn to discuss issues that can arise in the literature due to the angle dependent terms.

\subsection{Cancellations and Super-Sample Covariance issues}\label{Sect:context-SSC-bisp-proj-issues}

From the previous subsection, we see that projection methods (either with or without Limber's approximation) can be applied straightforwardly if the 3D bispectrum only depends on the moduli of the Fourier modes. One tempting way to reduce to that case is to use trigonometry with the law of cosines --since $(\kk_{1},\kk_{2},\kk_{3})$ form a closed triangle-- to write the angles in terms of the moduli\footnote{For instance $k_3^2=(\kk_1+\kk_2)^2$ leads to $\kk_1\cdot\kk_2 = (k_3^2-k_2^2-k_1^2)/2$.}.
We call this the trigonometric approach from now on. It works well for most triangle configurations. In Limber's approximation, since $k_{\ell_i}=(\ell_i+\sfrac{1}{2})/x$, the shape of the triangle $(k_1,k_2,k_3)$ is the same as that of the triangle $(\lu+\sfrac{1}{2},\ld+\sfrac{1}{2},\lt+\sfrac{1}{2})$, and it suffices to evaluate the angle-dependent kernels $F_2$ and $S_2$ on the latter triangle.

In the next two subsubsections we show that this trigonometric approach leads to high cancellations which can make Limber's approximation fail in the squeezed limit. This leads to results highly different from the exact projection and to Super-Sample Covariance literature. For this, we focus on the bispectrum term coming from the $F_2$ kernel.

\subsubsection{Cancellation issues}\label{Sect:context-cancellation-issues}
Using the law of cosines, in an isosceles triangle with $k_2=k_3$, the kernel $F_2(\kk_1,\kk_2)$ takes the form
\ba\label{Eq:28_F_2_iso}
28 F_2^\mr{iso} = 13 - 5 \left(\frac{k_1}{k_2}\right)^2 
\ea
where the first term takes contribution from $n=0$ and $n=1$ while the second term takes contribution from $n=1$ and $n=2$, and we multiplied by a factor 28 to avoid fractions and simplify the equations for discussion.\\
For a general triangle with $k_2\neq k_3$ we find
\ba\label{Eq:28_F_2_grl}
28 F_2(\kk_1,\kk_2) = \left(3 k_3^2 - 5 k_2^2 +2\frac{k_3^4}{k_2^2}\right) \frac{1}{k_1^2} + 28 F_2^\mr{iso} + 3 \left(\frac{k_3^2}{k_2^2} - 1 \right)
\ea
where we isolated in front the three terms proportional to $\frac{1}{k_1^2}$, and the equation properly reduces to the previous one when $k_2=k_3$. Let us examine this equation in the squeezed limit: $k_1 \ll k_2 \simeq k_3$, i.e. $k_2/k_1\gg 1$. The triangle is close to isosceles so one would expect $28 F_2$ to have value close to $28 F_2^\mr{iso}$. This is however severely not the case. Indeed, $28 F_2^\mr{iso}$ is a finite number of order $\mathcal{O}(1)$, whereas the first three terms are each separately large numbers of order $\mathcal{O}((k_2/k_1)^2)$. These three numbers cancel each other exactly in the isosceles case, but in the general case they only cancel to order $\mathcal{O}((k_2/k_1)^2)$, not necessarily to the lower orders $\mathcal{O}(k_2/k_1)$ and $\mathcal{O}(1)$. For instance in the case of a flattened triangle, $k_3=k_2+k_1$, some algebra shows that the sum of the three terms does leave a diverging term, $14 k_2/k_1$, as well as a constant term, $15$, and other subleading terms. In other words, in the squeezed limit, the $F_2$ kernel (and thus the associated 3D bispectrum) violently depends on the angles, it may be highly different from the (squeezed) isosceles value.

This issue can affect the projection of the angular bispectrum. Indeed, let us examine an angular configuration in the isosceles squeezed limit: $\lu \ll \ld=\lt \gg 1$. With Limber's approximation, $k_i\propto\ell_i$ and the 3D triangle is also squeezed isosceles. Thus one will infer a value for the angular bispectrum proportional to the isosceles squeezed  $F_2^\mr{iso,sqz}=\frac{13}{28}$. However in the exact projection with Bessel functions, Eq.~\ref{Eq:bispectrum-3D-to-angular-n=0}, the integral will sample $k_i$ values (slightly) away from the Limber values and thus 3D triangle that are not isosceles, for which the three terms will not cancel and the $F_2$ value will be highly different from $F_2^\mr{iso,sqz}$. After integration, one can thus expect that the exact angular bispectrum will differ from the Limber value, i.e. that Limber's approximation failed.

This is not only an expectation, and in fact we are not the first to discuss this cancellation issue. It has indeed been discussed and demonstrated numerically by \cite{Lee2020}. In that article they were studying the related but more complex case of the angular trispectrum (4-point function) in the squeezed-diagonal limit of relevance for power spectrum covariances. In 3D this corresponds to four Fourier modes $\kk_1 \cdots \kk_4$ which form a quadrilateron ($\kk_1 + \kk_2 + \kk_3 + \kk_4=0$) where one diagonal is much smaller than the sides: $s=|\kk_1 + \kk_2| \ll k_1\simeq k_2$ and $s\ll k_3\simeq k_4$. Effectively this is a quadrilateron made of two squeezed triangles joined on the short leg. In this case indeed, the result from Limber's approximation is vastly different from the exact projection performed with FFTlog (see for instance Fig. 11 of \cite{Lee2020}), and it is reported that this failure came from terms sensitive to the short leg that scale as $1/s^4$ or $1/s^2$. They specifically showed that this came from the cancellation issue: as shown in their Table 3 in the exact projection the three terms do not cancel exactly but leave a remnant 4-5 orders of magnitude smaller than each separate term. In other words this is a numerical issue of subtle cancellations.

In conclusion, the trigonometric approach is problematic for both projection methods. For the exact projection, it leads to subtle numerical cancellations which require high precision calculations and thus high numerical cost. This is particularly problematic because, even for other bispectrum terms, the exact methods are already more numerically demanding than Limber's approximation. And for Limber's approximation, the trigonometric approach yields inadequate cancellations and gives a plain wrong result in total.\\
It is thus desirable to find a projection approach which circumvents these cancellation issues, yielding equations that require less numerical precision for the exact projection and where Limber's approximation can be applied without issue.

\subsubsection{Inconsistency with Super-Sample Covariance literature}\label{Sect:context-SSC-issues}

There is another, related, context where the trigonometric + Limber approach is problematic: Super-Sample Covariance. Here the difference between approaches may not be as dramatic as in the previous section, i.e. orders of magnitude, but it is still or order $\mathcal{O}(1)$ as we will see. This is problematic enough since SSC is the main source of statistical error beyond the simple Gaussian covariance and thus mandate precise calculations.

To study this, let us examine the covariance associated to the squeezed bispectrum. Indeed, \cite{Lacasa2016} showed that the covariance between cluster counts and galaxy angular power spectra is proportional to the angular bispectrum $b^\mr{hgg}$ in the isosceles squeezed limit. More generally and with the notations of this article, in full sky the covariance between the counts $N_\mr{t_1}$ of a first tracer and the power spectrum $C_\ell^\mr{n_{t_{23}}}$ of two other tracers is
\be
\Cov\left(N_\mr{t_1},C_\ell^\mr{n_{t_{23}}}\right) = \frac{b_{0,\ell,\ell}^\mr{t_{123}}}{4\pi}. \label{Eq:bispectrum-to-Cov-N-Cl}
\ee
Here we must digress into a technical point: since the first multipole is the monopole, $\lu=0$, one cannot apply Limber's approximation on it. Instead one must use a mixed projection that is exact on $k_1$ and uses Limber on $k_2$ and $k_3$. In the angle-independent case, this changes Eq.~\ref{Eq:bispectrum-3D-to-angular-n=0-Limber} into
\ba
b_{0,\ell,\ell}^{n_\mr{t_{123}}}  =& \int  r_1^2 \, \dd r_1 \ k_1^2 \, \dd k_1 \ x^2 \, \dd x \ B_\mr{n_{t_{123}}}(k_1,k_{\ell},k_{\ell}|r_1,x,x) \ j_{0}(k_1 r_1) \, j_{0}(k_1 x). \label{Eq:bispectrum-3D-to-angular-n=0-mixed}
\ea
Using this, \cite{Lacasa2016} showed the appearance of covariance terms known as Super-Sample Covariance with the form
\be
\Cov_\mr{SSC}\left(N_\mr{t_1},C_\ell^\mr{n_{t_{23}}}\right) = \int r^2 \, \dd r \ x^2 \, \dd x \ \frac{\partial n_\mr{t_1}}{\partial \delta_b} \frac{\partial P_\mr{t_{23}}(k_\ell)}{\partial \delta_b} \ \sigma^2(r,x) \label{Eq:SSC-deltab-general-form}
\ee
with
\ba
\sigma^2(r,x) = \frac{1}{2\pi^2} \int k^2 \, \dd k \ P(k|r,x) \ j_{0}(k r) \, j_{0}(k x).
\ea
Now to reach this point and find the values of the response $\frac{\partial P(k)}{\partial \delta_b}$ we need to tackle the angle-dependent terms in the kernel $F_2$.

If we use the trigonometric approach, we evaluate the kernel on the multipoles' triangle shape i.e. the isosceles squeezed limit: $F_2^\mr{iso,sqz} = \frac{13}{28}$, where we got contributions from $n=0$ and $n=1$. Plugging this into the covariance equation (Eq.~\ref{Eq:bispectrum-to-Cov-N-Cl}), we do find a covariance of the form Eq.~\ref{Eq:SSC-deltab-general-form},  with a power spectrum response
\be
\left(\frac{\partial P(k)}{\partial \delta_b}\right)_{\mr{trigo}} = 4 \ F_2^\mr{iso,sqz} \ P(k) = \frac{13}{7} P(k),
\ee
or equivalently
\be\label{Eq:ln-power-spectrum-response-trigonometric}
\left(\frac{\partial \ln P(k)}{\partial \delta_b}\right)_{\mr{trigo}} = \frac{13}{7} .
\ee
And here lies the problem. Indeed, this power spectrum response prediction is in $\mathcal{O}(1)$ disagreement with the seminal 3D SSC result \cite{Takada2013,Li2014}: 
\be\label{Eq:ln-power-spectrum-response-seminal}
\left(\frac{\partial \ln P(k)}{\partial \delta_b}\right)_{\mr{seminal}} = \frac{68}{21}-\frac{1}{3}\frac{\dd \ln k^3 P(k)}{\dd \ln k}.
\ee
A closer look at the derivation in \cite{Takada2013} shows that the $\frac{68}{21}$ (the so-called growth-only part) comes from taking the azimuthal-angle average of the kernel:
\be
F_2^\mr{avg} = \int_0^1 F_2(\kk_1,\kk_2) \ \dd \cos\theta_{12} = \frac{5}{7} + 0 + \frac{2}{7} \times \frac{1}{3} = \frac{17}{21}
\ee
which takes contribution from $n=0$ and $n=2$, while the derivative term (the so-called dilation term) derived in \cite{Li2014} comes from $n=1$.

Furthermore, later literature \cite{Akitsu2017,Li2018} showed the appearance of another 3D SSC term: the tidal term that comes from $n=2$. This term is not sensitive to the covariance of the background density $\delta$ 
but to that of the traceless tidal tensor $s_{i,j} = (\partial_i\partial_j -\frac{1}{3} \delta^K_{i,j}) \Phi$ (where $\Phi$ is the gravitational potential).

In summary, the trigonometric approach yields a constant power spectrum response that takes contribution from $n=0$ and $n=1$, but the 3D SSC literature yields a different and much more complex response that takes contribution from $n=0,1,2$. The $n=0$ (angle-independent) contribution is the same but the $n=1,2$ (angle-dependent) contributions differ.

From this and results FROM \cite{Lee2020}, one could be tempted to just discard the trigonometric approach in favour of the 3D SSC results. However, we know that the 3D SSC results are not sufficient to predict the covariance of real observables that sit on the light cone. Indeed, the seminal 3D results projected in the flat sky limit yield a SSC equation that is a single redshift integral \cite{Takada2013}, while \cite{Lacasa2016} showed that in full sky this is instead a double redshift integral as in Eq.~\ref{Eq:SSC-deltab-general-form} \footnote{And more generally, \cite{Lacasa2018} derived the SSC equation in arbitrary partial sky coverage, also finding a double redshift integral.}.\\
Moreover, applying the dilation term in the 3D seminal response, Eq.~\ref{Eq:ln-power-spectrum-response-seminal}, to light cone observables raises two questions: \\
(i) Why do we only have a spatial derivative and not some temporal derivative ? This is suspicious in the context where this term was derived in 3D, i.e. on an hyper-surface of constant time. \\
(ii) Since $n=1$ yields this first derivative term, why is there no second derivative term coming from $n=2$ ? \\
Furthermore, the tidal term from $n=2$ raises an additional question : why do only two independent background variables come into play (the background density and tidal tensor) ? In other words, why does $n=2$ gives rise to a new background variable but not $n=1$ ?

In conclusion of this subsection, we need a rigorous computation of the $n=1-2$ terms of the angular bispectrum. This computation cannot use the trigonometric approach, it should remain valid in the squeezed limit of relevance for SSC, and it should account for the fact that observables sit on the light cone. It should also find how to deal with the $n=1-2$ contributions in such a way that we can apply Limber approximation at the end, both for numerical speed and for comparison with the SSC literature.

This is exactly what we perform in the sections below, assuming full sky geometry. We start with the analytical computation of the angular bispectrum (section~\ref{Sect:way-spin-spherical-harmonics}), finding numerically tractable forms with Limber approximation (section~\ref{Sect:tractable-forms}), and apply this to the covariance (section~\ref{Sect:application-covariance}).

\section{Angular bispectrum}\label{Sect:way-spin-spherical-harmonics}

In this section we derive the projection of the 3D bispectrum from Sect.~\ref{Sect:context-bispectrum} into the 2D angular bispectrum. To this end, we first develop in Sect.~\ref{Sect:2+1-decomposition} an analytical computation using spin-weighted spherical harmonics that effectively removes the $\kk_i\cdot\kk_j$ dot products and leaves Bessel derivatives instead. Then in Sect.~\ref{Sect:tractable-forms} we treat these Bessel derivatives through integrations by parts and find equations that are tractable numerically with or without Limber's approximation. 

\subsection{Bispectrum from the 2+1 decomposition} \label{Sect:2+1-decomposition}
\subsubsection{General strategy}
The angular bispectrum is the projection of the 3D bispectrum on the celestial sphere. It is defined through Eq.~\ref{Eq:bispectrum-3D-to-angular} that we recall here after replacing the Dirac distribution $\delta_D(\kk_1+\kk_2+\kk_3)$ by an auxiliary integral $\int \dd^3\xx \, e^{i(\kk_1+\kk_2+\kk_3)\cdot \xx}$ :
\ba
\nonumber \lbra a_{\boldsymbol{\ell}_1} a_{\boldsymbol{\ell}_2} a_{\boldsymbol{\ell}_3} \rbra =& \  i^{\lu+\ld+\lt} \, (4\pi)^3 \!\int r^2_{123} \, \dd r_{123}  \ \frac{\dd^3\kk_{123}}{(2\pi)^9} \ j_{\lu}(k_1 r_1) \, j_{\ld}(k_2 r_2) \, j_{\lt}(k_3 r_3)Y^*_{\boldsymbol{\ell}_1}(\hk_1) \, Y^*_{\boldsymbol{\ell}_2}(\hk_2) \, Y^*_{\boldsymbol{\ell}_3}(\hk_3) \\
& \times\!2 \sum_{n=0}^2 \int\dd^3\xx\Big[\tilde{B}_n(k_{123},r_{123})\frac{\left(\kk_1\cdot\kk_2\right)^n}{k_1^nk_2^n}+2 \ \mr{perm.}\Big]\, e^{i(\kk_1+\kk_2+\kk_3)\cdot \xx} . \label{eq:bispecgen}
\ea
The coefficients $\tilde{B}_n$ are given in the previous section and depend on the norms $k_1$, $k_2$, and $k_3$ of the wave-vectors, and on the comoving distances $r_i$.  

Our strategy then is to work order-by-order in $n$ and introduce a systematic way to perform all the angular integrals either in real space, i.e. integrals on $\hat\xx$, or in Fourier space, i.e. integrals on $\hat\kk$. The key difficulty is the angular integration of the wave-vectors because of the inner-dot products, i.e. $(\kk_1\cdot\kk_2)^n$. The usual way to treat them is the trigonometric approach which expresses $(\kk_1\cdot\kk_2)$ as a function of the three norms $k_1$, $k_2$ and $k_3$. The new route that we develop here consists in replacing $\kk e^{i\kk\cdot\xx}$ by $(-i)\boldsymbol{\nabla} e^{i\kk\cdot\xx}$, where the gradient $\boldsymbol{\nabla}$ is acting on $\xx$. One then plugs the Rayleigh expansion\footnote{We remind that the Rayleigh expansion reads
\ba
    e^{i\kk\cdot\xx}=4\pi\displaystyle\sum_{\ell,m}i^\ell j_\ell(kx)Y_{\ell m}(\hat{\kk})Y^\star_{\ell m}(\hat{\xx}). \label{eq:rayleigh}
\ea} and lets the gradient operates on it. We have to work here in the spherical coordinates system $(x,\theta,\varphi)$ and use the spherical basis vectors $(\mathbf{e}_x,\mathbf{e}_\theta,\mathbf{e}_\varphi)$, where $\mathbf{e}_x$ is along the radial direction aligned with $\xx$, and $(\mathbf{e}_\theta,\mathbf{e}_\varphi)$ are the basis vectors of the plane orthogonal to $\xx$ (or equivalently the plane which is tangent to the sphere of radius $x$). Gradient along the radial direction will operate on the Bessel function $j_\ell(k_ix)$ of the Rayleigh expansion. Gradient in the plane orthogonal to the radial direction will operate on the spherical harmonics $Y_{\ell m}(\hat\xx)$ of the Rayleigh expansion. Since gradients never operate on the wave-vectors, one can re-express $\kk e^{i\kk\cdot\xx} $ as follows
\ba
	-i\kk_\mu e^{i\kk_\nu\xx^\nu} = &(4\pi)\ds\sum_{\boldsymbol{L}_i}(i)^LY_{\boldsymbol{L}_i}(\hat\kk_i) \, \nabla_\mu\left[j_{L_i}(k_ix)Y^\star_{\boldsymbol{L}_i}(\hat\xx)\right].
\ea
In the above, we introduce Einstein's notations where $\mu$ runs over the three coordinates $(x,\theta,\varphi)$. In this coordinate system, the metric tensor is $g_{\mu\nu}=\mathrm{diag}\left[1, \,x^2,x^2\sin^2\theta\right]$ and $\nabla_\mu$ stands for the components of the covariant derivatives.
	
Using this, one obtains
\ba
	\!\int\dd^3\xx \left(\kk_1\cdot\kk_2\right)^ne^{i(\kk_1+\kk_2+\kk_3)\cdot \xx} = &(4\pi)^3\ds\sum_{\boldsymbol{L}_{123}}(i)^{L_1+L_2+L_3}Y_{\boldsymbol{L}_1}(\hat\kk_1) Y_{\boldsymbol{L}_2}(\hat\kk_2) Y_{\boldsymbol{L}_3}(\hat\kk_3) \nonumber \\
	& \times\!\int\dd^3\xx  (-1)^n F^{(n)}_{\boldsymbol{L}_{123}}\left(k_{123},\xx\right), \label{eq:innnerdotexp}
\ea 
where the functions $F^{(n)}_{\boldsymbol{L}_{123}}\left(k_{123},\xx\right)$ do not depend anymore on the angular part of the wave-vectors. Using Einstein's notation and covariant derivatives, they are given by
\ba
	F^{(0)}_{\boldsymbol{L}_{123}} = & \left[j_{L_1}(k_1x) \ j_{L_2}(k_2x) \ j_{L_1}(k_2x) \right] \left[Y_{\boldsymbol{L}_1}^\star(\hat{x}) \ Y_{\boldsymbol{L}_2}^\star(\hat{x}) \ Y_{\boldsymbol{L}_3}^\star(\hat{x}) \right], \label{eq:Fzero}\\
	F^{(1)}_{\boldsymbol{L}_{123}} = & g^{\mu\nu}\nabla_\mu\left[j_{L_1}(k_1x)Y_{\boldsymbol{L}_1}^\star(\hat{x})\right]\nabla^\nu\left[j_{L_2}(k_2x)Y_{\boldsymbol{L}_2}^\star(\hat{x})\right] \, \left[j_{L_3}(k_3x)Y_{\boldsymbol{L}_3}^\star(\hat{x})\right], \label{eq:Fone} \\
	F^{(2)}_{\boldsymbol{L}_{123}} = & g^{\mu\lambda}g^{\nu\kappa}\nabla_\mu\nabla_\nu\left[j_{L_1}(k_1x)Y_{\boldsymbol{L}_1}^\star(\hat{x})\right]\nabla_\lambda\nabla_\kappa\left[j_{L_2}(k_2x)Y_{\boldsymbol{L}_2}^\star(\hat{x})\right] \, \left[j_{L_3}(k_3x)Y_{\boldsymbol{L}_3}^\star(\hat{x})\right]. \label{eq:Ftwo}
\ea
We stress that for the first-order term given in the second line, the covariant derivatives operate on a scalar function. Hence they reduce to partial derivatives. For the second-order term however, given in the third line, two covariant derivatives operate successively on a scalar function. It is thus mandatory to keep the full covariant derivative in this case.

By plugging Eq. (\ref{eq:innnerdotexp}) in the general expression of the angular bispectrum, Eq. (\ref{eq:bispecgen}), we can easily perform the three integrals over the angular part of each wave-vector, $\hk_i$. Indeed, each integral reduces to the orthonormality conditions for the spherical harmonics and thus leads to $\delta_{\ell_i,L_i}\, \delta_{m_i,M_i}$ for $i=1,\,2$ and 3. Hence, the angular bispectrum boils down to
\ba
\nonumber \lbra a_{\boldsymbol{\ell}_1} a_{\boldsymbol{\ell}_2} a_{\boldsymbol{\ell}_3} \rbra = &  \left(\frac{2}{\pi}\right)^3 \!\int r^2_{123} \, \dd r_{123}  \ {k^2_{123}\dd k_{123}} \ j_{\lu}(k_1 r_1) \, j_{\ld}(k_2 r_2) \, j_{\lt}(k_3 r_3) \\
&  \times2\sum_{n=0}^2 \Big[\left(\frac{-1}{k_1k_2}\right)^n\tilde{B}_n(k_{123},r_{123})\!\int\dd^3\xx \ F^{(n)}_{\boldsymbol{\ell}_{123}}\left(k_{123},\xx\right)+2 \ \mr{perm.}\Big]. \label{eq:3ptsell}
\ea
We stress that in the first line, the integral over the three wave-vectors (involving 9 integrals) is now reduced to three integrals over each norm of the wave-vectors. 

As it is, the whole difficulty has been transferred to the calculations of $f^{(n)}_{\boldsymbol{\ell}_{123}}(k_{123})\equiv\!\int\dd^3\xx \, F^{(n)}_{\boldsymbol{\ell}_{123}}\left(k_{123},\xx\right)$. In the next subsection we show order-by-order that the $f^{(n)}_{\boldsymbol{\ell}_{123}}(k_{123})$'s can be expressed using Gaunt integrals, resulting from the angular integral over $\hat{\xx}$, multiplied by a remaining integral over the radial direction $x$.
To do so, we work in the spherical coordinate system, and split the 3-dimensional covariant derivative into a radial part and an angular part. We then express angular derivatives using spin-raising and spin-lowering operators on the sphere. This makes spin-weighted spherical harmonics arise, with spin-$(\pm1)$ and spin-($\pm2$) since we go up to second derivatives.
The mathematical materials needed for such computations are presented in App.~\ref{app:math}. In details, the appendix contains a presentation of geometry on the sphere, the computation of (double) covariant derivatives in spherical coordinate system of $\mathbb{R}^3$, a presentation of spin-operators, spin-weighted spherical harmonics, and Gaunt integrals. Finally, the presence of spin-weighted spherical harmonics in the calculation leads to Gaunt integrals with non-zero spins. However, the angular bispectrum is defined with a Gaunt integral with spins zero. Hence, in App.~\ref{apps:spinto0} we derive two identities relating Gaunt integrals with spin $s=\pm1$ and $s=\pm2$ to Gaunt integrals with spins zero. 

\subsubsection{Angular integrals}\label{Sect:angular-integrals}
The derivation of the $f^{(n)}_{\boldsymbol{\ell}_{123}}(k_{123})$ functions is rather lengthy. The details of the calculation are deferred to App. \ref{app:Fn} and we provide here a summary. 

The case $n=0$ is rather straightforward. Starting from Eq. (\ref{eq:Fzero}) one directly recognizes spin-0 Gaunt integral. It yields 
\ba
    f^{(0)}_{\boldsymbol{\ell}_{123}}(k_{123})=\Gamma_{\ell_1;\ell_2;\ell_3}\left(\begin{array}{ccc}
		\ell_1 & \ell_2 & \ell_3 \\
		0 & 0 & 0
	\end{array}\right)\left(\begin{array}{ccc}
		\ell_1 & \ell_2 & \ell_3 \\
		m_1 &m_2 & m_3
	\end{array}\right)\displaystyle\int x^2\dd x \, j_{\ell_1}(k_1x)j_{\ell_2}(k_2x)j_{\ell_3}(k_3x),
\ea
where $\Gamma_{\ell_1;\ell_2;\ell_3}=\sqrt{(2\ell_1+1)(2\ell_2+1)(2\ell_3+1)/4\pi}$. 

For $n=1$, Eq. (\ref{eq:Fone}) is split into two pieces. The first piece implies radial derivatives only. Hence we are left with a spin-0 Gaunt integral. It leads to a contribution to $f^{(1)}$ reading
\ba
    f^{(1),S}_{\boldsymbol{\ell}_{123}}(k_{123})=\Gamma_{\ell_1;\ell_2;\ell_3}\left(\begin{array}{ccc}
		\ell_1 & \ell_2 & \ell_3 \\
		0 & 0 & 0
	\end{array}\right)\left(\begin{array}{ccc}
		\ell_1 & \ell_2 & \ell_3 \\
		m_1 &m_2 & m_3
	\end{array}\right)\displaystyle\int x^2\dd x\,\left[{\partial_x j_{\ell_1}(k_1x)}\right]\,\left[{\partial_x j_{\ell_2}(k_2x)}\right]\,j_{\ell_3}(k_3x).
\ea
This first piece will be referred to as a {\it scalar} contribution in the sense that it derives from spherical harmonics with spin zero (note that with this terminology, the $n=0$ case is a scalar too). Hence, it is to be understood as a scalar contribution on the celestial sphere. The second piece implies angular derivatives only which thus operate on the spherical harmonics. This gives rise to a Gaunt integral with spins $s=\pm1$. Because of this, this contribution will be dubbed as a {\it vector} contribution where the vectorial nature is to be understood as vectors on the celestial sphere. We use the identities presented in App. \ref{apps:spinto0} to turn the spin-$(\pm1)$ Gaunt integrals into spin-0 Gaunt integrals. This yields a contribution of the form
\ba
    f^{(1),V}_{\boldsymbol{\ell}_{123}}(k_{123})=\Gamma_{\ell_1;\ell_2;\ell_3}\left(\begin{array}{ccc}
		\ell_1 & \ell_2 & \ell_3 \\
		0 & 0 & 0
	\end{array}\right)\left(\begin{array}{ccc}
		\ell_1 & \ell_2 & \ell_3 \\
		m_1 &m_2 & m_3
	\end{array}\right)\Lambda^\mathrm{V}_{\ell_{123}}\displaystyle\int x^2\dd x\,\left[\frac{ j_{\ell_1}(k_1x)}{x}\right]\,\left[\frac{ j_{\ell_2}(k_2x)}{x}\right]\,j_{\ell_3}(k_3x),
\ea
where the coefficient $\Lambda^V_{\ell_{123}}$ can be found in Tab. \ref{Table:ellcoeff}. We note that there is no term mixing radial derivative and angular derivative for $n=1$ because the metric is diagonal. 

\begin{table}
\begin{equation}
\begin{array}{ll} \hline\hline
    & \\
    \Delta^{(s)}_\ell =& \displaystyle\frac{(\ell+s)!}{(\ell-s)!} \\
    \\
    \hline
	& \\
	\Lambda^{\mathrm{V}}_{\ell_{123}} = & \displaystyle\frac{1}{2}\left(\Delta^{(1)}_{\ell_1}+\Delta^{(1)}_{\ell_2}-\Delta^{(1)}_{\ell_3}\right)\\
	& \\
	\Lambda^{\mathrm{T}}_{\ell_{123}} = & \displaystyle\frac{1}{8}\left(\frac{\Delta^{(2)}_{\ell_1}}{\Delta^{(1)}_{\ell_1}}+\frac{\Delta^{(2)}_{\ell_2}}{\Delta^{(1)}_{\ell_2}}+\Delta^{(1)}_{\ell_1}+\Delta^{(1)}_{\ell_2}-2\Delta^{(1)}_{\ell_3}\right)\left(\Delta^{(1)}_{\ell_1}+\Delta^{(1)}_{\ell_2}-\Delta^{(1)}_{\ell_3}\right) \\
	& \\ \hline\hline
\end{array} \nonumber
\end{equation} 
\caption{List of $\ell_{123}$-dependent coefficients involved in the radial integrands of the reduced bispectrum, $f^{(n)}_{\boldsymbol{\ell}_{123}}(k_{123})$. We note that $\Delta^{(1)}_\ell=\ell(\ell+1)$ and $\Delta^{(2)}_\ell/\Delta^{(1)}_\ell=(\ell-1)(\ell+2)$. Hence $\Lambda^{\mathrm{V}}_{\ell_{123}}$ is quadratic in $\ell_{123}$ as expected from the application of two angular derivatives, while $\Lambda^{\mathrm{T}}_{\ell_{123}}$ is quartic in $\ell_{123}$ since it follows from four angular derivatives.}
\label{Table:ellcoeff}
\end{table}

The case $n=2$ is much more involved. We start by splitting Eq. (\ref{eq:Ftwo}) into four terms. The first one solely implies radial derivative. Hence, we are left with a spin-0 Gaunt integral. It gives a first contribution to $f^{(2)}$ which is of scalar-type and reads
\ba
    f^{(2),S,a}_{\boldsymbol{\ell}_{123}}(k_{123})= \Gamma_{\ell_1;\ell_2;\ell_3}\left(\begin{array}{ccc}
		\ell_1 & \ell_2 & \ell_3 \\
		0 & 0 & 0
	\end{array}\right)\left(\begin{array}{ccc}
		\ell_1 & \ell_2 & \ell_3 \\
		m_1 &m_2 & m_3
	\end{array}\right)\displaystyle\int x^2\dd x\,\left[\partial^2_x j_{\ell_1}(k_1x)\right]\,\left[\partial^2_x j_{\ell_2}(k_2x)\right]\,j_{\ell_3}(k_3x). \label{eq:2scalrad}
\ea
The second and the third terms mix one radial derivative with one angular derivative. These give rise to a vector contribution to $f^{(2)}$ which reads
\ba
    f^{(2),V}_{\boldsymbol{\ell}_{123}}(k_{123})= &\Gamma_{\ell_1;\ell_2;\ell_3}\left(\begin{array}{ccc}
		\ell_1 & \ell_2 & \ell_3 \\
		0 & 0 & 0
	\end{array}\right)\left(\begin{array}{ccc}
		\ell_1 & \ell_2 & \ell_3 \\
		m_1 &m_2 & m_3
	\end{array}\right) \\
 &\times\Lambda^\mathrm{V}_{\ell_{123}}\displaystyle\int x^2\dd x\left(\frac{2}{x^2}\right)\left[\frac{\partial j_{\ell_1}(k_1x)}{\partial x}-\frac{ j_{\ell_1}(k_1x)}{x}\right]\left[\frac{\partial j_{\ell_2}(k_2x)}{\partial x}-\frac{ j_{\ell_2}(k_2x)}{x}\right]j_{\ell_3}(k_3x), \nonumber 
\ea
which can be further simplified into
\ba
    f^{(2),V}_{\boldsymbol{\ell}_{123}}(k_{123})= &\Gamma_{\ell_1;\ell_2;\ell_3}\left(\begin{array}{ccc}
		\ell_1 & \ell_2 & \ell_3 \\
		0 & 0 & 0
	\end{array}\right)\left(\begin{array}{ccc}
		\ell_1 & \ell_2 & \ell_3 \\
		m_1 &m_2 & m_3
	\end{array}\right) \\
 &\times2\Lambda^\mathrm{V}_{\ell_{123}}\displaystyle\int x^2\dd x\,\left\{\partial_x \left[\frac{j_{\ell_1}(k_1x)}{x}\right]\right\}\,\left\{\partial_x \left[\frac{j_{\ell_2}(k_2x)}{x}\right]\right\}\,j_{\ell_3}(k_3x). \nonumber 
\ea
We note that the vector contribution arising at $n=2$ corresponds to the radial derivative of the vector contribution arising at $n=1$.\\
The fourth term implies 3D covariant derivatives all projected along angular directions. However, it is important to keep in mind that in full generality the 3-dimensional covariant derivative projected on the celestial sphere is not equal to the 2-dimensional covariant derivative on the sphere. Indeed, it is only true for covariant derivative of a scalar function but here we have two successive covariant derivatives on a scalar function, hence one covariant derivative of a vector field. The Christoffel symbols give rise to additional derivatives along the radial direction in addition to covariant derivatives on the sphere [see Eq. (\ref{eq:covang})]. Moreover, the application of two successive covariant derivatives on the sphere yields a symmetric tensor whose pure-trace part is made of spin-$0$ spherical harmonics and whose traceless part is made of spin-$(\pm2)$ spherical harmonics. Because of these reasons, the last term will have two contributions. The first one is of scalar type since it implies a spin-0 Gaunt integral. It reads
\ba
    &f^{(2),S,b}_{\boldsymbol{\ell}_{123}}(k_{123}) =\Gamma_{\ell_1;\ell_2;\ell_3}\left(\begin{array}{ccc}
		\ell_1 & \ell_2 & \ell_3 \\
		0 & 0 & 0
	\end{array}\right)\left(\begin{array}{ccc}
		\ell_1 & \ell_2 & \ell_3 \\
		m_1 &m_2 & m_3
	\end{array}\right)  \label{eq:2scalang} \\
 &\ \times \displaystyle\int x^2\dd x \left(\frac{2}{x^2}\right) \left[\partial_x j_{\ell_1}(k_1x)-\frac{\ell_1(\ell_1+1)}{2}\frac{j_{\ell_1}(k_1x)}{x}\right] \left[\partial_x j_{\ell_2}(k_2x)-\frac{\ell_2(\ell_2+1)}{2}\,\frac{j_{\ell_2}(k_2x)}{x}\right] j_{\ell_3}(k_3x).\nonumber
\ea
Derivatives of the Bessel functions come from the additional radial derivatives while the term proportional to $\ell(\ell+1)j_\ell$ comes from the pure-trace part of two successive covariant derivatives on the sphere. The second contribution implies spin-$(\pm2)$ Gaunt integrals and it will be dubbed as a {\it tensor} contribution where the tensorial nature applies on the celestial sphere. We further make use of the identity derived in App. \ref{apps:spinto0} to express spin-$(\pm2)$ Gaunt integrals as functions of spin-0 Gaunt integrals. The tensor contribution to $f^{(2)}$ finally reads
\ba
    f^{(2),T}_{\boldsymbol{\ell}_{123}}(k_{123}) = \Gamma_{\ell_1;\ell_2;\ell_3}\left(\begin{array}{ccc}
		\ell_1 & \ell_2 & \ell_3 \\
		0 & 0 & 0
	\end{array}\right)\left(\begin{array}{ccc}
		\ell_1 & \ell_2 & \ell_3 \\
		m_1 &m_2 & m_3
	\end{array}\right)\Lambda^\mathrm{T}_{\ell_{123}}\displaystyle\int x^2\dd x\,\left[\frac{j_{\ell_1}(k_1x)}{x^2}\right]\,\left[\frac{j_{\ell_2}(k_2x)}{x^2}\right]\,j_{\ell_3}(k_3x),
\ea
where the expression of $\Lambda^\mathrm{T}_{\ell_{123}}$ is given in Tab. \ref{Table:ellcoeff}.

\subsubsection{Reduced spherical bispectrum}

The reduced bispectrum, $b_{\ell_1,\ell_2,\ell_3}$, is defined through 
\ba
	\left<a_{\boldsymbol{\ell}_1}a_{\boldsymbol{\ell}_2}a_{\boldsymbol{\ell}_3}\right> =  \Gamma_{\ell_1;\ell_2;\ell_3}\left(\begin{array}{ccc}
		\ell_1 & \ell_2 & \ell_3 \\
		0 & 0 & 0
	\end{array}\right)\left(\begin{array}{ccc}
		\ell_1 & \ell_2 & \ell_3 \\
		m_1 &m_2 & m_3
	\end{array}\right) \ b_{\ell_1,\ell_2,\ell_3}.
\ea
Starting from Eq. (\ref{eq:3ptsell}) and noticing that for all $n$, the quantities $\int\dd^3\xx \ F^{(n)}_{\boldsymbol{\ell}_{123}}$ can be written as
\ba
	\!\int\dd^3\xx \ F^{(n)}_{\boldsymbol{\ell}_{123}}=\Gamma_{\ell_1;\ell_2;\ell_3}\left(\begin{array}{ccc}
		\ell_1 & \ell_2 & \ell_3 \\
		0 & 0 & 0
	\end{array}\right)\left(\begin{array}{ccc}
		\ell_1 & \ell_2 & \ell_3 \\
		m_1 &m_2 & m_3
	\end{array}\right)\!\int x^2\dd x\, I^{(n)}_{\ell_{123}}(k_{123},x)
\ea
where $I^{(n)}_{\ell_{123}}(k_{123},x)$'s can be found in App. \ref{app:Fn}, the reduced bispectrum boils down to
\ba
	b_{\ell_1,\ell_2,\ell_3} = & \left(\frac{2}{\pi}\right)^3\!\int \left(r^2_{123} \, \dd r_{123}\right)  \ \left({k^2_{123}\dd k_{123}}\right) \ j_{\lu}(k_1 r_1) \, j_{\ld}(k_2 r_2) \, j_{\lt}(k_3 r_3) \nonumber \\
& \times 2\sum_{n=0}^2 \Big[(-1)^n \tilde{B}_n(k_{123},r_{123})\!\int x^2\dd x \, \mathcal{I}^{(n)}_{\ell_{123}}(k_{123},x)+2 \ \mr{perm.}\Big], \label{eq:redbispectrum}
\ea
where $\mathcal{I}^{(n)}_{\ell_{123}}(k_{123},x) \equiv {I}^{(n)}_{\ell_{123}}(k_{123},x)/(k_1k_2)^n$. We note that the permutations on the second line operates on the $k_i$'s, $r_i$'s and $\ell_i$'s. There are obviously many terms in the second line of the above. Hence we summarize them in Tab. \ref{Table:radkernels}. We note that $\mathcal{I}^{(2)\mr{S}}_{\ell_{123}}$ is given by the sum of the radial integrands in Eqs. (\ref{eq:2scalrad}) \& (\ref{eq:2scalang}). This is further simplified in Tab. \ref{Table:radkernels} by using the differential equation satisfied by Bessel functions, i.e. 
\ba\label{Eq:differential-equation-spherical-Bessel}
\left[\partial_x^2 + \frac{2}{x} \partial_x  - \frac{\ell(\ell+1)}{x^2}\right] j_\ell(kx)= -k^2 j_\ell(kx).
\ea
It is worth stressing that $\mathcal{I}^{(n)}_{\ell_{123}}(k_{123},x)$'s are functions of $k_1x$, $k_2x$ and $k_3x$ only.

\begin{table}
\begin{equation}
\begin{array}{ll} \hline\hline
	n=0\,\text{:} &\\ \hline 
	& \\
	\tilde{B}_0 = &   n_\mr{t_1} \, n_\mr{t_2} \, n_\mr{t_3}\ P(k_1|r_{13}) P(k_2|r_{23}) \ b^\mr{t_1}_1 \, b^\mr{t_2}_1 \, b^\mr{t_3}_{(0)} \\ 
		& \\
	\mathcal{I}^{(0)}_{\ell_{123}}= & j_{\ell_1}(k_1x) \ j_{\ell_2}(k_2x) \ j_{\ell_3}(k_3x) \\ 
	& \\ \hline\hline
	n=1\,\text{:} &\\ \hline
	& \\
	\tilde{B}_1 = &   n_\mr{t_1} \, n_\mr{t_2} \, n_\mr{t_3}\ P(k_1|r_{13}) P(k_2|r_{23}) \ b^\mr{t_1}_1 \, b^\mr{t_2}_1 \, b^\mr{t_3}_{(1)} \, q_{(1)} \\ 
	 & \\
	\mathcal{I}^{(1)}_{\ell_{123}}= & \ds \mathcal{I}^{(1),\mathrm{S}}_{\ell_{123}}+\Lambda^{\mathrm{V}}_{\ell_{123}}\,\mathcal{I}^{(1),\mathrm{V}}_{\ell_{123}}  \\ 
	& \\ 
	\text{where\,:} & \\
	& \\
    \mathcal{I}^{(1),\mathrm{S}}_{\ell_{123}}= &\ds\left(\frac{1}{k_1k_2}\right)\left[\partial_x j_{\ell_1}(k_1x)\right]\left[\partial_x j_{\ell_2}(k_2x)\right]j_{\ell_3}(k_3x)  \\ 
    & \\
    \mathcal{I}^{(1),\mathrm{V}}_{\ell_{123}}= &\ds\left(\frac{1}{k_1k_2}\right)\left[\frac{j_{\ell_1}(k_1 x)}{x}\right]\left[\frac{j_{\ell_2}(k_2x)}{x}\right]j_{\ell_3}(k_3x)  \\ 
	& \\ \hline\hline
	n=2\,\text{:} &\\ \hline
	& \\
	\tilde{B}_2 = &   n_\mr{t_1} \, n_\mr{t_2} \, n_\mr{t_3}\ P(k_1|r_{13}) P(k_2|r_{23}) \ b^\mr{t_1}_1 \, b^\mr{t_2}_1 \, b^\mr{t_3}_{(2)} \\
	& \\
	\mathcal{I}^{(2)}_{\ell_{123}}= & \ds \mathcal{I}^{(2),\mathrm{S}}_{\ell_{123}}+\Lambda^{\mathrm{V}}_{\ell_{123}}\,\mathcal{I}^{(2),\mathrm{V}}_{\ell_{123}} +\Lambda^{\mathrm{T}}_{\ell_{123}}\,\mathcal{I}^{(2),\mathrm{T}}_{\ell_{123}}  \\ 
	& \\ 
	\text{where\,:} & \\
	& \\
    \mathcal{I}^{(2),\mathrm{S}}_{\ell_{123}}= & \ds \frac{1}{2}\left[j_{\ell_1}(k_1x) + \frac{1}{k_1^2}\,\partial^2_x j_{\ell_1}(k_1x)\right]\left[j_{\ell_2}(k_2x) + \frac{1}{k_2^2}\,\partial^2_x j_{\ell_2}(k_2x)\right]j_{\ell_3}(k_3x) \\
    & \ds+\frac{1}{(k_1k_2)^2}\left[\partial^2_x j_{\ell_1}(k_1x)\right]\left[\partial^2_x j_{\ell_2}(k_2x)\right]j_{\ell_3}(k_3x) \\
	& \\
	\mathcal{I}^{(2),\mathrm{V}}_{\ell_{123}} = & \ds\frac{2}{(k_1k_2)^2}\left\{\partial_x\left[\frac{j_{\ell_1}(k_1x)}{x}\right]\right\}\left\{\partial_x\left[\frac{j_{\ell_2}(k_2x)}{x}\right]\right\}j_{\ell_3}(k_3x) \\
    & \\
	\mathcal{I}^{(2),\mathrm{T}}_{\ell_{123}} = &\ds\frac{1}{(k_1k_2)^2}\left[\frac{j_{\ell_1}(k_1x)}{x^2}\right]\, \left[\frac{j_{\ell_2}(k_2x)}{x^2}\right]j_{\ell_3}(k_3x) \\
	& \\ \hline\hline
\end{array} \nonumber
\end{equation} 
\caption{List of all the terms and radial integrand involved in the reduced bispectrum, Eq. (\ref{eq:redbispectrum}). The coefficients $\Lambda^{\mathrm{V}}_{\ell_{123}}$ and $\Lambda^{\mathrm{T}}_{\ell_{123}}$ are given in Table~\ref{Table:ellcoeff}.}
\label{Table:radkernels}
\end{table}

\subsection{Numerically tractable forms with and without Limber's approximation}\label{Sect:tractable-forms}

\subsubsection{Projecting Bessel derivatives}\label{Sect:proj-Bessel}

All bispectrum terms are of the form
\ba
b_{\ell_1,\ell_2,\ell_3} = & \left(\frac{2}{\pi}\right)^3\!\int x^2\dd x \ r^2_{123} \, \dd r_{123}  \ k_{123}^2 \, \dd k_{123} \ F(k_{123},r_{123},x) \ j_{\lu}(k_1 r_1) \, j_{\ld}(k_2 r_2) \, j_{\lt}(k_3 r_3) \nonumber \\
& \times \partial_x^{n_1} j_{\lu}(k_1 x) \, \partial_x^{n_2} j_{\ld}(k_2 x) \, \partial_x^{n_3} j_{\lt}(k_3 x)
\ea
for a suitable choice of $F, n_1,n_2,n_3$. To obtain a numerically tractable form we first need to get rid of the derivatives of Bessel functions. This problem is tackled by Appendix~\ref{App:remove-bessel-derivs-IPP} for a general form of the bispectrum, which may be of interest beyond this article. However, the present problem can be further simplified by noting that the bispectrum terms found in Sect.~\ref{Sect:way-spin-spherical-harmonics} have a separable form: $F(k_{123},r_{123},x) = F_{(1)}(k_1,r_1) \, F_{(2)}(k_2,r_2) \, F_{(3)}(k_3,r_3) \, F_{(x)}(x)$, with $F_{(i)}$ taking specific forms. This problem is tackled in Appendix~\ref{App:proj-LSS}. Specifically we find that a bispectrum term with ingredients
\ba
\tilde{B}_n(k_{123},r_{123}) &= n_\mr{t_1} \, n_\mr{t_2} \, n_\mr{t_3}\ P(k_1|r_{13}) \, P(k_2|r_{23}) \ b^\mr{t_1}_1 \, b^\mr{t_2}_1 \, b^\mr{t_3}_{a} \ k_1^\mu \, k_2^\nu \\
\mathcal{I}^{(n),A}_{\ell_{123}} &= \frac{B}{x^b}\frac{\partial^c_x j_{\ell_1}}{k_1^\xi} \ \frac{\partial^d_x j_{\ell_2}}{k_2^\rho} \ j_{\ell_3}.
\ea
yields an angular bispectrum
\ba \label{Eq:proj-LSS-noLimber-bispectrum}
\nonumber b_{\lu,\ld,\lt}^{n,A} = 
\frac{(-1)^{c+d} \, B}{(\lu+\sfrac{1}{2})^\alpha \, (\ld+\sfrac{1}{2})^\beta} \int & x^2 \, \dd x \ n_\mr{t_1} \, n_\mr{t_2} \, n_\mr{t_3} \ b^\mr{t_1}_1 \, b^\mr{t_2}_1 \, b^\mr{t_3}_{a} \ P_1 \, P_2 \\
& \times \mathscr{L}^{t_1,(c)}_{\alpha}(\lu,x) \, \mathscr{L}^{t_2,(d)}_{\beta}(\ld,x) \, \mathscr{G}^{t_3}_{a}(\ell_3,x).
\ea
where $P_i = P(k_i|x)$, $\alpha=\xi-\mu$, $\beta=\rho-\nu$, and
\ba
\mathscr{G}^{t}_{a}(\ell,x) &\equiv \frac{2/\pi}{n_{t}(x) \, b_a^{t}(k_{\ell},x) \, G(x)^2}\int k^2 \, \dd k \ r^2 \, \dd r \ n_{t}(r) \, b_a^{t}(k,r) \, G(r)^2 \ j_{\ell}(k x) \, j_{\ell}(k r).
\ea
As argued in Appendix \ref{App:proj-LSS-LimberCorrespondence} we can assume $\mathscr{G}^{t}_{a}(\ell,x)\approx 1$ at all multipoles.\\
Furthermore in Eq.~\ref{Eq:proj-LSS-noLimber-bispectrum} we have
\ba
\mathscr{L}^{t,(c)}_{\alpha}(\ell,x) &= \frac{2/\pi}{n_{t}(x) \, b_1^{t}(k_{\ell},x) \, G(x) \, P(k_{\ell}) / k_{\ell}^\alpha} \int k^2 \, \dd k \ r^2 \, \dd r \ n_{t} \, b_1^{t} \, G(r) \, P(k) / k^\alpha \nonumber \\
& \quad \times L^{t,(c)}_{\alpha}(k,r) \ j_{\ell}(k x) \, j_{\ell}(k r) \label{Eq:def-scrL^t,c_alpha}
\ea
with
\ba
L^{t,(c)}_{\alpha}(k,r) &= \frac{1}{n_{t} \, b_1^{t} \, G(r) \, P(k) / k^\alpha} \sum_{m_c=0}^c \!\begin{pmatrix}c \\ m_c\end{pmatrix}\! \frac{(-1)^{m_c}}{r^2 \, k^{2+m_c}} \nonumber \\
& \quad \times \frac{\partial^{c}}{\partial k^{c-m_c} \, \partial r^{m_c}} \left[ n_{t} \, b_1^{t} \, G(r) \, P(k)/ k^\alpha \ k^{c+2} \, r^{m_c+2}\right] \label{Eq:def-L^t,c_alpha}
\ea

\subsubsection{Limber's approximation}\label{Sect:proj-Limber}

Here we find the reduction of the previous equations when using Limber's approximation Eq.~\ref{Eq:Limber}, which is of interest for faster numerical computations. The case of a general non-separable bispectrum is treated in Appendix~\ref{App:naive-Limber}. For the bispectrum of interest here, the computation is simplified by separability and is treated in Appendix~\ref{App:proj-LSS}. Specifically, we find that applying Limber's approximation on all multipoles in Eq.~\ref{Eq:proj-LSS-noLimber-bispectrum} yields
\ba \label{Eq:proj-LSS-Limber-bispectrum}
\nonumber b_{\lu,\ld,\lt}^{n,A} \approx 
\frac{(-1)^{c+d} \, B}{(\lu+\sfrac{1}{2})^\alpha \, (\ld+\sfrac{1}{2})^\beta} \int & x^2 \, \dd x \ n_\mr{t_1} \, n_\mr{t_2} \, n_\mr{t_3} \ b^\mr{t_1}_1 \, b^\mr{t_2}_1 \, b^\mr{t_3}_{a} \ P_1 \, P_2 \\
& \times \mathcal{L}^{t_1,(c)}_{\alpha}(\lu,x) \, \mathcal{L}^{t_2,(d)}_{\beta}(\ld,x),
\ea
where again $P_i = P(k_i|x)$, $\alpha=\xi-\mu$, $\beta=\rho-\nu$, and we have the dimensionless quantity
\be
\mathcal{L}^{t,(c)}_{\alpha}(\ell,x) = \left[\frac{k^c}{n_{t}(r_\ell) \, b_1^{t}(k,r_\ell) \, G(r_\ell) \, \frac{P(k)}{k^\alpha} k^{c-1}}\frac{\dd^c}{\dd k^c} \left[ n_{t}(r_\ell) \, b_1^{t}(k,r_\ell) \, G(r_\ell) \, \frac{P(k)}{k^\alpha} k^{c-1}\right]  \right]_{k=(\ell+\sfrac{1}{2})/x},
\label{Eq:def-calL^t,c_alpha}
\ee
with $r_\ell = (\ell+\sfrac{1}{2})/k$.
$\mathcal{L}^{t,(c)}_{\alpha}$ is linked to the quantity $L^{t,(c)}_{\alpha}$ from Eq.~\ref{Eq:def-L^t,c_alpha} by
\ba
\mathcal{L}^{t,(c)}_{\alpha}(\ell,x) = L^{t,(c)}_{\alpha}(k_\ell,x). \label{Eq:relation-calL-L}
\ea
It has recursion properties detailed in Appendix~\ref{App:props-L-symbols}. For our purpose we will need only the orders $c=0, 1$ and 2. In these cases, $\mathcal{L}^{t,(c)}_{\alpha}$ can be split into dilation terms, where the derivative act on spatial wavenumber, and evolution terms, where the derivative acts on comoving distance and hence redshift/time.
\ba
\mathcal{L}^{t,(0)}_{\alpha}(\ell,x) & = 1 \\
\mathcal{L}^{t,(1)}_{\alpha}(\ell,x) & = g_\mr{dil,\alpha}^{t,(1)}(k_\ell,x) - g_\mr{evo}^{t,(1)}(k_\ell,x) \\
\mathcal{L}^{t,(2)}_{\alpha}(\ell,x) & = g_\mr{dil,\alpha}^{t,(2)}(k_\ell,x) - 2 g_\mr{dil-evo,\alpha}^{t,(2)}(k_\ell,x) + g_\mr{evo}^{t,(2)}(k_\ell,x)
\ea
with
\ba
g_\mr{dil,\alpha}^{t,(c)}(k,r) &= \frac{k^c}{b_1^t(k,r) \, \frac{P(k)}{k^\alpha} k^{c-1}}\frac{\partial^c \left[b_1^t(k,r) \, \frac{P(k)}{k^\alpha} k^{c-1}\right]}{\partial k^c} \\
g_\mr{evo}^{t,(c)}(k,r) &= \frac{r^c}{n_t(r) \, b_1^t(k,r) \, G(r)} \frac{\partial^c \left[n_t(r) \, b_1^t(k,r) \, G(r)\right]}{\partial r^c} \\
g_\mr{dil-evo,\alpha}^{t,(2)}(k,r) &= \frac{k r}{n_t(r) \, b_1^t(k,r) \, G(r) \frac{P(k)}{k^\alpha}} \frac{\partial^2 \left[n_t(r) \, b_1^t(k,r) \, G(r) \frac{P(k)}{k^\alpha} \right]}{\partial k \ \partial r}
\ea
The equations for these dilation and evolution terms allow the practical numerical implementation of Eq.~\ref{Eq:proj-LSS-Limber-bispectrum}, as well as later physical interpretation of the results.

\begin{figure}[ht!]
    \begin{center}
        \includegraphics[width=\linewidth]{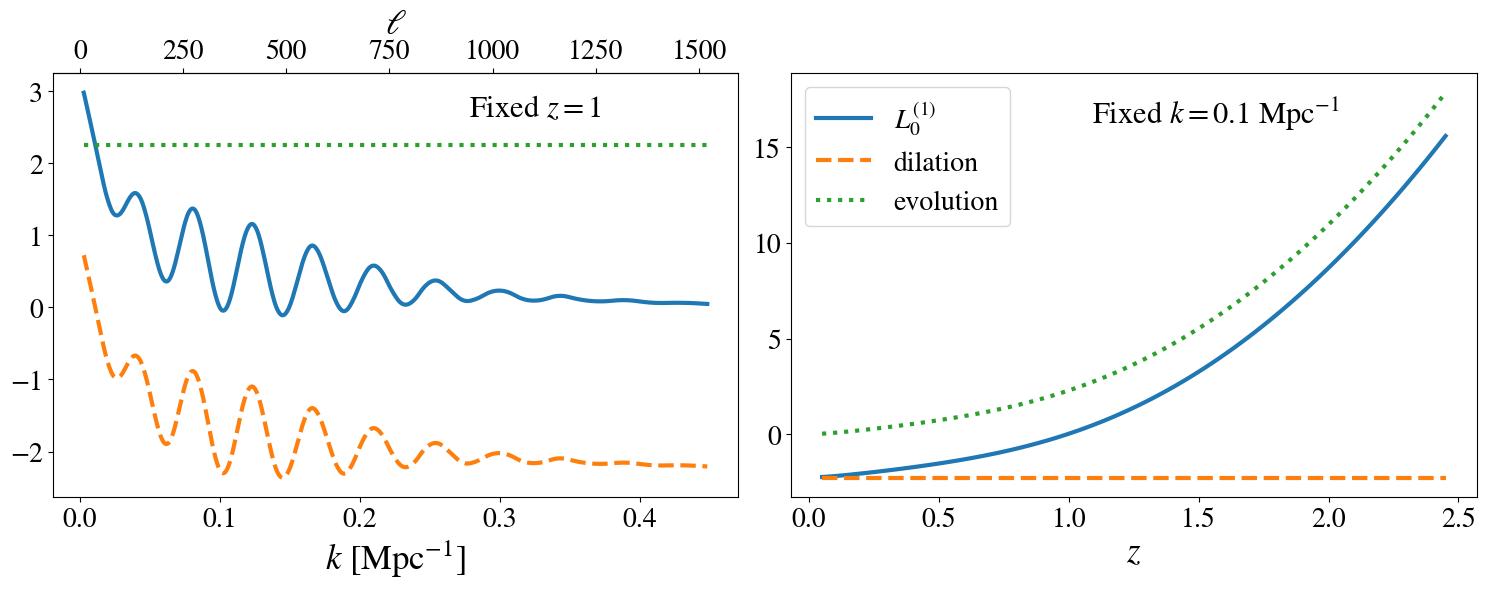}
         \caption{$L_0^{(1)}(k,z)$ (solid blue) and its components: the dilation term $g_\mr{dil,0}^{(1)}(k)$ (dashed orange), and the evolution term $-g_\mr{evo}^{(1)}(z)$ (dotted green). \textit{Left:} plot as a function of Fourier mode $k$ with fixed redshift $z=1$; the equivalent Limber multipole $\ell=kr$ is shown in the top x axis. \textit{Right:} plot as a function of redshift $z$ with fixed Fourier mode $k=0.1$ Mpc$^{-1}$. The tracer chosen here for illustration is the expected sample of Euclid photometric galaxies. The details of the specifications and modeling are covered in Appendix~\ref{App:specs-models-phot}.
         }
        \label{Fig:Lsymbol^(1)}
    \end{center}
\end{figure}

\begin{figure}[ht!]
    \begin{center}
        \includegraphics[width=\linewidth]{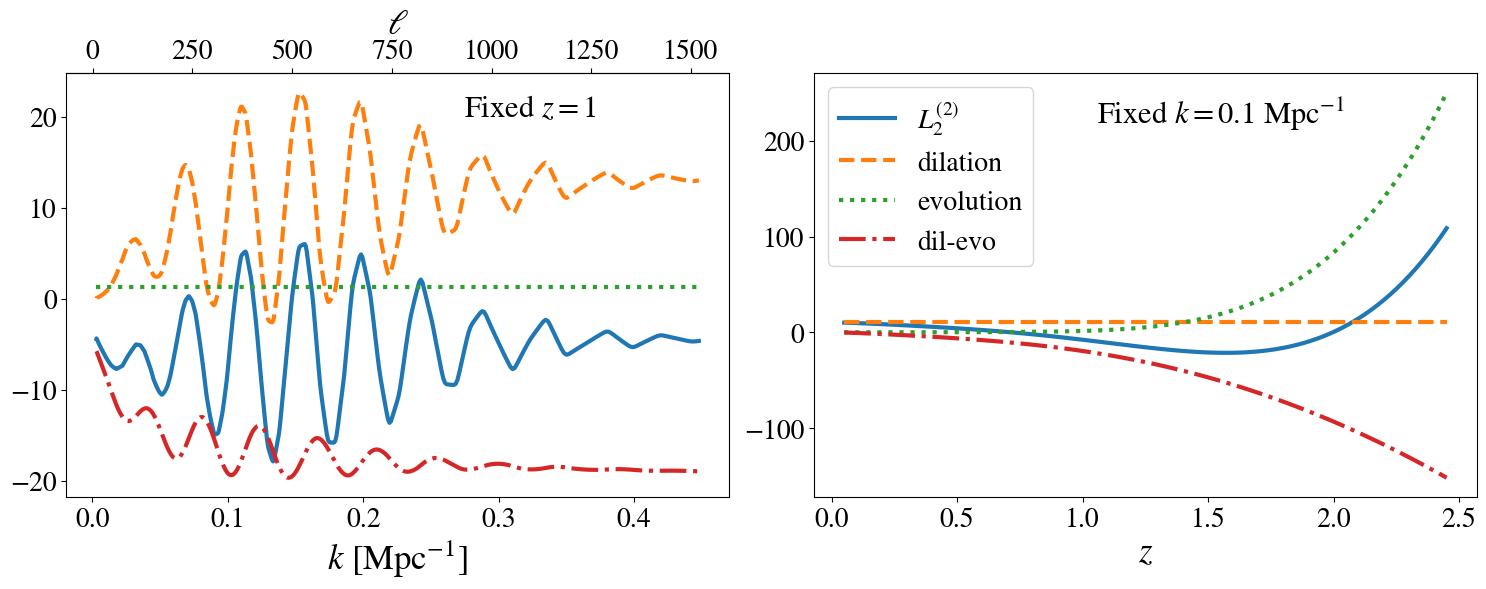}
         \caption{$L_2^{(2)}(k,z)$ (solid blue) and its components: the dilation term $g_\mr{dil,2}^{(2)}(k)$ (dashed orange), the evolution term $g_\mr{evo}^{(2)}(z)$ (dotted green), and the mixed term $g_\mr{dil-evo,2}^{t,(2)}(k,z)$ (dash-dotted red). \textit{Left:} plot as a function of Fourier mode $k$ with fixed redshift $z=1$; the equivalent Limber multipole $\ell=kr$ is shown in the top x axis. \textit{Right:} plot as a function of redshift $z$ with fixed Fourier mode $k=0.1$ Mpc$^{-1}$. The tracer chosen here for illustration is the expected sample of Euclid photometric galaxies. The details of the specifications and modeling are covered in Appendix~\ref{App:specs-models-phot}.}
        \label{Fig:Lsymbol^(2)}
    \end{center}
\end{figure}

On large scales, the first order bias $b_1^t$ depends only on $r$ (i.e. time/redshift) and not on $k$ (scale). This simplifies the computations, as then $g_\mr{dil,\alpha}^{t,(c)}$ only depends on $k$, $g_\mr{evo}^{t,(c)}$ only depends on $r$, and we have
\ba
g_\mr{dil-evo,\alpha}^{t,(2)}(k,r) = g_\mr{dil,\alpha}^{t,(1)}(k) \times g_\mr{evo}^{t,(1)}(r).
\ea
For illustration, we chose as tracer $t$ the expected sample of Euclid photometric galaxies, with specifications and modeling detailed in Appendix~\ref{App:specs-models-phot}, and show the $L$ symbol and its terms in Fig.~\ref{Fig:Lsymbol^(1)} for the first order $L_0^{(1)}$ and in Fig.~\ref{Fig:Lsymbol^(2)} for the second-order $L_2^{(2)}$ \footnote{We chose to plot $L_2^{(2)}$ instead of $L_0^{(2)}$, as it is the former symbol that is relevant to the computation of Super-Sample Covariance later in this article.}.

From Fig.~\ref{Fig:Lsymbol^(1)} we see oscillations in $k$ space for the dilation term of the first order $L_0^{(1)}$. These oscillations are also visible in Fig.~\ref{Fig:Lsymbol^(2)} for the dilation and mixed terms of the second-order $L_2^{(2)}$. These oscillations come from the Baryon Acoustic Oscillations (BAO) present in the power spectrum $P(k)$; they are in fact more visible than in the raw power spectrum because we are looking essentially at derivatives of $P(k)$. Next, the evolution term has a smooth behaviour, increasing with redshift as the density of galaxies goes down. It becomes dominant over the dilation term approximately above the median redshift of the survey ($z\simeq1$). As perhaps expected, the second-order mixed term (dil-evo) has a behaviour mixed between the dilation and evolution terms: it has oscillations in $k$ space and grows (in absolute value) with redshift. 

\subsection{Outcome: angular bispectrum of LSS tracers}\label{Sect:result-for-angular-bispectrum}
In this section we essentially collect the results of the previous sections to derive the angular bispectrum $b_{\lu\ld\lt}$ of LSS tracers. For this, we recall the result of section~\ref{Sect:way-spin-spherical-harmonics}:
\ba
b_{\ell_1,\ell_2,\ell_3} = & \left(\frac{2}{\pi}\right)^3\!\int x^2 \, \dd x \ r^2_{123} \, \dd r_{123}  \ k^2_{123}\dd k_{123} \ j_{\lu}(k_1 r_1) \, j_{\ld}(k_2 r_2) \, j_{\lt}(k_3 r_3) \nonumber \\
& \times \sum_{n=0}^2 \Big[(-1)^n \, 2 \, \tilde{B}_n(k_{123},r_{123}) \ \mathcal{I}^{(n)}_{\ell_{123}}(k_{123},x)+2 \ \mr{perm.}\Big],
\ea
where the functions $\tilde{B}_n$ and $\mathcal{I}^{(n)}$ are defined in Table~\ref{Table:radkernels}.\\
These integrals involve derivatives of Bessel functions and are thus evaluated using the techniques of sections~\ref{Sect:proj-Bessel} and \ref{Sect:proj-Limber}, in particular Eq.~\ref{Eq:proj-LSS-noLimber-bispectrum} in the beyond-Limber case and Eq.~\ref{Eq:proj-LSS-Limber-bispectrum} in the Limber case. We spare the reader the careful derivation of the six bispectrum terms: a scalar term for $n=0$, scalar and vector terms for $n=1$, and scalar, vector and tensor terms for $n=2$. Instead we summarise the results in tables below. These involve the functions $\mathscr{L}^{t,(c)}_{\alpha}$ (beyond-Limber case) and $\mathcal{L}^{t,(c)}_{\alpha}$ (Limber case) which are defined respectively in Eq.~\ref{Eq:def-scrL^t,c_alpha} and Eq.~\ref{Eq:def-calL^t,c_alpha}. In the following, we shorten the notations by leaving implicit the arguments of these functions, e.g. $\mathcal{L}^{t_i,(c)}_{\alpha} \equiv \mathcal{L}^{t_i,(c)}_{\alpha}(\ell_i,x)$, where the multipole can be inferred from the tracer's subscript. The equations also involve the biases summed over all second-order contributions which are defined in Eqs~\ref{Eq:def-bias_(0)}--\ref{Eq:def-bias_(2)} and restated here for convenience
\ba\label{Eq:b_(n)-q_(n)-definition-for-summary}
b^\mr{t_3}_{(0)} &= \frac{5}{7} b^\mr{t_3}_1 + \frac{1}{2} b^\mr{t_3}_2 - \frac{1}{3} b^\mr{t_3}_{s^2} \qquad & q_{(0)}(k_1,k_2) &= 1 \\ 
b^\mr{t_3}_{(1)} &= b^\mr{t_3}_1  \qquad & q_{(1)}(k_1,k_2) &= \frac{1}{2} \left(\frac{k_1}{k_2}+\frac{k_2}{k_1}\right) \\
b^\mr{t_3}_{(2)} &= \frac{2}{7} b^\mr{t_3}_1 + b^\mr{t_3}_{s^2}  \qquad & q_{(2)}(k_1,k_2) &= 1 
\ea
The angular bispectrum results are summarised in two tables. Table~\ref{Table:bisp-summary-noLimber} gives the equations in the beyond-Limber case, and Table~\ref{Table:bisp-summary-Limber} gives the equations in the Limber case.
These tables are some of the main results of the article and constitutes the reference equations for potential numerical implementation.

\begin{table}
\begin{equation}
\begin{array}{ll}
\hline\hline
\mr{General} & \mr{equation:} \\ \hline
  & \\
b_{\ell_1,\ell_2,\ell_3}^{(n)} &=  2 \int \dd V \ n_\mr{t_1} \, n_\mr{t_2} \, n_\mr{t_3} \ b^\mr{t_1}_1 \, b^\mr{t_2}_1 \ P_1 \, P_2 \times C^\mr{t_3}_n \quad + \quad 2 \ \mr{perm.}\\
   & \\
\hline\hline
n=0\,\text{:} &\\ \hline 
   & \\
C^\mr{t_3}_0 & =  C_0^S = b^{t_3}_{(0)}  \ \mathscr{L}^{t_1,(0)}_{0} \, \mathscr{L}^{t_2,(0)}_{0}\\ 
	& \\
\hline\hline
n=1\,\text{:} &\\ \hline
	& \\
C^\mr{t_3}_1 & = C_1^V + C_1^S \\
   & \\
C_1^V & = - \Lambda^{\mathrm{V}}_{\ell_{123}} \ b^{t_3}_{1} \ \frac{1}{2}\left(\frac{\mathscr{L}^{t_1,(0)}_{2}}{(\lu+\sfrac{1}{2})^2} \mathscr{L}^{t_2,(0)}_{0} + \mathscr{L}^{t_1,(0)}_{0} \frac{\mathscr{L}^{t_2,(0)}_{2}}{(\ld+\sfrac{1}{2})^2}\right) \\
   & \\
C_1^S & = - b^{t_3}_{1} \ \frac{1}{2} \left(\frac{\mathscr{L}^{t_1,(1)}_{2}}{(\lu+\sfrac{1}{2})^2} \mathscr{L}^{t_2,(1)}_{0} + \mathscr{L}^{t_1,(1)}_{0} \frac{\mathscr{L}^{t_2,(1)}_{2}}{(\ld+\sfrac{1}{2})^2} \right)  \\
	& \\
\hline\hline
n=2\,\text{:} &\\ \hline
	& \\
C^\mr{t_3}_2 & = C_2^T + C_2^V + C_2^S \\
   & \\
C_2^T & = \Lambda^T_{\ell_{123}} \ b^{t_3}_{(2)} \ \frac{\mathscr{L}^{t_1,(0)}_{2}}{(\lu+\sfrac{1}{2})^2} \frac{\mathscr{L}^{t_2,(0)}_{2}}{(\ld+\sfrac{1}{2})^2}\\
   & \\
C_2^V & = \Lambda^{\mathrm{V}}_{\ell_{123}} \ b^{t_3}_{(2)} \ 2 \frac{\mathscr{L}^{t_1,(1)}_{2} + \mathscr{L}^{t_1,(0)}_{2}}{(\lu+\sfrac{1}{2})^2} \ \frac{\mathscr{L}^{t_2,(1)}_{2} + \mathscr{L}^{t_2,(0)}_{2}}{(\ld+\sfrac{1}{2})^2}  \\
   & \\
C_2^S & = b^{t_3}_{(2)} \ \frac{1}{2} \left(\mathscr{L}^{t_1,(0)}_{0} \mathscr{L}^{t_2,(0)}_{0} + \frac{\mathscr{L}^{t_1,(2)}_{2}}{(\lu+\sfrac{1}{2})^2} \mathscr{L}^{t_2,(0)}_{0} + \mathscr{L}^{t_1,(0)}_{0} \frac{\mathscr{L}^{t_2,(2)}_{2}}{(\ld+\sfrac{1}{2})^2} + 3 \frac{\mathscr{L}^{t_1,(2)}_{2}}{(\lu+\sfrac{1}{2})^2} \frac{\mathscr{L}^{t_2,(2)}_{2}}{(\ld+\sfrac{1}{2})^2} \right) \\
	& \\
\hline\hline
\end{array} \nonumber
\end{equation} 
\caption{Summary of the angular bispectrum terms in the beyond-Limber (exact) case. The coefficients $\Lambda^{\mathrm{V}}_{\ell_{123}}$ and $\Lambda^{\mathrm{T}}_{\ell_{123}}$ are defined in Table~\ref{Table:ellcoeff}. The function $\mathscr{L}^{t,(c)}_{\alpha}$ is defined in Eq.~\ref{Eq:def-scrL^t,c_alpha} and the biases $b^{t_3}_{(n)}$ are defined in Eq.~\ref{Eq:b_(n)-q_(n)-definition-for-summary}.} 
\label{Table:bisp-summary-noLimber}
\end{table}

\begin{table}
\begin{equation}
\begin{array}{ll}
\hline\hline
\mr{General} & \mr{equation:} \\ \hline
  & \\
b_{\ell_1,\ell_2,\ell_3}^{(n)} &=  2 \int \dd V \ n_\mr{t_1} \, n_\mr{t_2} \, n_\mr{t_3} \ b^\mr{t_1}_1 \, b^\mr{t_2}_1 \ P_1 \, P_2 \times C^\mr{t_3}_n \quad + \quad 2 \ \mr{perm.}\\
   & \\
\hline\hline
n=0\,\text{:} & \\ \hline 
   & \\
C^\mr{t_3}_0 & =  C_0^S = b^{t_3}_{(0)} \\ 
	& \\
\hline\hline
n=1\,\text{:} &\\ \hline
	& \\
C^\mr{t_3}_1 & = C_1^V + C_1^S \\
   & \\
C_1^V & = - \Lambda^{\mathrm{V}}_{\ell_{123}} \ b^{t_3}_{1} \ \frac{1}{2}\left(\frac{1}{(\lu+\sfrac{1}{2})^2}+\frac{1}{(\ld+\sfrac{1}{2})^2}\right) \\
   & \\
C_1^S & = - b^{t_3}_{1} \ \frac{1}{2} \left(\frac{\mathcal{L}^{t_1,(1)}_{2}}{(\lu+\sfrac{1}{2})^2} \mathcal{L}^{t_2,(1)}_{0} + \mathcal{L}^{t_1,(1)}_{0} \frac{\mathcal{L}^{t_2,(1)}_{2}}{(\ld+\sfrac{1}{2})^2} \right) \\
	& \\
\hline\hline
n=2\,\text{:} &\\ \hline
	& \\
C^\mr{t_3}_2 & = C_2^T + C_2^V + C_2^S \\
   & \\
C_2^T & = \Lambda^T_{\ell_{123}} \ b^{t_3}_{(2)} \ \frac{1}{(\lu+\sfrac{1}{2})^2} \frac{1}{(\ld+\sfrac{1}{2})^2}\\
   & \\
C_2^V & = \Lambda^{\mathrm{V}}_{\ell_{123}} \ b^{t_3}_{(2)} \ 2 \frac{\mathcal{L}^{t_1,(1)}_{2} +1}{(\lu+\sfrac{1}{2})^2} \frac{\mathcal{L}^{t_2,(1)}_{2} +1}{(\ld+\sfrac{1}{2})^2} \\
   & \\
C_2^S & = b^{t_3}_{(2)} \ \frac{1}{2} \left(1 + \frac{\mathcal{L}^{t_1,(2)}_{2}}{(\lu+\sfrac{1}{2})^2} + \frac{\mathcal{L}^{t_2,(2)}_{2}}{(\ld+\sfrac{1}{2})^2} + 3 \frac{\mathcal{L}^{t_1,(2)}_{2}}{(\lu+\sfrac{1}{2})^2}\frac{\mathcal{L}^{t_2,(2)}_{2}}{(\ld+\sfrac{1}{2})^2} \right) \\
	& \\
\hline\hline
\end{array} \nonumber
\end{equation} 
\caption{Summary of the angular bispectrum terms with Limber approximation. The coefficients $\Lambda^{\mathrm{V}}_{\ell_{123}}$ and $\Lambda^{\mathrm{T}}_{\ell_{123}}$ are defined in Table~\ref{Table:ellcoeff}. The function $\mathcal{L}^{t,(c)}_{\alpha}$ is defined in Eq.~\ref{Eq:def-scrL^t,c_alpha} and the biases $b^{t_3}_{(n)}$ are defined in Eq.~\ref{Eq:b_(n)-q_(n)-definition-for-summary}.} 
\label{Table:bisp-summary-Limber}
\end{table}

\section{Application to the covariance}\label{Sect:application-covariance}

The covariance between the number counts and power spectrum of tracers of the LSS is related to their bispectrum. In the full sky case we get (see e.g. \cite{Lacasa2016} for clusters and galaxies):
\be\label{Eq:Cov=bisp-sqz}
\Cov\left[N_{t_1}, C_\ell^{t_2,t_3}\right] = \frac{b_{0\ell\ell}^{t_1,t_2,t_3}}{4\pi}.
\ee
This computation is thus a special case of the bispectrum computation, specifically in the squeezed configuration ($\lu=0$, $\ld=\lt=\ell$). In practice, this non-Gaussian covariance is of interest at high enough multipoles $\ell$ that Limber approximation is warranted for $C_\ell^{t_2,t_3}$. However, Limber approximation is not warranted for $\lu=0$. This is why Sect.~\ref{Sect:mixed-Limber} presents the mixed Limber equations for the bispectrum, where Limber approximation is applied to $k_2$ and $k_3$ but we keep the exact integration on $k_1$.

Then in section~\ref{Sect:ISC} we present the covariance term sensitive only to $k_2$ and $k_3$: intra-survey covariance. Finally in section~\ref{Sect:SSC} we present the covariance terms sensitive to $P(k_1)$: Super-Sample Covariance.

\subsection{Prelude: mixed Limber}\label{Sect:mixed-Limber}

Appendix~\ref{App:proj-LSS-mixedLimber} shows that for a LSS bispectrum term of the usual form
\ba
\tilde{B}_n(k_{123},r_{123}) &= n_\mr{t_1} \, n_\mr{t_2} \, n_\mr{t_3}\ P(k_1|r_{13}) \, P(k_2|r_{23}) \ b^\mr{t_1}_1 \, b^\mr{t_2}_1 \, b^\mr{t_3}_{a} \ k_1^\mu \, k_2^\nu \\
\mathcal{I}^{(n),A}_{\ell_{123}} &= \frac{B}{x^b}\frac{\partial^c_x j_{\ell_1}}{k_1^\xi} \ \frac{\partial^d_x j_{\ell_2}}{k_2^\rho} \ j_{\ell_3}.
\ea
Applying Limber approximation on $k_2,k_3$ but not on $k_1$ gives the following projection of the three permutation terms
\ba
b_{\lu,\ld,\lt}^{n,(23)} &= \frac{(-1)^{c+d} \, B}{(\ld+\sfrac{1}{2})^\alpha \, (\lt+\sfrac{1}{2})^\beta} \int \dd V \ n_\mr{t_1} \, n_\mr{t_2} \, n_\mr{t_3} \ b^\mr{t_1}_a \, b^\mr{t_2}_1 \, b^\mr{t_3}_1 \ P_2 \, P_3 \ \mathcal{L}^{t_2,(c)}_{\alpha} \, \mathcal{L}^{t_3,(d)}_{\beta}. \label{Eq:proj-LSS-mixedLimber-(23)-bispectrum} \\
b_{\lu,\ld,\lt}^{n,(12)} &= \frac{(-1)^{c+d} \, B}{(\lu+\sfrac{1}{2})^\alpha \, (\ld+\sfrac{1}{2})^\beta} \int \dd V \ n_\mr{t_1} \, n_\mr{t_2} \, n_\mr{t_3} \ b^\mr{t_1}_1 \, b^\mr{t_2}_1 \, b^\mr{t_3}_a \ P_1 \, P_2 \ \mathscr{L}^{t_1,(c)}_{\alpha} \, \mathcal{L}^{t_2,(d)}_{\beta} \label{Eq:proj-LSS-mixedLimber-(12)-bispectrum} \\
b_{\lu,\ld,\lt}^{n,(13)} &= \frac{(-1)^{c+d} \, B}{(\lu+\sfrac{1}{2})^\alpha \, (\lt+\sfrac{1}{2})^\beta} \int \dd V \ n_\mr{t_1} \, n_\mr{t_2} \, n_\mr{t_3} \ b^\mr{t_1}_1 \, b^\mr{t_2}_a \, b^\mr{t_3}_1 \ P_1 \, P_3 \ \mathscr{L}^{t_1,(c)}_{\alpha} \, \mathcal{L}^{t_3,(d)}_{\beta} \label{Eq:proj-LSS-mixedLimber-(13)-bispectrum}
\ea
For the squeezed configuration needed for the covariance (Eq.~\ref{Eq:Cov=bisp-sqz}), $k_2$ and $k_3$ alias into the multipole $\ell$ and are thus intra-survey modes (also known as "hard" modes in the literature), while $k_1$ aliases into the monopole and is thus a super-sample mode (or "super-survey" or "soft" mode). \\
From the three permutations above, we see that the first one involves only the power spectrum of intra-survey modes, $P_2$ and $P_3$. Furthermore, this permutation only depends on $k_1$ through the scale dependence of the bias $b^\mr{t_1}_a(k_1,r_1)$ which can be neglected on large scales. Hence this permutation is only sensitive to intra-survey modes and we dub this part intra-survey covariance. We study this term in greater detail in Sect.~\ref{Sect:ISC}. \\
Then the last two permutations involve the power spectrum of the super-sample mode $P_1$; they thus contribute to the part of covariance called Super-Sample Covariance (SSC). We study these terms in greater detail in Sect.~\ref{Sect:SSC}. 

\subsection{Intra-survey covariance (ISC)}\label{Sect:ISC}

Collecting the results from sections \ref{Sect:result-for-angular-bispectrum} \& \ref{Sect:mixed-Limber}, we find that the ISC takes contribution from bispectrum terms for $n=0,1$ and 2, with
\ba
\Cov_\mr{ISC}^{(n)} &= \frac{1}{4\pi} \int \dd V \ n_\mr{t_1} \, n_\mr{t_2} \, n_\mr{t_3} \ b^\mr{t_2}_1 \, b^\mr{t_3}_1 \ P_\ell^2 \times C^\mr{t_1}_n
\ea
where $C^\mr{t_1}_n$ is taken in the Limber approximation (Table~\ref{Table:bisp-summary-Limber}). \\
For the $n=0$ term, we have
\ba
\Cov_\mr{ISC}^{(n=0)} &= \frac{1}{4\pi} \int \dd V \ n_\mr{t_1} \, n_\mr{t_2} \, n_\mr{t_3} \ b^\mr{t_1}_{(0)} \, b^\mr{t_2}_1 \, b^\mr{t_3}_1 \ P_\ell^2
\ea
For the $n=1$ term, we have a vector and a scalar contribution:
\ba
\Cov_\mr{ISC}^{(n=1),V} &= - \frac{\Lambda^V_{\ell,\ell,0}}{4\pi} \int \dd V \ n_\mr{t_1} \, n_\mr{t_2} \, n_\mr{t_3} \ b^\mr{t_1}_{(1)} \, b^\mr{t_2}_1 \, b^\mr{t_3}_{1} \frac{P_\ell^2}{(\ell+\sfrac{1}{2})^2}
\ea
where $\Lambda^V_{\ell,\ell,0} = \ell(\ell+1)$, and
\ba
\Cov_\mr{ISC}^{(n=1),S} &= - \frac{1}{4\pi} \int \dd V \ n_\mr{t_1} \, n_\mr{t_2} \, n_\mr{t_3} \ b^\mr{t_1}_{(1)} \, b^\mr{t_2}_1 \, b^\mr{t_3}_{1} \ P_\ell^2
\ \frac{ \mathcal{L}^{t_2,(1)}_{2} \mathcal{L}^{t_3,(1)}_{0} + \mathcal{L}^{t_3,(1)}_{0} \mathcal{L}^{t_2,(1)}_{2} }{2 \ (\ell+\sfrac{1}{2})^2}
\ea
For the $n=2$ term, we have tensor, vector and scalar contributions:
\ba
\Cov_\mr{ISC}^{(n=2),T} &= \frac{\Lambda^T_{\ell,\ell,0}}{4\pi} \int \dd V \ n_\mr{t_1} \, n_\mr{t_2} \, n_\mr{t_3} \ b^\mr{t_1}_{(2)} \, b^\mr{t_2}_1 \, b^\mr{t_3}_1 \frac{P_\ell^2}{(\ell+\sfrac{1}{2})^4}
\ea
where $\Lambda^T_{\ell,\ell,0} = (\ell^2+\ell-1)\ell(\ell+1)$, and 
\ba
\Cov_\mr{ISC}^{(n=2),V} &= \frac{\Lambda^V_{\ell,\ell,0}}{4\pi} \int \dd V \ n_\mr{t_1} \, n_\mr{t_2} \, n_\mr{t_3} \ b^\mr{t_1}_{(2)} \, b^\mr{t_2}_1 \, b^\mr{t_3}_1 \ P_\ell^2 \times 2 \frac{\left(\mathcal{L}^{t_2,(1)}_{2} +1\right) \left(\mathcal{L}^{t_3,(1)}_{2} +1\right)}{(\ell+\sfrac{1}{2})^4}
\ea
and
\ba
\Cov_\mr{ISC}^{(n=2),S} &= \frac{1}{4\pi} \int \dd V \ n_\mr{t_1} \, n_\mr{t_2} \, n_\mr{t_3} \ b^\mr{t_1}_{(2)} \, b^\mr{t_2}_1 \, b^\mr{t_3}_1 \ P_\ell^2 \times \frac{1}{2} \left(1 + \frac{\mathcal{L}^{t_2,(2)}_{2} + \mathcal{L}^{t_3,(2)}_{2}}{(\ell+\sfrac{1}{2})^2} + 3 \frac{\mathcal{L}^{t_2,(2)}_{2} \, \mathcal{L}^{t_3,(2)}_{2}}{(\ell+\sfrac{1}{2})^4} \right)
\ea

In total, we find
\ba
\Cov_\mr{ISC} &= \frac{1}{4\pi} \int \dd V \, \left[ n_\mr{t_1} \right] \, \left[ n_\mr{t_2} \, n_\mr{t_3} \, b^\mr{t_2}_1 \, b^\mr{t_3}_1 \ P_\ell^2 \right] \ X_\mr{ISC}
\ea
with
\ba
X_\mr{ISC} &=
b_{(0)}^\mr{t_1} 
- \frac{\Lambda^V_{\ell,\ell,0}}{(\ell+\sfrac{1}{2})^2} b_{(1)}^\mr{t_1} 
+ \frac{\Lambda^T_{\ell,\ell,0}}{(\ell+\sfrac{1}{2})^4} b_{(2)}^\mr{t_1} 
+ \frac{1}{2} b_{(2)}^\mr{t_1} \nonumber \\ 
& - \frac{ \mathcal{L}^{t_2,(1)}_{2} \mathcal{L}^{t_3,(1)}_{0} + \mathcal{L}^{t_3,(1)}_{0} \mathcal{L}^{t_2,(1)}_{2} }{2 \ (\ell+\sfrac{1}{2})^2} b_{(1)}^\mr{t_1} 
+ 2 \Lambda^V_{\ell,\ell,0} \frac{\left(\mathcal{L}^{t_2,(1)}_{2} +1\right) \left(\mathcal{L}^{t_3,(1)}_{2} +1\right)}{(\ell+\sfrac{1}{2})^4} b^\mr{t_1}_{(2)} \nonumber \\ 
& + \frac{1}{2}\frac{\mathcal{L}^{t_2,(2)}_{2} + \mathcal{L}^{t_3,(2)}_{2}}{(\ell+\sfrac{1}{2})^2} b^\mr{t_1}_{(2)} 
+ \frac{3}{2} \frac{\mathcal{L}^{t_2,(2)}_{2} \, \mathcal{L}^{t_3,(2)}_{2}}{(\ell+\sfrac{1}{2})^4} b^\mr{t_1}_{(2)}  
\ea
where we ordered the contribution in terms of scaling with multipole: the four terms on the first line are the dominant ones, then the next three terms on the second and third line fall as $1/\ell^2$, and finally the last term on the third line falls as $1/\ell^4$.

\begin{figure}[ht!]
    \begin{center}
        \includegraphics[width=.7\linewidth]{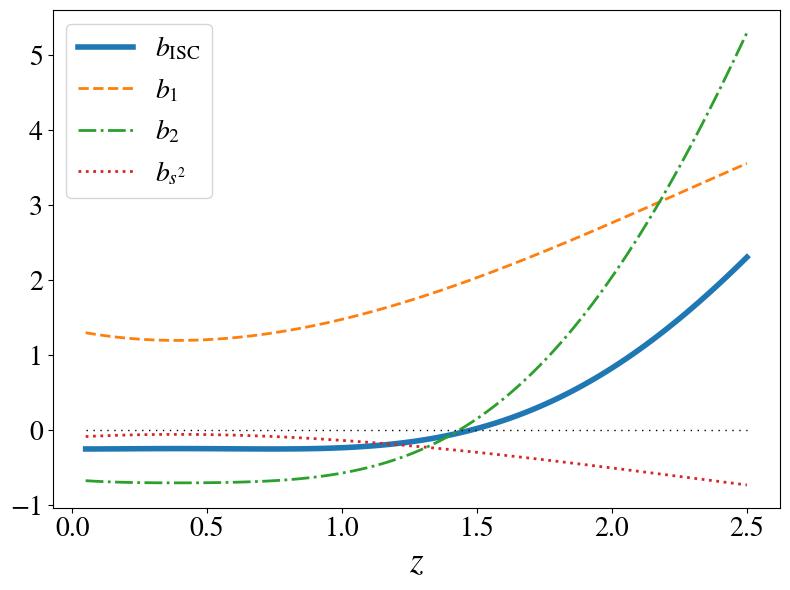}
         \caption{$b_\mr{ISC}$ for the Euclid photometric galaxy specifications. First and second-order biases are also shown for comparison.}
        \label{Fig:b_ISC}
    \end{center}
\end{figure}

Interestingly, in the high multipole limit $\ell\gg1$, $X_\mr{ISC}$ simplifies greatly and becomes independent of $\ell$, $t_2$ and $t_3$. We get
\ba
\Cov_\mr{ISC,\ell\gg1} &= \frac{1}{4\pi} \int \dd V \, \left[ n_\mr{t_1} \, b_\mr{ISC}^\mr{t_1} \right] \, \left[ n_\mr{t_2} \, n_\mr{t_3} \, b^\mr{t_2}_1 \, b^\mr{t_3}_1 \ P_\ell^2 \right] 
\ea
with
\ba
b_\mr{ISC}^\mr{t_1} \equiv \lim_{\ell\rightarrow\infty} X_\mr{ISC} &=
b_{(0)}^\mr{t_1} 
- b_{(1)}^\mr{t_1} 
+ \frac{3}{2} b_{(2)}^\mr{t_1} \\ 
&= \frac{1}{7} b^\mr{t_1}_{1} + \frac{1}{2} b^\mr{t_1}_{2} + \frac{7}{6} b^\mr{t_1}_{s^2} .
\ea
In this limit, the intra-survey covariance is thus fairly straightforward to compute. Indeed it only involves elements already known in the literature (tracer density, first order bias, power spectrum) except for one new element: $b_\mr{ISC}^\mr{t_1}$ that is a combination of first-order and second-order biases and that we plot for illustration in Fig.~\ref{Fig:b_ISC} for the Euclid photometric galaxy specifications.

We see that $b_\mr{ISC}$ has a monotonic behaviour with redshift, increasing from negative values at low redshift $z\lesssim 1.5$ to positive values at higher redshift. This evolution is driven by the contribution from the quadratic bias $b_2$ which we found dominant. At low redshift the considered galaxy sample is hosted preferentially by low-mass haloes which have a negative $b_2$. On the contrary, at higher redshifts the sample is hosted preferentially in massive haloes with a positive $b_2$. There is no mathematical issue with the intra-survey covariance being negative, it just tends to make error bars anti-correlated, the joint covariance matrix is still well-behaved.  

\subsection{Super-sample covariance (SSC)}\label{Sect:SSC}

Collecting the previous results, we see that the SSC takes contribution from bispectrum terms for $n=0,1$ and 2, with
\ba
\Cov_\mr{SSC}^{(n)} &= \frac{2}{4\pi} \int \dd V \ n_\mr{t_1} \, n_\mr{t_2} \, n_\mr{t_3} \ b^\mr{t_1}_1 \ P_0 \, P_\ell \times \left( b^\mr{t_2}_1 \, C^\mr{t_3}_n + b^\mr{t_3}_1 \, C^\mr{t_2}_n \right)
\ea
where
\ba
&P_0 = P(k_{0}|x) = P\left(\frac{\sfrac{1}{2}}{x}\middle|x\right)  &P_\ell = P(k_{\ell}|x) = P\left(\frac{\ell+\sfrac{1}{2}}{x}\middle|x\right)
\ea
and $C^\mr{t}_n$ in the mixed Limber case can be deduced from Tables~\ref{Table:bisp-summary-noLimber}~and~\ref{Table:bisp-summary-Limber}.

An equation, proved in Appendix~\ref{App:props-scrL}, that will be useful in this section is
\ba\label{Eq:relation-scrL-sigma2_ell,alpha}
n_\mr{t}(x) \, b^\mr{t}_1(k_\ell,x) \, G(x) \frac{P(k_\ell)}{k_{\ell}^\alpha} \ \mathscr{L}^{t,(c)}_{\alpha}(\ell,x) = 4\pi \int r^2 \, \dd r \ n_{t}(r) \, b_1^{t}(r) \, G(r) \ (-x)^c \, \partial^c_x \left[\sigma^2_{\ell,\alpha}(r,x)\right]
\ea
with
\ba
\sigma^2_{\ell,\alpha}(r,x) = \frac{1}{2\pi^2} \int k^2 \, \dd k \ P(k)/k^\alpha \ j_\ell(kr) \, j_\ell(kx)
\ea

\subsubsection{Individual terms}\label{Sect:SSC-computation}

A first remark is that we have
\ba
\Lambda^V_{0,\ell,\ell} = \Lambda^T_{0,\ell,\ell} = 0
\ea
which means that vector and tensor contributions to SSC vanish. This result might be specific to the symmetry of the full sky coverage that we are considering. We leave the investigation of partial sky to future works.

\paragraph{First term: $n=0$}\ \\
For the $n=0$ term, we have
\ba
\Cov_\mr{SSC}^{(n=0)} &= \frac{2}{4\pi} \int \dd V \ n_\mr{t_1} \, n_\mr{t_2} \, n_\mr{t_3} \ b^\mr{t_1}_1 \ P_0 \, P_\ell \times \left( b^\mr{t_2}_1 \, b^{t_3}_{(0)} \, \mathscr{L}^{t_1,(0)}_{0} + b^\mr{t_3}_1 \, b^{t_2}_{(0)} \, \mathscr{L}^{t_1,(0)}_{0} \right).
\ea
Using Eq.~\ref{Eq:relation-scrL-sigma2_ell,alpha} for the $\mathscr{L}$ symbol, we get
\ba
\Cov_\mr{SSC}^{(n=0)} =& \int r^2 \, \dd r \ x^2 \, \dd x \ \Big[n_\mr{t_1}(r) \, b^\mr{t_1}_1(r) \Big] \ \Big[n_\mr{t_2}(x) \, n_\mr{t_3}(x) \nonumber \\
& \times 2\left( b^\mr{t_2}_1(k_{\ell},x) \, b^{t_3}_{(0)}(k_{\ell},x) + b^\mr{t_3}_1(k_{\ell},x) \, b^{t_2}_{(0)}(k_{\ell},x) \right) P(k_{\ell}|x) \Big] \, G(r) \, G(x) \ \sigma^2_{0,0}(r,x) \\
=& \int r^2 \, \dd r \ x^2 \, \dd x \ \Big[n_\mr{t_1}(r) \, b^\mr{t_1}_1(r) \Big] \ \Big[n_\mr{t_2}(x) \, n_\mr{t_3}(x) \nonumber \\
& \times 2\left( b^\mr{t_2}_1(k_{\ell},x) \, b^{t_3}_{(0)}(k_{\ell},x) + b^\mr{t_3}_1(k_{\ell},x) \, b^{t_2}_{(0)}(k_{\ell},x) \right) P(k_{\ell}|x) \Big] \times \sigma^2_{\delta_b,\delta_b}(r,x)
\ea
where
\ba
\sigma^2_{\delta_b,\delta_b}(r,x) \equiv G(r) G(x) \, \sigma^2_{0,0}(r,x) &= \frac{1}{2\pi^2} \int k^2 \, \dd k \ P(k|r,x) \ j_0(kr) \, j_0(kx)
\ea
is the covariance of the background density shift: $\sigma^2_{\delta_b,\delta_b}(r,x)=\lbra \delta_b(r) \, \delta_b(x) \rbra$.\\
This can be rewritten in a compact form as
\ba
\Cov_\mr{SSC}^{n=0} = \int r^2 \, \dd r \ x^2 \, \dd x \ \frac{\partial n_\mr{t_1}(r)}{\partial \delta_b} \ \partial_0 P_{n_{t_2},n_{t_3}}(k_{\ell}|x) \ \sigma^2_{\delta_b,\delta_b}(r,x) \label{Eq:Cov_SSC^n=0}
\ea
where 
\ba
\frac{\partial n_\mr{t_1}}{\partial \delta_b} &= n_\mr{t_1} \ b^\mr{t_1}_1
\ea
and
\ba
\partial_0 P_{n_{t_2},n_{t_3}}(k_{\ell}|x) &= 2 \ n_\mr{t_2} \, n_\mr{t_3} \left( b^\mr{t_2}_1 \, b^\mr{t_3}_{(0)} + b^\mr{t_2}_{(0)} \, b^\mr{t_3}_1 \right) P(k_{\ell}|x) \label{Eq:d0P-shortform}\\
&= n_\mr{t_2} \, n_\mr{t_3} \ P(k_{\ell}|x) \times \left[
\frac{20}{7} b^\mr{t_2}_1 \, b^\mr{t_3}_1 + b^\mr{t_2}_1 \, b^\mr{t_3}_2 + b^\mr{t_2}_2 \, b^\mr{t_3}_1 - \frac{2}{3} \left(b^\mr{t_2}_1 \, b^\mr{t_3}_{s^2} + b^\mr{t_2}_{s^2} \, b^\mr{t_3}_1\right) \label{Eq:d0P-longform}
\right]
\ea

\paragraph{Second term: $n=1$}\ \\
For the $n=1$ term, we have the scalar contribution
\ba
\Cov_\mr{SSC}^{(n=1),S} =& -\frac{2}{4\pi} \int \dd V \ n_\mr{t_1} \, n_\mr{t_2} \, n_\mr{t_3} \ b^\mr{t_1}_1 \, b^\mr{t_2}_1 \, b^\mr{t_3}_1 \ P_0 \, P_\ell \times \frac{1}{2} \left(\frac{\mathscr{L}^{t_1,(1)}_{2} \, \mathcal{L}^{t_2,(1)}_{0}}{(0+\sfrac{1}{2})^2} + \frac{\mathscr{L}^{t_1,(1)}_{0} \, \mathcal{L}^{t_2,(1)}_{2}}{(\ell+\sfrac{1}{2})^2} \right) \nonumber \\
&+ (2\leftrightarrow 3).
\ea
which is the only contribution in full sky.\\
Using Eq.~\ref{Eq:relation-scrL-sigma2_ell,alpha} for the $\mathscr{L}$ symbols, we get two subterms 
\ba
\Cov_\mr{SSC}^{(n=1)}
=& \int r^2 \, \dd r \ x^2 \, \dd x \ \Big[n_\mr{t_1}(r) \, b^\mr{t_1}_1(r) \Big] \ \Big[n_\mr{t_2}(x) \, n_\mr{t_3}(x) \ b^\mr{t_2}_1(k_{\ell},x) \, b^\mr{t_3}_1(k_{\ell},x) \ P(k_{\ell}|x) \Big] \nonumber \\
& \times G(r) \, G(x) \ \Bigg( \left( \mathcal{L}^{t_2,(1)}_{0}(\ell_2,x) + \mathcal{L}^{t_3,(1)}_{0}(\ell_3,x) \right) \, \frac{1}{x} \, \partial_x \, \sigma^2_{0,2}(r,x) \nonumber \\
& \qquad + \frac{\mathcal{L}^{t_2,(1)}_{2}(\ell_2,x) + \mathcal{L}^{t_3,(1)}_{2}(\ell_3,x)}{(\ell+\sfrac{1}{2})^2} \ x \, \partial_x \, \sigma^2_{0,0}(r,x) \Bigg)
\ea
Now if we define the following new kernels (whose interpretation we postpone to Sect.~\ref{Sect:SSC-total})
\ba
\upsilon(r,x) \equiv G(r) \, G(x) \ \frac{1}{x} \partial_x \, \sigma^2_{0,2} &= \frac{1}{2\pi^2} \int k^2 \, \dd k \ P(k|r,x) \ j_0(kr) \, \frac{j'_0(kx)}{kx} \\
\tau(r,x) \equiv G(r) \, G(x) \ x \, \partial_x \, \sigma^2_{0,0}(r,x) &= \frac{1}{2\pi^2} \int k^2 \, \dd k \ P(k|r,x) \ j_0(kr) \, kx \, j'_0(kx)
\ea
we can put the $n=1$ term in a form similar to Eq.~\ref{Eq:Cov_SSC^n=0}:
\ba
\Cov_\mr{SSC}^{n=1} = \int r^2 \, \dd r \ x^2 \, \dd x \ \frac{\partial n_\mr{t_1}(r)}{\partial \delta_b} \ \partial_1^A P_{n_{t_2},n_{t_3}}(k_{\ell}|x) \ \upsilon(r,x) \label{Eq:Cov_SSC^n=1} \\
+ \int r^2 \, \dd r \ x^2 \, \dd x \ \frac{\partial n_\mr{t_1}(r)}{\partial \delta_b} \ \partial_1^B P_{n_{t_2},n_{t_3}}(k_{\ell}|x) \ \tau(r,x)
\ea
where
\ba
\partial_1^A P_{n_{t_2},n_{t_3}}(k_{\ell}|x) &= n_\mr{t_2} \, n_\mr{t_3} \   b^\mr{t_2}_1 \, b^\mr{t_3}_{1} \left(\mathcal{L}^{t_2,(1)}_{0}+\mathcal{L}^{t_3,(1)}_{0} \right) P(k_{\ell}|x) ,  \label{Eq:d1AP} \\
\partial_1^B P_{n_{t_2},n_{t_3}}(k_{\ell}|x) &= n_\mr{t_2} \, n_\mr{t_3} \   b^\mr{t_2}_1 \, b^\mr{t_3}_{1} \frac{\mathcal{L}^{t_2,(1)}_{2} + \mathcal{L}^{t_3,(1)}_{2}}{(\ell+\sfrac{1}{2})^2} P(k_{\ell}|x) . \label{Eq:d1BP}
\ea
Here we note that the $B$ contribution is suppressed by a $\frac{1}{\ell^2}$ factor, so that we expect it to be of importance only at low multipoles. In other words, this is a wide-angle correction.

\paragraph{Third term: $n=2$}\ \\
For the $n=2$ term, we have
\ba
\Cov_\mr{SSC}^{(n=2),S} =& \frac{2}{4\pi} \int \dd V \ n_\mr{t_1} \, n_\mr{t_2} \, n_\mr{t_3} \ b^\mr{t_1}_1 \, b^\mr{t_2}_1 \, b^\mr{t_3}_{(2)} \ P_0 \, P_\ell \times \frac{1}{2} \Bigg(\mathscr{L}^{t_1,(0)}_{0} + \frac{\mathscr{L}^{t_1,(2)}_{2}}{(0+\sfrac{1}{2})^2} \nonumber \\
& + \mathscr{L}^{t_1,(0)}_{0} \frac{\mathcal{L}^{t_2,(2)}_{2}}{(\ell+\sfrac{1}{2})^2} + 3 \frac{\mathscr{L}^{t_1,(2)}_{2}}{(0+\sfrac{1}{2})^2} \frac{\mathcal{L}^{t_2,(2)}_{2}}{(\ell+\sfrac{1}{2})^2} \Bigg) \quad + \quad (2\leftrightarrow 3).
\ea
Using Eq.~\ref{Eq:relation-scrL-sigma2_ell,alpha} for the $\mathscr{L}$ symbols, we get four subterms
\ba
\Cov_\mr{SSC}^{(n=2)}
&= \int r^2 \, \dd r \ x^2 \, \dd x \ \Big[n_\mr{t_1}(r) \, b^\mr{t_1}_1(r) \Big] \ \Big[n_\mr{t_2}(x) \, n_\mr{t_3}(x) \ b^\mr{t_2}_1(k_{\ell},x) \, b^\mr{t_3}_{(2)}(k_{\ell},x)  P(k_{\ell}|x) \Big]  \nonumber \\
& \quad \times G(r) \, G(x) \left( \sigma^2_{0,0} + \partial_x^2 \, \sigma^2_{0,2} + \frac{\mathcal{L}^{t_2,(2)}_{2}}{(\ell+\sfrac{1}{2})^2} \left( \sigma^2_{0,0} + 3 \ \partial_x^2 \, \sigma^2_{0,2}\right) \right) \quad + (2\leftrightarrow 3) \label{Eq:Cov_SSC^n=2_betaversion}
\ea
Here it is tempting to define
\ba
\kappa(r,x) \equiv G(r) \, G(x) \ \partial^2_x \, \sigma^2_{0,2} &= \frac{1}{2\pi^2} \int k^2 \, \dd k \ P(k|r,x) \ j_0(kr) \, j''_0(kx) 
\ea
and use that to write the $n=2$ term in the $(\sigma^2_{\delta_b,\delta_b},\kappa)$ basis. Using this basis is interesting for comparison with previous literature and we thus compute the results in Appendix~\ref{App:alt-base-SSC}. However, Fig.~\ref{Fig:SSC-kernels-suk} shows that this basis is highly degenerate. Indeed, $\kappa \simeq -\sigma^2_{\delta_b,\delta_b}$ which leads to a high cancellations of the first two terms in Eq.~\ref{Eq:Cov_SSC^n=2_betaversion}. By comparison, the $\upsilon$ defined previously for $n=1$ has a shape (and amplitude) significantly different from $\sigma^2_{\delta_b,\delta_b}$.

\begin{figure}[ht!]
    \begin{center}
        \includegraphics[width=.7\linewidth]{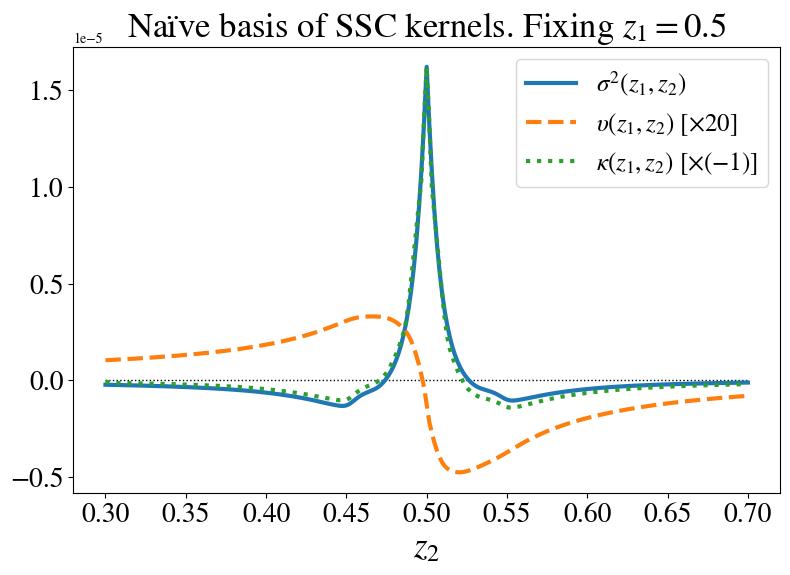}
         \caption{Comparison of some SSC kernels in the first naive basis. \textit{Solid blue:} the standard $\sigma^2_{\delta_b,\delta_b}$ that contributes to the term at $n=0$. \textit{Dashed orange:} $\upsilon$ that contributes to the term at $n=1$. \textit{Dotted green:} $-\kappa$, that contributes to the term at $n=2$. We included a $(-1)$ factor to make it more evident that $\kappa$ is essentially degenerate with $\sigma^2_{\delta_b,\delta_b}$.}
        \label{Fig:SSC-kernels-suk}
    \end{center}
\end{figure}

So, instead of using the $(\sigma^2_{\delta_b,\delta_b},\kappa)$ basis, we use the differential equation that defines spherical Bessel functions,
\ba
t^2 \, j''_0(t) + 2t \, j'_0(t) + t^2 \, j_0(t) = 0, 
\ea
to write $\kappa$ in terms of quantities defined previously, namely
\ba\label{Eq:kappa=-sig2-2upsilon}
\kappa = - \sigma^2_{\delta_b,\delta_b} -2 \upsilon.
\ea
This allows to write the $n=2$ term in the $(\sigma^2_{\delta_b,\delta_b},\upsilon)$ basis, in a form similar to Eq.~\ref{Eq:Cov_SSC^n=0}:
\ba
\Cov_\mr{SSC}^{n=2} = - 2 \int r^2 \, \dd r \ x^2 \, \dd x \ \frac{\partial n_\mr{t_1}(r)}{\partial \delta_b} \ \partial_2^A P_{n_{t_2},n_{t_3}}(k_{\ell}|x) \ \upsilon(r,x)  \\
- 2 \int r^2 \, \dd r \ x^2 \, \dd x \ \frac{\partial n_\mr{t_1}(r)}{\partial \delta_b} \ \partial_2^B P_{n_{t_2},n_{t_3}}(k_{\ell}|x) \ \sigma^2_{\delta_b,\delta_b}(r,x) \label{Eq:Cov_SSC^n=2}
\ea
where 
\ba
\partial_2^A P_{n_{t_2},n_{t_3}}(k_{\ell}|x) &= n_\mr{t_2} \, n_\mr{t_3} \left[ b^\mr{t_2}_{1} \, b^\mr{t_3}_{(2)} \left(1+3 \frac{\mathcal{L}^{t_2,(2)}_{2}}{(\ell+\sfrac{1}{2})^2}\right) + b^\mr{t_2}_{(2)} \, b^\mr{t_3}_{1} \left(1+3 \frac{\mathcal{L}^{t_3,(2)}_{2}}{(\ell+\sfrac{1}{2})^2}\right)\right] P(k_{\ell}|x) \label{Eq:d2AP-shortform} \\
&= n_\mr{t_2} \, n_\mr{t_3} \ P(k_{\ell}|x) \times \left[
\frac{4}{7} b^\mr{t_2}_{1} \, b^\mr{t_3}_{1} + b^\mr{t_2}_{1} \, b^\mr{t_3}_{s^2} + b^\mr{t_2}_{s^2} \, b^\mr{t_3}_{1} \right] \nonumber \\
& + \frac{3 \ n_\mr{t_2} \, n_\mr{t_3} \ P(k_{\ell}|x)}{(\ell+\sfrac{1}{2})^2} \left[ \frac{2}{7} b^\mr{t_2}_{1} \, b^\mr{t_3}_{1} \left(\mathcal{L}^{t_2,(2)}_{2} \!+\! \mathcal{L}^{t_3,(2)}_{2}\right) + b^\mr{t_2}_{1} \, b^\mr{t_3}_{s^2} \, \mathcal{L}^{t_2,(2)}_{2} + b^\mr{t_2}_{s^2} \, b^\mr{t_3}_{1} \, \mathcal{L}^{t_3,(2)}_{2} \right] \label{Eq:d2AP-longform} \\
\partial_2^B P_{n_{t_2},n_{t_3}}(k_{\ell}|x) &= \frac{n_\mr{t_2} \, n_\mr{t_3} \ P(k_{\ell}|x) }{(\ell+\sfrac{1}{2})^2} \left[ b^\mr{t_2}_{1} \, b^\mr{t_3}_{(2)} \ \mathcal{L}^{t_2,(2)}_{2} + b^\mr{t_2}_{(2)} \, b^\mr{t_3}_{1} \ \mathcal{L}^{t_3,(2)}_{2}\right] \label{Eq:d2BP-shortform} \\
&= \frac{n_\mr{t_2} \, n_\mr{t_3} \ P(k_{\ell}|x)}{(\ell+\sfrac{1}{2})^2} \left[ \frac{2}{7} b^\mr{t_2}_{1} \, b^\mr{t_3}_{1} \left(\mathcal{L}^{t_2,(2)}_{2} \!+\! \mathcal{L}^{t_3,(2)}_{2}\right) + b^\mr{t_2}_{1} \, b^\mr{t_3}_{s^2} \, \mathcal{L}^{t_2,(2)}_{2} + b^\mr{t_2}_{s^2} \, b^\mr{t_3}_{1} \, \mathcal{L}^{t_3,(2)}_{2} \right]. \label{Eq:d2BP-longform}
\ea

\subsubsection{Total}\label{Sect:SSC-total}
\begin{figure}[ht!]
    \begin{center}
        \includegraphics[width=.7\linewidth]{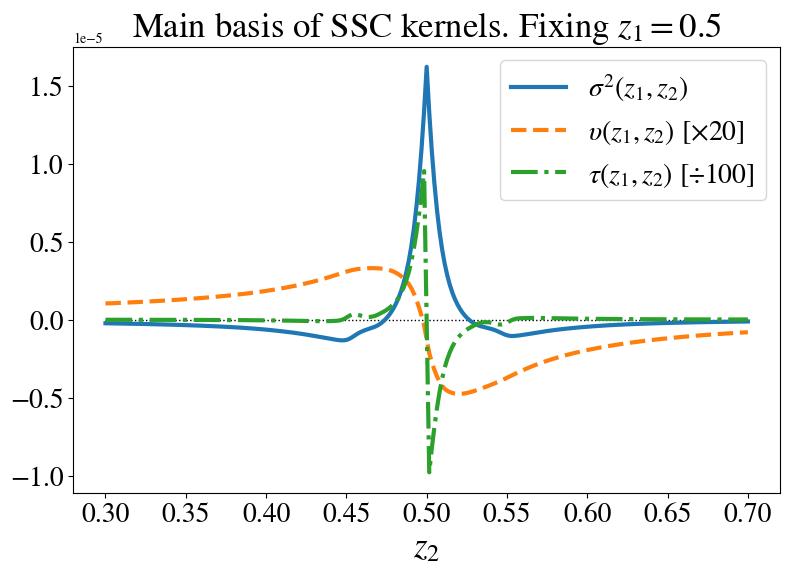}
         \caption{Comparison of the three SSC kernels in the final basis. \textit{Solid blue:} the standard $\sigma^2_{\delta_b,\delta_b}$. \textit{Dashed orange:} $\upsilon$ ; it is significantly smaller than $\sigma^2_{\delta_b,\delta_b}$, we multiply it by a factor 20 to fit on the same plot. \textit{Dash-dotted green:} $\tau$ ; it is significantly larger than $\sigma^2_{\delta_b,\delta_b}$, we divide it by a factor 100 to fit on the same plot.}
        \label{Fig:SSC-kernels-sut}
    \end{center}
\end{figure}

When we sum the $n=0,1,2$ terms, we get
\ba
\Cov_\mr{SSC} &= \int r^2 \, \dd r \ x^2 \, \dd x \ \frac{\partial n_\mr{t_1}(r)}{\partial \delta_b} \ \frac{\partial P_{n_{t_2},n_{t_3}}(k_{\ell}|x)}{\partial \delta_b} \ \sigma^2_{\delta_b,\delta_b}(r,x) \nonumber \\
&+ \int r^2 \, \dd r \ x^2 \, \dd x \ \frac{\partial n_\mr{t_1}(r)}{\partial \delta_b} \ \frac{\partial P_{n_{t_2},n_{t_3}}(k_{\ell}|x)}{\partial x_b} \ \upsilon(r,x) \\
&+ \int r^2 \, \dd r \ x^2 \, \dd x \ \frac{\partial n_\mr{t_1}(r)}{\partial \delta_b} \ \frac{\partial P_{n_{t_2},n_{t_3}}(k_{\ell}|x)}{\partial y_b} \ \tau(r,x)
\label{Eq:Cov_SSC_total}
\ea
where the kernels are
\ba
\sigma^2_{\delta_b,\delta_b}(r,x) &= \frac{1}{2\pi^2} \int k^2 \, \dd k \ P(k|r,x) \ j_0(kr) \, j_0(kx) \\
\upsilon(r,x) &= \frac{1}{2\pi^2} \int k^2 \, \dd k \ P(k|r,x) \ j_0(kr) \, \frac{j'_0(kx)}{kx} \\
\tau(r,x) &= \frac{1}{2\pi^2} \int k^2 \, \dd k \ P(k|r,x) \ j_0(kr) \, kx \, j'_0(kx)
\ea
and are plotted in Fig.~\ref{Fig:SSC-kernels-sut}.

Next, the responses are
\ba
\frac{\partial P_{n_{t_2},n_{t_3}}}{\partial \delta_b} &\equiv \partial_0 P_{n_{t_2},n_{t_3}} - 2 \, \partial_2^B P_{n_{t_2},n_{t_3}} \label{Eq:dP/ddeltab} \\
\frac{\partial P_{n_{t_2},n_{t_3}}}{\partial x_b} &\equiv \partial_1^A P_{n_{t_2},n_{t_3}} - 2 \, \partial_2^A P_{n_{t_2},n_{t_3}} \label{Eq:dP/dxb} \\
\frac{\partial P_{n_{t_2},n_{t_3}}}{\partial y_b} &\equiv  \partial_1^B P_{n_{t_2},n_{t_3}} \label{Eq:dP/dyb}
\ea
with explicit expressions:
\ba
\frac{\partial P_{n_{t_2},n_{t_3}}}{\partial \delta_b} 
=& \ n_\mr{t_2} \, n_\mr{t_3} \ P(k_{\ell}|x) \times \left[
\frac{20}{7} b^\mr{t_2}_1 \, b^\mr{t_3}_1 + b^\mr{t_2}_1 \, b^\mr{t_3}_2 + b^\mr{t_2}_2 \, b^\mr{t_3}_1 - \frac{2}{3} \left(b^\mr{t_2}_1 \, b^\mr{t_3}_{s^2} + b^\mr{t_2}_{s^2} \, b^\mr{t_3}_1\right)
\right] \nonumber \\
& -2 \ \frac{n_\mr{t_2} \, n_\mr{t_3} \ P(k_{\ell}|x)}{(\ell+\sfrac{1}{2})^2} \left[ \frac{2}{7} b^\mr{t_2}_{1} \, b^\mr{t_3}_{1} \left(\mathcal{L}^{t_2,(2)}_{2} \!+\! \mathcal{L}^{t_3,(2)}_{2}\right) + b^\mr{t_2}_{1} \, b^\mr{t_3}_{s^2} \, \mathcal{L}^{t_2,(2)}_{2} + b^\mr{t_2}_{s^2} \, b^\mr{t_3}_{1} \, \mathcal{L}^{t_3,(2)}_{2} \right] \\
\frac{\partial P_{n_{t_2},n_{t_3}}}{\partial x_b} 
=& \ n_\mr{t_2} \, n_\mr{t_3} \ P(k_{\ell}|x) \times \left[
-\frac{8}{7} b^\mr{t_2}_{1} \, b^\mr{t_3}_{1} - 2 b^\mr{t_2}_{1} \, b^\mr{t_3}_{s^2} - 2 b^\mr{t_2}_{s^2} \, b^\mr{t_3}_{1} 
+ b^\mr{t_2}_1 \, b^\mr{t_3}_{1} \left(\mathcal{L}^{t_2,(1)}_{0} + \mathcal{L}^{t_3,(1)}_{0} \right) \right] \nonumber \\
& - \frac{6 \ n_\mr{t_2} \, n_\mr{t_3} \ P(k_{\ell}|x)}{(\ell+\sfrac{1}{2})^2} \left[ \frac{2}{7} b^\mr{t_2}_{1} \, b^\mr{t_3}_{1} \left(\mathcal{L}^{t_2,(2)}_{2} \!+\! \mathcal{L}^{t_3,(2)}_{2}\right) + b^\mr{t_2}_{1} \, b^\mr{t_3}_{s^2} \, \mathcal{L}^{t_2,(2)}_{2} + b^\mr{t_2}_{s^2} \, b^\mr{t_3}_{1} \, \mathcal{L}^{t_3,(2)}_{2} \right] \\
\frac{\partial P_{n_{t_2},n_{t_3}}}{\partial y_b} 
=& \ \frac{n_\mr{t_2} \, n_\mr{t_3} \ P(k_{\ell}|x)}{(\ell+\sfrac{1}{2})^2} \times  b^\mr{t_2}_1 \, b^\mr{t_3}_{1} \left(\mathcal{L}^{t_2,(1)}_{2} + \mathcal{L}^{t_3,(1)}_{2} \right)
\ea

\begin{figure}[ht!]
    \begin{center}
        \includegraphics[width=\linewidth]{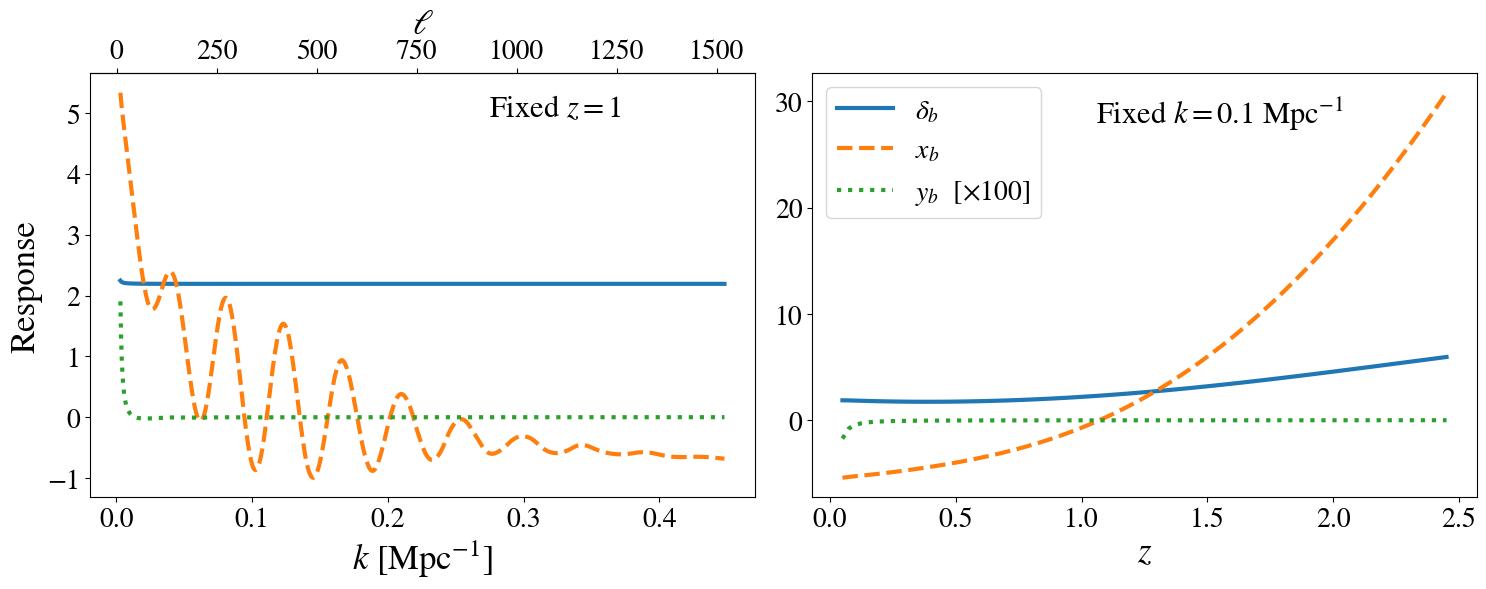}
         \caption{The logarithmic responses $\frac{\partial\ln P}{\partial v}(k,z)$ for the three variables $v=\delta_b$ (solid blue), $v=x_b$ (dashed orange) and $v=y_b$ (dotted green). \textit{Left:} plot as a function of Fourier mode $k$ with fixed redshift $z=1$; the equivalent Limber multipole $\ell=kr$ is shown in the top x axis. \textit{Right:} plot as a function of redshift $z$ with fixed Fourier mode $k=0.1$ Mpc$^{-1}$. The tracer chosen here for illustration is the expected sample of Euclid photometric galaxies. The details of the specifications and modeling are covered in Appendix~\ref{App:specs-models-phot}.}
        \label{Fig:Responses}
    \end{center}
\end{figure}

Recalling $P_{n_{t_2},n_{t_3}} = n_\mr{t_2} \, n_\mr{t_3} \ b_1^\mr{t_2} \, b_1^\mr{t_3} \ P(k_{\ell}|x)$, we can write equivalently the logarithmic responses:
\ba
\frac{\partial \ln P_{n_{t_2},n_{t_3}}}{\partial \delta_b} 
=& \ \frac{20}{7} + \frac{b^\mr{t_2}_2}{b^\mr{t_2}_1} - \frac{2}{3} \frac{b^\mr{t_2}_{s^2}}{b^\mr{t_2}_1}
- \ \frac{2}{(\ell+\sfrac{1}{2})^2} \left[ \frac{2}{7} \mathcal{L}^{t_2,(2)}_{2} + \frac{b^\mr{t_2}_{s^2}}{b^\mr{t_2}_1} \, \mathcal{L}^{t_3,(2)}_{2} \right] + (2\leftrightarrow 3)
\label{Eq:dlnP/ddeltab} \\
\frac{\partial \ln P_{n_{t_2},n_{t_3}}}{\partial x_b} 
=& \ -\frac{8}{7} - 2 \frac{b^\mr{t_2}_{s^2}}{b^\mr{t_2}_1} + \mathcal{L}^{t_2,(1)}_{0} 
- \frac{6}{(\ell+\sfrac{1}{2})^2} \left[ \frac{2}{7} \mathcal{L}^{t_2,(2)}_{2} + \frac{b^\mr{t_2}_{s^2}}{b^\mr{t_2}_1} \, \mathcal{L}^{t_3,(2)}_{2} \right] + (2\leftrightarrow 3) 
\label{Eq:dlnP/dxb} \\
\frac{\partial \ln P_{n_{t_2},n_{t_3}}}{\partial y_b} 
=& \ \frac{\mathcal{L}^{t_2,(1)}_{2} + \mathcal{L}^{t_3,(1)}_{2}}{(\ell+\sfrac{1}{2})^2}.
\label{Eq:dlnP/dyb} 
\ea
These are dimensionless and simpler, and allow comparison with literature. We illustrate these responses in Fig.~\ref{Fig:Responses} in the case of the auto-spectrum ($t_2=t_3$) of the expected sample of Euclid photometric galaxies.

First, in Fig.~\ref{Fig:Responses} we can look at the wide-angle corrections (terms with a $1/(\ell+\sfrac{1}{2})^2$ prefactors). These are the $y_b$ response and the tiny spike visible in the $\delta_b$ response at low $k$. We see that these corrections have a negligible impact: the $y_b$ response reaches percent level only at the lowest $k$ and $z$ ; for the first two responses ($\delta_b$ and $x_b$) the correction is barely visible, we found it reaches a few percent only at the lowest $k$.\\
Second, we examine the $\delta_b$ response. On one hand it is effectively constant with $k$, as the wide-angle correction is negligible and we assumed scale-independent biases, as standard on large scales. On the other hand it increases slightly with redshift ; we found it is due to the term $b_2/b_1$ which increases with redshift: high redshift galaxies are more biased and hence more non-linear than low redshift ones.\\
Third, we examine the $x_b$ response. It has both a scale dependence and redshift dependence. We found that this is essentially due to the $\mathcal{L}^{(1)}_{0}$ term, while the $b_{s^2}/b_1$ is roughly constant. The scale dependence is due to the dilation sub-term in $\mathcal{L}^{(1)}_{0}$, which is a derivative of the matter power spectrum. Specifically, the oscillations visible in the response come from (derivatives of) the Baryon Acoustic Oscillations. The redshift dependence is due to the evolution sub-term in $\mathcal{L}^{(1)}_{0}$. Specifically, the large increase with redshift is due to the fact that the comoving density of galaxies falls steeper and steeper as we approach the faint end of the sample.

Finally, the new kernels and responses can be interpreted physically similarly to the growth-only term where $\sigma^2_{\delta_b,\delta_b}(r_1,r_2) = \Cov\left( \delta_b(z_1) , \delta_b(z_2) \right)$ with the so-called background density perturbation $\delta_b(z)=\lbra \delta(\hn,z)\rbra_\mr{survey}$ the spatial average of $\delta$ over the survey footprint (full sky here). Indeed, we find that $\upsilon(r_1,r_2) = \Cov\left( \delta_b(z_1) , x_b(z_2) \right)$ where $x_b$ is the (rescaled) background radial velocity:
\ba
x_b(z) = G(z) \lbra \vv_{(1),r}\rbra_\mr{survey}
\ea
with $\vv_{(1),r}$ the radial projection of the (time-independent part of the) first order velocity field.\\
Furthermore we find that $\tau(r_1,r_2)=\Cov\left( \delta_b(z_1) , y_b(z_2) \right)$ where $y_b$ is the (rescaled) background radial gradient density perturbation:
\ba
y_b(z) = G(z) \lbra \nabla_{\ln r} \, \delta_{(1)} \rbra_\mr{survey}.
\ea
with $\delta_{(1)}$ the (time-independent part of the) first order density field.\\
The rescaling by the growth function is there to provide the correct time dependence to the end result. Indeed, as shown in Sect.~\ref{Sect:perturbation-theory}, perturbation theory relates the time-independent parts of the $n$-th order density contrast or velocity fields, the time dependence being taken care of by powers of the growth factor or its derivative. Our end result (the covariance) has a $G^4$ time dependence ; for backward compatibility with previous literature we want this dependence to be absorbed by the kernels, so we have to rescale by the correct growth factors.

\section{Discussion and conclusions}\label{Sect:discussion}
Here we summarise and discuss our findings.

\subsection*{A new formalism for angular polyspectra}
We have developed an alternative formalism to compute the projected angular bispectrum from the 3D bispectrum in perturbation theory. Here, derivative couplings induced by the fluid dynamics are directly reprojected on the celestial sphere. It allows us to treat angle-dependent terms in the 3D bispectrum without resorting to a trigonometric method, i.e. where angles are expressed as functions of the triangle sides. It results in an angular bispectrum which is organized as the sum of a scalar, a vector and a tensor contribution on the plane tangent to the line-of-sight. Each of these contributions finds its origin in the fluid dynamics. First, the scalar part is sourced by the growth-only, the transport and the advection contributions. Second, the vector part is sourced by both the transport and the advection contribution and it vanishes for $\ell=0$. Third, the tensor part is solely sourced by advection and it vanishes for $\ell=0$ and $\ell=1$. 

At a technical level, it leads to an expression of the reduced angular bispectrum summarized in Table~\ref{Table:bisp-summary-noLimber} allowing for a fast and accurate evaluation of it.  Indeed, it solely involves Bessel functions (no derivatives, no Wigner symbols) and as such, it is amenable to beyond-Limber numerical methods \cite[e.g.][]{Assassi2017,Grasshorn2017,Fang2019,LSST2022,Feldbrugge2023,Raccanelli2023,Chiarenza2024}. Moreover, its expression does not involve subtle cancellations in any triangle configuration, in contrast with the trigonometric approach in the squeezed limit. It should relax precision requirements on integration in the aforementioned numerical methods, speeding up the computations as necessary for the analysis of the future high volume data of stage IV surveys of the Large Scale Structures. Furthermore, our equations are amenable to the Limber approximation in any triangle configuration, including the squeezed limit where trigonometric Limber fails; hence making the computation of the bispectrum even faster. This is important because the squeezed limit is of particular interest for searches of primordial non-Gaussianity \citep[e.g.][]{Chudaykin2025,Finelli2025} and tests of consistency relations of the Large Scale Structure \cite{Kehagias2013,Creminelli2013}.

Our method is generic and could be extended to higher order polyspectra. In future, it should be applied to the trispectrum (4-point function) which is needed to compute the covariance of the angular power spectrum. We believe this would cure the problem noted by \cite{Lee2020} who found that Limber's approximation performed poorly for this trispectrum-induced covariance, due to subtle cancellations. Indeed, they used the trigonometric method, i.e. writing angles as a function of the quadrilateral's sides, and we have shown here for the bispectrum that this method does not work in tandem with Limber. Our formalism cures this issue at the bispectrum level as it yields equations free of such subtle cancellations and we expect that the same will apply to the trispectrum.

\subsection*{A new view on Super-Sample Covariance}
Our projection results also allow new calculations of the impact of non-Gaussianity on the covariance of LSS observables. In particular we found new important results on Super-Sample Covariance (SSC). 
For this discussion it is helpful to recall that the $F_2$ kernel from perturbation theory is obtained by solving to second order the fluid equations of motion Eqs~\ref{Eq:Poisson}-\ref{Eq:Navier-Stokes}. It is organized as a sum of powers $\left(\kk_i\cdot\kk_j\right)^n$ with $n=0,1,2$. $n=0$ is the growth-only term and comes from the $(\delta_{(1)})^2$ contribution to $\delta_{(2)}$. $n=1$ is the transport term and comes from the $\nabla \delta_{(1)}\cdot \vv_{(1)}$ contribution. $n=2$ is the advection term and comes from the $\partial_i v_j^{(1)} \times \partial_j v_i^{(1)}$ contribution.

\paragraph*{Answering decade-old questions}
Flat-sky derivations of SSC in the literature have found terms beyond the growth-only one which are called the dilation \cite{Li2014} and tidal \cite{Akitsu2017,Li2018} terms. The naming and interpretation of these terms come from a 3D perspective on the impact of the super-sample modes. As such, after the first full-sky derivation of the growth-only part of SSC by \cite{Lacasa2016}, it was an open question whether these dilation and tidal terms could come out from a full-sky light-cone perspective, and how this would happen. Furthermore, the functional form of these terms leads to other questions, presented in the introduction and that we recap here: why do we only have a spatial derivative and not some temporal derivative? Since $n=1$ yields a first derivative term, why is there no second derivative term coming from $n=2$? Why do only two independent background variables (the background density and tidal tensor) come into play? In other words, why does $n=2$ gives rise to a new background variable but not $n=1$?

The results of this article answer all these questions.\\
Let us start with the ones easiest to answer: the derivatives. We do have the derivative terms intuited in the questions above: (i) a first order temporal derivative that comes in addition to the spatial derivative and is sensitive to the tracers' evolution, it is present in Eq.\ref{Eq:d1AP} through the symbol $\mathscr{L}^{t,(1)}$ defined by Eqs~\ref{Eq:def-scrL^t,c_alpha}-\ref{Eq:def-L^t,c_alpha}, and (ii) second order derivatives (spatial, time and mixed) coming from $n=2$, that are present in Eqs~\ref{Eq:d2AP-shortform}-\ref{Eq:d2BP-shortform}. The reasons why these terms were not found in the 3D + flatsky derivations in the literature are (i) because the derivation implicitly sits at constant time, so it cannot have time derivatives, and (ii) the second order-derivative terms are suppressed by a factor $1/(\ell+\sfrac{1}{2})^2$, i.e. they are wide-angle effects not captured by the flat-sky approximation.\\
Regarding the number of background variables, or equivalently the number of SSC kernels, we find that this number is three, instead of the two found in previous literature. However, one of these kernels ($\tau$) comes into play only for terms suppressed by a factor $1/(\ell+\sfrac{1}{2})^2$, i.e. again wide-angle effects not captured by the flat-sky approximation.\\
Finally, the (hardest) question on how the dilation and tidal effects appear in curved sky. From a technical (but simplified) point of view, it is the factors $\kk$ in the $F_2$ kernel that are transformed into derivatives, and these derivatives later\footnote{After being decomposed into transversal and radial parts, and then being transformed by integrations by parts.} hit the tracer power spectrum or Bessel functions. We get the dilation term when a first derivative hits the power spectrum, and the tidal term when a second derivative hits a Bessel function. From a more physical point of view, everything comes back to the fundamental fluid equations Eqs~\ref{Eq:Continuity}-\ref{Eq:Navier-Stokes} in terms of the perturbations of density $\delta$ and velocity $\vv$ \footnote{The gravitational potential $\Phi$ is trivially solved via the Poisson equation $\Delta \Phi \propto \rho$ so in Fourier space $k^2 \Phi(\kk) \propto \delta(\kk)$}. The velocity can be seen as a derivative (through the vector $\kk$ in Eq.~\ref{Eq:first-order-velocity}) and the question is whether its orientation is transversal or radial. When the second order density contrast $\delta_{(2)}$ gets contribution from a single radial velocity we have the dilation term, when it gets contribution from two radial velocities we have the tidal term.

\paragraph*{New effects I: wide-angle}
Compared to previous flat-sky literature, we find several new SSC terms. A class of them are wide-angle effects: they contain a factor $1/(\ell+\sfrac{1}{2})^2$ and thus become irrelevant when $\ell \gg 1$. We give below an interpretation of the wide-angle term coming from $n=1$, and only sketch the case for the $n=2$ terms which are more complex.

For $n=1$ (transport term), recall that this comes from $\nabla\delta_{(1)}\cdot\vv_{(1)}$ in the fluid equations. For simplicity we drop the subscript $(1)$ from now on. In Fourier space we have $\vv(\kk)\propto \delta(\kk) \, \kk/k^2$, so the permutations of $\kk_1$ and $\kk_2$ give two contributions: (i) $\nabla\delta(\kk_1)\cdot\vv(\kk_2)$ yields a $1/k_2^2$ and hence a wide-angle effect, (ii) $\nabla\delta(\kk_2)\cdot\vv(\kk_1)$ yields a $1/k_1^2$ and is dominant if $\ell\gg 1$. Indeed, it can be seen that the $1/k_2^2$ in (i) propagates into a $1/(\ell_2+\sfrac{1}{2})^2$ in the angular bispectrum (precisely in the second term of $C_1^S$ in Table~\ref{Table:bisp-summary-Limber}) and then into a $1/(\ell+\sfrac{1}{2})^2$ in the SSC.\\
There is another way to see that contribution (i) is negligible compared to (ii) in the high multipole limit. Indeed the two differ by a factor $(k_1/k_2)^2$. Recall that in the Limber approximation $k=(\ell+\sfrac{1}{2})/r$. If $\ell\gg 1$ we have $\ell_1=0 \ll \ell=\ell_2$, thus $k_1\ll k_2$ and $(k_1/k_2)^2 \ll 1$.

The $n=2$ term is more complex\footnote{For several reasons. First the calculations are more complex because we now have velocity squared so second derivatives which can be either radial or transverse. Second, it leads to many subterms. Third, some of these subterms recombine together so we cannot give a single origin to a final subterm. Fourth and finally, some subterms were transformed during the calculations, e.g. when we used the differential equation of Bessel function Eq.\ref{Eq:differential-equation-spherical-Bessel}, or when rewrote the new $\kappa$ kernel in terms of the already introduced SSC kernels, Eq.~\ref{Eq:kappa=-sig2-2upsilon}}. At first glance it should only provide wide-angle effects since it contains a $1/k_2^2$ factor, which lead to a wide-angle effect for $n=1$. This intuition allows us to understand the last term of $C_2^S$ in Table~\ref{Table:bisp-summary-Limber}, which leads to a wide-angle effect in SSC. This term is the most intuitive because it contains a $1/(\ell_1+\sfrac{1}{2})^2 (\ell_2+\sfrac{1}{2})^2$ coming from the $1/k_1^2 k_2^2$ in $F_2$, and second derivatives (in the $\mathcal{L}^{t,(2)}$ factors) coming from the $(\kk_1\cdot\kk_2)^2$. Since this term is transparent, the question becomes why are there more? \\
The reason is second derivatives. Technically, as explained in Section~\ref{Sect:angular-integrals}, a spherical covariant derivative (of a scalar function) is the same as the projection of the 3D derivative, but this is no longer true when we take two successive derivatives (because the derivative of a scalar is a vector). We did not have this issue for $n=1$ because we had only single derivatives. So for $n=2$ we have new terms that involve the Christoffel symbols of the covariant derivative and lead to extra terms in $C_2^S$ in Table~\ref{Table:bisp-summary-Limber} which can be traced back to the first line of $\mathcal{I}^{(2),\mathrm{S}}$ in Table~\ref{Table:radkernels}. For instance, one of these terms has no Bessel derivatives and then propagates in the equations exactly as $n=0$.\\
Beyond this technical explanation, let us remark that something analogous happens, though with simpler maths, in the seminal 3D SSC derivation by \cite{Takada2013}: the constant (growth-only) part of SSC not only takes contribution from $n=0$ but also from $n=2$ due to the azimuthal-angle average and $\int \dd \mu_{12} (\kk_1\cdot\kk_2)^2 = \frac{1}{3} \neq 0$. In other words, the 3D literature allows to expect a contribution from $n=2$ beyond wide-angle, and the detailed explanation is found in the technical calculations of Section~\ref{Sect:angular-integrals}.

\paragraph*{New effects II: key differences on seminal terms}
Beyond the wide-angle effects, we want to compare our results with the 3D literature. To do so, we first need to neglect the already-discussed wide-angle effects, and thus neglect the impact of $\tau$. Then we need to remap our results to the $(\sigma^2,\kappa)$ basis of SSC kernels instead of the $(\sigma^2,\upsilon)$ basis of Section~\ref{Sect:SSC-total}. Indeed, $\kappa$ is the kernel for the so-called tidal term in the 3D literature, see e.g. \cite{Barreira2018}. This basis change is a linear mapping detailed in Appendix~\ref{App:alt-base-SSC}. The result of this appendix that we want to discuss here is Eq.~\ref{App:Eq:dlnP/ddeltab-altbasis} that gives the power spectrum response to a change of background density, $\frac{\partial \ln P}{\partial \delta_b}$. We recall here that equation for convenience:
\ba
\frac{\partial \ln P_{n_{t_2},n_{t_3}}}{\partial \delta_b} 
=&  \frac{24}{7} + \frac{b^\mr{t_2}_2 + \frac{1}{3} b^\mr{t_2}_{s^2}}{b^\mr{t_2}_1}
+ \frac{b^\mr{t_3}_2 + \frac{1}{3} b^\mr{t_3}_{s^2}}{b^\mr{t_3}_1} 
- \frac{1}{2} \left( \mathcal{L}^{t_2,(1)}_{0} + \mathcal{L}^{t_3,(1)}_{0} \right) \nonumber \\
& - \frac{1}{(\ell+\sfrac{1}{2})^2} \left( \frac{\frac{2}{7} b^\mr{t_2}_1 + b^\mr{t_2}_{s^2}}{b^\mr{t_2}_1} \mathcal{L}^{t_3,(2)}_{2} + \frac{\frac{2}{7} b^\mr{t_3}_1 + b^\mr{t_3}_{s^2}}{b^\mr{t_3}_1} \mathcal{L}^{t_2,(2)}_{2} \right).
\ea
This response can be compared to the seminal result from \cite{Takada2013}. Specifically we compare with the equation added in a post-publication note that includes the dilation effect found in \cite{Li2014}, that is Eq.44 found in the acknowledgements in the arXiv version v3. In our notations, and neglecting the small scale 1-halo term, this equation reads
\ba
\frac{\partial \ln P_{n_{t_2},n_{t_3}}}{\partial \delta_b} 
=  \frac{68}{21} - \frac{1}{3} \frac{\dd \ln (k^3 P_{n_{t_2},n_{t_3}}(k))}{\dd \ln k} 
\ea
Our result has several key differences:
\begin{itemize}
    \item Secondary differences
    \begin{itemize}
        \item We have a term sensitive to the quadratic bias $b_2$. This was already found in \cite{Lacasa2016} and arises when studying the case of a biased tracer instead of pure matter that \cite{Takada2013} studied.\\
        On the topic of second-order bias terms, we also have a new term sensitive to the tidal bias $b_{s^2}$ that we discuss more later.
        \item We have a new term with a $\frac{1}{(\ell+\sfrac{1}{2})^2}$ prefactor. This is a wide angle effect, already discussed in the previous paragraph
    \end{itemize}
    \item Primary differences: 2+1 is not 3
    \begin{itemize}
        \item Constant term: we have $\frac{24}{7} = 4 \times \left( \frac{5}{7}+\frac{1}{2}\times\frac{2}{7}\right)$ while \cite{Takada2013} has $\frac{68}{21} = 4 \times \left( \frac{5}{7}+\frac{1}{3}\times\frac{2}{7}\right)$, where the $\frac{5}{7}$ and $\frac{2}{7}$ are the numerical coefficients appearing in the $F_2$ kernel.\\
        This factor $\frac{1}{2}$ instead of $\frac{1}{3}$ is not a happy coincidence, and we will find it again later. It comes from the dimension: we have 2 transversal dimensions while \cite{Takada2013} operates in 3D. Specifically the $\frac{1}{3}$ appears in \cite{Takada2013} when they compute the $n=2$ part of the azimuthal-angle average of $F_2(\kk_1,\kk_2)$. Indeed in 3D they compute $\lbra\cos^2\theta\rbra=\frac{1}{2}\int_0^\pi \sin\theta\dd\theta \, \cos^2\theta=\frac{1}{3}$, where $\theta$ is the angle between $\kk_1$ and $\kk_2$. If instead one computes this average for 2D vectors, one finds $\lbra\cos^2\theta\rbra=\frac{1}{2}$ \footnote{More generally, in dimension $d$ this average is $1/d$.}. This gives intuition for the appearance of the $\frac{1}{2}$, although we stress that that this is not how we derived it: the factor comes from the full derivation with spherical harmonics.
        \item Tidal bias term: we have $\frac{1}{3}\frac{b^\mr{t_2}_{s^2}}{b^\mr{t_2}_1}$ while the 3D derivation in \cite{Lacasa2016} (which generalises \cite{Takada2013} to biased tracers) finds that this contribution cancels out.\\
        This happens because our computation features $b^\mr{t_2}_{(0)} + \frac{1}{2} b^\mr{t_2}_{(2)}$ \footnote{Specifically, what enters the response is $2 \times \left(b^\mr{t_2}_{(0)} + \frac{1}{2} b^\mr{t_2}_{(2)}\right) + (2\leftrightarrow 3)$} while the 3D computation features $b^\mr{t_2}_{(0)} + \frac{1}{3} b^\mr{t_2}_{(2)}$. In the former there is a relic tidal bias contribution while in the latter the tidal contributions from $n=0$ and $n=2$ cancel each other.
        \item Dilation+evolution term: we have $-\frac{1}{2} \mathcal{L}^{t_2,(1)}_{0}$ while \cite{Takada2013} has $- \frac{1}{3} \frac{\dd \ln (k^3 P_{n_{t_2},n_{t_3}}(k))}{\dd \ln k}$. First, we again have a factor $\frac{1}{2}$ instead of $\frac{1}{3}$. Second, the $\mathcal{L}^{t_2,(1)}_{0}$ term does contain a spatial derivative as in \cite{Takada2013}: this is the dilation part. However it also contains a new temporal derivative that we call the evolution part (see e.g. Appendix~\ref{App:props-L-symbols}). This comes from the 2+1 splitting: the last dimension is the light-cone coordinate. This effect is completely new and means that Super-Sample Covariance is sensitive to the redshift evolution of the tracer we consider.
    \end{itemize}
\end{itemize}

\paragraph*{New and potentially simpler total SSC}
On top of finding new effects in the power spectrum response, we have also expanded SSC into a new kernel basis, $(\sigma^2,\upsilon,\tau)$, instead of the $(\sigma^2,\kappa)$ basis used in the 3D literature. We think our new basis has several key advantages. \\
First, we think this basis is more natural because it can be interpreted in terms of fluid quantities that appear directly in the starting fluid equations (continuity and Navier-Stokes), whereas the 3D basis involves the tidal tensor which is not present in those equations. It is well known that $\sigma^2$ can be interpreted as $\Cov\left(\delta_b,\delta_b\right)$ where $\delta_b$ is the background density perturbation, i.e. $\delta_b(z)=\lbra \delta\rbra_\mr{survey}$ is the average of $\delta$ over the survey in the considered redshift slice. In a similar manner, we found that $\upsilon$ can be interpreted in terms of the (rescaled) background radial velocity: $x_b(z) = G(z) \lbra \vv_{(1),r}\rbra_\mr{survey}$ with $\vv_{(1),r}$ the radial projection of the first order velocity field. Furthermore we found that $\tau$ can be interpreted in terms of the background radial gradient density perturbation: $y_b(z) = G(z) \lbra \nabla_{\ln r} \, \delta \rbra_\mr{survey}$. This is a more natural basis because $\delta$, $\vv$ and $\nabla\delta$ are quantities that appear directly in the fluid equations. And indeed we used these quantities for the interpretation of our new effects in the previous paragraphs. \\
Second, $\sigma^2$ and $\upsilon$ have significantly different shapes, as seen in Fig.~\ref{Fig:SSC-kernels-suk}. In contrast, $\sigma^2$ and $\kappa$ are near perfectly degenerate and cancel each other out, meaning that expensive precision is necessary in numerical applications.
Third, the dilation + evolution effect is isolated to the SSC term containing $\upsilon$. This links with the fourth advantage: we found numerically that the $\upsilon$ kernel is negligible with respect to $\sigma^2$. Thus, at least for the covariance we have studied, it is possible to effectively neglect the complicated dilation + evolution term. If we couple this approximation with that of neglecting wide-angle effects for $\ell\gg 1$, we arrive at a compact equation for the SSC:
\ba
\Cov_\mr{SSC} &\simeq \int r^2 \, \dd r \ x^2 \, \dd x \ \frac{\partial n_\mr{t_1}(r)}{\partial \delta_b} \ \frac{\partial P_{n_{t_2},n_{t_3}}(k_{\ell}|x)}{\partial \delta_b} \ \sigma^2_{\delta_b,\delta_b}(r,x)
\label{Eq:Cov_SSC_total_simpler}
\ea
with the response
\ba
\frac{\partial P_{n_{t_2},n_{t_3}}}{\partial \delta_b} 
&\simeq \ n_\mr{t_2} \, n_\mr{t_3} \ P(k_{\ell}|x) \times \left[
\frac{20}{7} b^\mr{t_2}_1 \, b^\mr{t_3}_1 + b^\mr{t_2}_1 \, b^\mr{t_3}_2 + b^\mr{t_2}_2 \, b^\mr{t_3}_1 - \frac{2}{3} \left(b^\mr{t_2}_1 \, b^\mr{t_3}_{s^2} + b^\mr{t_2}_{s^2} \, b^\mr{t_3}_1\right)
\right] \label{Eq:dP/ddeltab_simpler}
\ea
or equivalently
\ba
\frac{\partial \ln P_{n_{t_2},n_{t_3}}}{\partial \delta_b} 
=& \ \frac{20}{7} + \frac{b^\mr{t_2}_2}{b^\mr{t_2}_1} - \frac{2}{3} \frac{b^\mr{t_2}_{s^2}}{b^\mr{t_2}_1} + (2\leftrightarrow 3)
\label{Eq:dlnP/ddeltab_simpler}
\ea
This is effectively much simpler than past SSC equations in the literature: there is no tidal effect, no dilation effect, and on large scales the response is independent of scale.

\paragraph*{Opening}
It is an open question whether similar results hold for the more complex case of the power spectrum covariance $\Cov\left(C_\ell,C_{\ell'}\right)$ which is sensitive to the trispectrum (4-point function). This will be the subject of future works. Another open question raised by our results, is on how the SSC equations transition from these full-sky results to the seminal results derived in the flat-sky limit. This means studying the complex case of partial-sky coverage. In the case of the growth-only (angle-independent) SSC term, \cite{Lacasa2018} have shown how this transition happen, with the flat-sky equation (which is a single redshift integral) being the limit of the partial-sky equation (which is a double redshift integral) for small enough sky coverage. \cite{Lacasa2018} also showed that the flat-sky limit is effectively a poor approximation for modern surveys covering more than 25 deg$^2$. On the contrary, \cite{Beauchamps2021} showed that, after a simple rescaling by the sky fraction, the full-sky covariance equation gives a very good approximation of SSC for modern surveys. This lets us expect that the results derived in this article will be the most relevant ones for coming surveys of the Large Scale Structures, such as Euclid, LSST and Roman.

\section*{Acknowledgements}
FL thanks Matteo Rizzato for inspiring discussions on Super-Sample Covariance.\\ 
The calculations in this article made use of \texttt{\href{https://class-code.net}{class}} \citep{Blas2011}, \texttt{matplotlib} \citep{Hunter2007}, \texttt{scipy} \citep{scipy}, and \texttt{numpy} \citep{numpy}. 

\appendix
\addtocontents{toc}{\protect\setcounter{tocdepth}{1}} 

\section{Mathematical materials}
\label{app:math}
\subsection{Geometry on the sphere}
\label{sapp:sphere}
We use the polar and azimuthal angles, $(\theta,\varphi)$, as the general coordinate system on the sphere. The components of the metric tensor in the natural basis $(\dd\theta,\dd\varphi)$ is
\ba
	\bar{g}_{ij}=\left(\begin{array}{cc}
		1 & 0 \\
		0 & \sin^2\theta
	\end{array}\right). \label{eq:spheremetric}
\ea
The covariant derivative which is compatible with the above metric is denoted hereafter as $\bar{\boldsymbol{\nabla}}$. Its components in the natural basis associated to $(\theta,\varphi)$ are obtained from the Christoffel symbols $\bar{\Gamma}^i_{jn}=\bar{g}^{il}\left(\partial_j\bar{g}_{ln}+\partial_{n}\bar{g}_{jl}-\partial_l\bar{g}_{jn}\right)/2$. Component of the covariant derivatives are $\bar{\nabla}_i f=\partial_if$ and $\bar{\nabla}_iA_j=\partial_iA_j-\bar{\Gamma}^{l}_{ij}A_l$, for a scalar $f$ and a vector $\boldsymbol{A}$, respectively.

We then introduce the set of vectors $(\bar{\mathbf{e}}^\theta,\bar{\mathbf{e}}^\varphi)$ which form an orthonormal basis in the plane tangent to the sphere since $\bar{g}^{ij}~\bar{\mathbf{e}}^{\theta}_{i}~\bar{\mathbf{e}}^{\theta}_{j}=1=\bar{g}^{ij}~\bar{\mathbf{e}}^{\varphi}_{i}~\bar{\mathbf{e}}^{\varphi}_{j}$ and $\bar{g}^{ij}~\bar{\mathbf{e}}^{\theta}_{i}~\bar{\mathbf{e}}^{\varphi}_{j}=0$. The components  of such diad vectors in the natural basis associated to $(\theta,\varphi)$ are given by
\ba
	\bar{\mathbf{e}}^\theta =(1, ~0) ~~~\mathrm{and}~~~ \bar{\mathbf{e}}^\varphi =(0, ~\sin\theta).
\ea
This set of vectors is used to define the helicity basis in the plane tangent to the sphere by $\bar{\mathbf{e}}^{\pm}=\left(\bar{\mathbf{e}}^\theta\pm i\bar{\mathbf{e}}^\varphi\right)/\sqrt{2}$. Their inner-dot product is given by
\ba
	\bar{g}^{ij}~\bar{\mathbf{e}}^{\pm}_{i}~\bar{\mathbf{e}}^{\pm}_{j}=0 ~~~\mathrm{and}~~~\bar{g}^{ij}~\bar{\mathbf{e}}^{\pm}_{i}~\bar{\mathbf{e}}^{\mp}_{j}=1,
\ea
meaning that the two tensors $\bar{\mathbf{e}}^{+} \otimes \bar{\mathbf{e}}^{+}$ and $\bar{\mathbf{e}}^{-} \otimes \bar{\mathbf{e}}^{-}$ are symmetric and traceless.

\subsection{Radial and angular decomposition of $\mathbb{R}^3$}
\label{app:radangdec}
The radial and angular decomposition of $\mathbb{R}^3$ is made by using the spherical coordinate system $(x,\theta,\varphi)$. Since it is a curvilinear system, the metric tensor reads
\ba
	g_{\mu\nu}=\left(\begin{array}{ccc}
		1 & 0 & 0 \\
		0 & x^2 & 0 \\
		0 & 0 & x^2\sin^2\theta
	\end{array}\right)
\ea
We then introduce the set of three orthonormal vectors $ (\mathbf{e}^x,\mathbf{e}^\theta,\mathbf{e}^\varphi)$, whose components in the natural basis associated to $(x,\theta,\varphi)$ are
\ba
	\mathbf{e}^{x}_{\mu} = & \left( 1, ~0,~0\right), \\
	\mathbf{e}^{\theta}_{\mu} = & \left( 0, ~x,~0\right), \\
	\mathbf{e}^{\varphi}_{\mu} = & \left( 0, ~0,~x\sin\theta\right).
\ea	
This set of vectors forms a triad since $g^{\mu\nu}~\mathbf{e}^a_\mu~\mathbf{e}^b_\nu=\gamma^{ab}$ where $a,~b$ run over $(x,~\theta,~\varphi)$ and $\gamma^{ab}$ is the flat metric tensor. Spheres $\Sigma_x$ at a radius $x$ are hyper-surfaces in $\mathbb{R}^3$ defined as the set of surfaces orthogonal to the vector $\mathbf{e}^x$. It is straightforward to see that the metric induced on each hyper-surface is indeed the metric on the sphere, $\bar{g}_{ij}$, up to  the conformal factor $x^2$. Hence the full metric can be decomposed as 
\ba
	g_{\mu\nu}=\mathbf{e}^x_\mu~\mathbf{e}^x_\nu+x^2\bar{g}_{\mu\nu},
\ea
where $\bar{g}_{x\mu}=0$ and $\bar{g}_{ij}$ is given by Eq. (\ref{eq:spheremetric}). The inverse metric is given by $g^{\mu\nu}=\mathbf{e}_x^\mu~\mathbf{e}_x^\nu+x^{-2}\bar{g}^{\mu\nu}$ where $\bar{g}^{x\mu}=0$ and $\bar{g}^{ij}$ is the inverse of $\bar{g}_{ij}$. Since we are now working in a specific $2+1$ decomposition of $\mathbb{R}^3$, we will use $x$ as the index for the radial component and $i,~j$ letters for components on the sphere $\Sigma_x$, i.e. Latin letters run over $\theta$ and $\varphi$.

We now write the covariant derivative using the above $2+1$ decomposition and we will write it using radial partial derivative, $\partial_x$, and covariant derivative on the sphere, $\bar{\nabla}$ (note that in full generality $\nabla_x\neq\partial_x$ and $\nabla_i\neq\bar{\nabla}_i$). A direct calculation of the Christoffel symbols gives 
\ba
	\Gamma^{x}_{xx}= 0,~~~\Gamma^{x}_{xj}=0,~~~\mathrm{and}~~~\Gamma^x_{ij}=-x\bar{g}_{ij},
\ea
and
\ba
	\Gamma^{i}_{xx}= 0,~~~\Gamma^{i}_{xj}=\frac{1}{x}\delta^i_j,~~~\mathrm{and}~~~\Gamma^{i}_{jl}=\bar{\Gamma}^{i}_{jl},
\ea
where $\bar{\Gamma}^{i}_{jl}$ are the Christoffel symbols on the sphere and which defines the covariant derivative on the sphere, $\bar{\nabla}_i$, introduced in App. \ref{sapp:sphere}.  Our derivation needs the quantities $\nabla_\mu f$ and $\nabla_\mu\nabla_\nu f$ where $f$ is a scalar function on $\mathbb{R}^3$. The $2+1$ decomposition of the gradient of a scalar is simply obtained by replacing covariant derivatives by partial derivatives, hence $\nabla_x f=\partial_x f$ and $\nabla_i f= \partial_i f = \bar{\nabla}_i f$, where we use that $\bar\nabla_i$ on a scalar function is also given by the partial derivative. The double gradient is written as $\nabla_\mu\nabla_\nu f=\left(\partial_\mu\partial_\nu-\Gamma_{\mu\nu}^\lambda\partial_\lambda\right)f$. From the expression of the Christoffel symbols, it boils down to
\ba
	\nabla_x\nabla_x f = & \partial^2_x f, \label{eq:covrad} \\
	\nabla_x\nabla_i f = & \left(\partial_x-\frac{1}{x}\right)\bar\nabla_i f, \label{eq:covmix} \\
	\nabla_i\nabla_j f = & \bar\nabla_i\bar\nabla_j f +x\bar{g}_{ij}\partial_x f, \label{eq:covang}
\ea
where we introduce the covariant derivative on the sphere. We note that $\bar\nabla_i$ and $\partial_x$ are commuting, hence  a direct calculation shows that $\nabla_x\nabla_i f = \nabla_i\nabla_x f $. 

\subsection{Spin operators and spin-weighted spherical harmonics}
\label{sapp:spin}
Properties of the spin-weighted spherical harmonics, spin-weighted functions on the sphere, and the way to build them from the spin-raising and spin lowering operators can be fund in many articles; see e.g. \cite{doi:10.1063/1.1931221,doi:10.1063/1.1705135}. Applications of these in the cosmological context are presented in e.g. \cite{1997PhRvD..55.1830Z,Kamionkowski:1996ks,Stebbins:1996wx,Hu:2000ee}, where most of their properties of interest for our calculations can also be found. Here, we do not provide an exhaustive description of spin-weighted functions on the sphere. Instead, we only briefly sketch how to build them and their properties which are of interests for our purpose.

\paragraph*{Spin-weighted function--}
A function on the sphere is said to be spin-weighted if its transformation under a rotation by an angle $\gamma$ of the basis $(\bar{\mathbf{e}}^\theta,\bar{\mathbf{e}}^\varphi)$ (or any orthonormal basis in the tangent plane) is $f_s\to f'_s=e^{is\gamma}f_s$. Spin functions are components of tensors expressed in the helicity basis. As examples, a spin-$(0)$ function is just a scalar. Spin-$(\pm1)$ functions are the two components of a vector, i.e. $\boldsymbol{A}_i=A_1~\bar{\mathbf{e}}^{-}_i+A_{-1}~\bar{\mathbf{e}}^{+}_i$. Similarly, spin-$(\pm2)$ functions are the components of a symmetric and traceless, rank-2 tensor, i.e. $\boldsymbol{T}_{ij}=T_2~\left(\bar{\mathbf{e}}^{-}_{i}~\bar{\mathbf{e}}^{-}_{j}\right)+T_{-2}~\left(\bar{\mathbf{e}}^{+}_{i}~\bar{\mathbf{e}}^{+}_{j}\right)$. The spin of a function can be raised or lowered using the so-called spin-raising and spin-lowering operators, $\bar\nabla_\pm$. These operators are just the projection of the covariant derivative on the helicity basis \cite{Hu:2000ee}:
 \ba
 	\bar\nabla_\pm=\bar{g}^{ij}~\bar{\mathbf{e}}^{\pm}_i\bar\nabla_j.
\ea
We note that these operators are also defined in the literature with an extra $\sqrt{2}$ factor, i.e.  $\bar\nabla_\pm=\sqrt{2}~\bar{g}^{ij}~\bar{\mathbf{e}}^{\pm}_i\bar\nabla_j$. Here we will follow the definition used in \cite{Hu:2000ee}. If applied to a scalar function, the covariant derivative can be replaced by partial derivatives. For non-zero spin however, this is the full covariant derivative which is projected on the helicity basis since spin-$s$ functions are component of tensors. 

\paragraph*{Differential and integral calculus with spin-operators--}
Let us first mention some integration and derivative calculus using the spin-raising and spin-lowering operators. We first note that the total spin of the product of spin functions is the sum of the spins, i.e. the spin $s_1$ of $h_{s_1}\equiv f_{s_2}\times g_{s_3}$ is $s_1=s_2+s_3$. Since we are working with spherical harmonics, we consider functions whose support is the entire sphere. Finally, the explicit expression of the spin-raising and spin-lowering operators applied to a spin-$s$ function is
\ba
	\nabla_\pm f_s=\frac{-1}{\sqrt{2}}(\sin\theta)^{\pm s}\left[\frac{\partial}{\partial\theta}\pm \frac{i}{\sin\theta}\frac{\partial}{\partial\varphi}\right](\sin\theta)^{\mp s} f_s,
\ea
which explicitly shows that these operators depend on the spin of the function on which they operate. Thanks to this explicit expression, it is straightforward to check that the Leibniz rules of differential calculus holds for spin-operators, i.e. $\nabla_\pm\left(f_s\times g_{s'}\right)=f_s\nabla_\pm g_{s'}+g_{s'}\nabla_\pm f_s$.

We can also show from the above that the integration by parts can be directly used with such operators if the integrand has zero total spin. Suppose indeed that we have to deal with the following integral
\ba
	\!\int_{4\pi} \sin\theta\, \dd\theta \, \dd\varphi \ f_{s} \, \left(\nabla_{\pm} g_{s'}\right) \nonumber
\ea
such that the integrand is a scalar, meaning that $s+s'\pm1=0$. Plugging the explicit expression of the spin-raising and spin-lowering operator in the above and performing integration by parts shows that\footnote{We mention that there is no contour integral (boundary term) here since we integrate over the entire celestial sphere and we consider that the support of $f_s$ and  $g_{s'}$ is the whole sphere.}
\ba
	\!\int_{4\pi} \sin\theta \, \dd \theta \, \dd\varphi \ f_{s}\left(\nabla_{\pm} g_{s'}\right)=-\!\int_{4\pi} \sin\theta \, \dd \theta \, \dd\varphi \ g_{s'}\left(\nabla_{\pm} f_{s}\right) - \frac{s+s'\pm1}{\sqrt{2}} \!\int_{4\pi} \cos\theta \, \dd \theta \, \dd\varphi \ f_{s} \, g_{s'}. \nonumber
\ea
Hence for a scalar integrand, $s+s'\pm1=0$, we obtain
\ba
	\!\int_{4\pi} \sin\theta \, \dd \theta \, \dd\varphi \ f_{s}\left(\nabla_{\pm} g_{s'}\right)=-\!\int_{4\pi} \sin\theta \, \dd \theta \, \dd\varphi \ g_{s'}\left(\nabla_{\pm} f_{s}\right). \label{eq:ippspin}
\ea

\paragraph*{Spin-weighted spherical harmonics--}
For our purpose, it is sufficient to present how those operators act on the spin-weighted spherical harmonics, denoted ${}_sY_{\ell m}$ in the following. These functions are the eigenvectors of the Laplacian operator for tensors on the celestial sphere. They satisfy the orthonormality and completeness relations
\ba
	\!\int\dd^2\hat\xx ~{}_sY^\star_{\ell m}(\hat\xx) \ {}_sY_{\ell' m'}(\hat\xx) &= \delta_{\ell,\ell'}~\delta_{m,m'}, \\
	\ds\sum_{\ell,m}~{}_sY^\star_{\ell m}(\hat\xx) \ {}_sY_{\ell m}(\hat\xx') &= \delta^2(\hat{\xx}-\hat{\xx}').
\ea
We note that in the above, both spherical harmonics need to have the same spin. The spin is raised and lowered as follows
\ba
	\nabla_+~{}_sY_{\ell m} = &\sqrt{\frac{(\ell-s)(\ell+s+1)}{2}}~{}_{s+1}Y_{\ell m}, \label{eq:spinrais}\\
	\nabla_-~{}_sY_{\ell m} = &-\sqrt{\frac{(\ell+s)(\ell-s+1)}{2}}~{}_{s-1}Y_{\ell m}. \label{eq:spinlow}
\ea
This allows to obtain the spin-weighted spherical harmonics by successive application of the spin-raising and spin-lowering operators on the scalar spherical harmonics $Y_{\ell m}$. From now on, $s$ will systematically be positive valued and negative spins will be explicitly denoted $(-s)$. For positive values of $s$ we have
\ba
	{}_sY_{\ell m}=\sqrt{\frac{(\ell-s)!}{(\ell+s)!}}\left(\sqrt{2}\nabla_+\right)^s~Y_{\ell m},
\ea
and for negative spins
\ba
	{}_{-s}Y_{\ell m}=(-1)^{s}\sqrt{\frac{(\ell-s)!}{(\ell+s)!}}\left(\sqrt{2}\nabla_-\right)^s~Y_{\ell m}.
\ea
The above yields
\ba
	\left(\nabla_+\right)^s\left(\nabla_-\right)^sY_{\ell m} = &\left(\frac{-1}{2}\right)^s\frac{(\ell+s)!}{(\ell-s)!}\,Y_{\ell m}.
\ea
We note that these two operators commute when applied to a scalar function, i.e. for any function $f$ with zero spin we have $\nabla_+\nabla_-f=\nabla_-\nabla_+f$.

Spin-weighted spherical harmonics can finally be used to express the successive application of covariant derivatives on the spherical harmonics. For our purpose, the two important expressions are \cite{Hu:2000ee}
\ba
	\bar{\nabla}_iY_{\ell m} = & \sqrt{\frac{\ell(\ell+1)}{2}}\left[{}_1Y_{\ell m} \bar{\mathbf{e}}^-_i - {}_{-1}Y_{\ell m} \bar{\mathbf{e}}^+_i\right], \label{eq:spin1dec} \\
	\bar{\nabla}_i\bar{\nabla}_jY_{\ell m} = & -\frac{\ell(\ell+1)}{2}Y_{\ell m}\bar{g}_{ij}+\frac{1}{2}\sqrt{\frac{(\ell+2)!}{(\ell-2)!}}\left[{}_2Y_{\ell m}\left(\bar{\mathbf{e}}^{-}_{i}~\bar{\mathbf{e}}^{-}_{j}\right)+{}_{-2}Y_{\ell m}\left(\bar{\mathbf{e}}^{+}_{i}~\bar{\mathbf{e}}^{+}_{j}\right)\right], \label{eq:spin2dec}
\ea
where in the second line the term proportional to $\bar{g}_{ij}$ is the pure trace part of the tensor $\bar{\nabla}_i\bar{\nabla}_jY_{\ell m}$, and the remaining terms are the two components of its traceless part. Since the tensors $\left(\bar{\mathbf{e}}^{\pm}_{i}~\bar{\mathbf{e}}^{\pm}_{j}\right)$ are traceless, we also derive the well-known expression of the Laplacian of the spherical harmonics:
\ba
	\bar{g}^{ij}\bar{\nabla}_i\bar{\nabla}_jY_{\ell m} =  -\ell(\ell+1)Y_{\ell m} \label{eq:lapylm}.
\ea
We note that the Laplacian of a scalar function can also be written using spin-operators as $\bar{g}^{ij}\bar{\nabla}_i\bar{\nabla}_jf=\nabla_+\nabla_-f=\nabla_-\nabla_+f$.

\subsection{Covariant derivative in the 2+1 decomposition} 
\label{sapp:diffbesselYlm}
Our calculations make heavy use of covariant derivatives and double covariant derivatives applied on the product $f_{k,\boldsymbol{\ell}}(\mathbf{x})\equiv j_\ell(kx)Y_{\ell m}(\hat{\mathbf{x}})$. Here we provide the expressions of these derivatives projected on the basis vectors $(\mathbf{e}^x,\mathbf{e}^+,\mathbf{e}^-)$ with $\mathbf{e}^\pm=(\mathbf{e}^\theta\pm i \mathbf{e}^\varphi)/\sqrt{2}$.

For the single derivative we have 
\begin{eqnarray}
    g^{\mu\nu}\mathbf{e}^x_\mu\nabla_\nu f_{k,\boldsymbol{\ell}}&=&\frac{\partial j_\ell(kx)}{\partial x}\,Y_{\ell m}(\hat{\mathbf{x}}), \\
    g^{\mu\nu}\mathbf{e}^\pm_\mu\nabla_\nu f_{k,\boldsymbol{\ell}}&=&\pm\frac{j_\ell(kx)}{x}\,\bar{g}^{ij}\bar{\mathbf{e}}^\pm_i\bar{\nabla}_jY_{\boldsymbol{\ell}}(\hat{\mathbf{x}})=\sqrt{\frac{\ell(\ell+1)}{2}}\,\left(\frac{j_\ell(kx)}{x}\right)\,{}_{\pm1}Y_{\ell m}(\hat{\mathbf{x}}).
\end{eqnarray}
For the double covariant derivative, we have
\begin{eqnarray}
    g^{\mu\nu}g^{\lambda\rho}\mathbf{e}^x_\mu\mathbf{e}^x_\lambda\nabla_\nu\nabla_\rho f_{k,\boldsymbol{\ell}}&=&\frac{\partial^2 j_\ell(kx)}{\partial x^2}\,Y_{\ell m}(\hat{\mathbf{x}}), \\
    g^{\mu\nu}g^{\lambda\rho}\mathbf{e}^x_\mu\mathbf{e}^\pm_\lambda\nabla_\nu\nabla_\rho f_{k,\boldsymbol{\ell}}&=&\frac{1}{x}\left[\frac{\partial j_\ell(kx)}{\partial x}-\frac{j_\ell(kx)}{x}\right]\bar{g}^{ij}\bar{\mathbf{e}}^\pm_i\bar{\nabla}_jY_{\ell m}(\hat{\mathbf{x}}) \nonumber \\
    &=&\sqrt{\frac{\ell(\ell+1)}{2}}\left(\frac{1}{x}\right)\left[\frac{\partial j_\ell(kx)}{\partial x}-\frac{j_\ell(kx)}{x}\right]\,{}_{\pm1}Y_{\ell m}(\hat{\mathbf{x}}), \\
    g^{\mu\nu}g^{\lambda\rho}\mathbf{e}^\pm_\mu\mathbf{e}^\pm_\lambda\nabla_\nu\nabla_\rho f_{k,\boldsymbol{\ell}}&=&\frac{j_\ell(kx)}{x^2}\bar{g}^{il}\bar{g}^{jn}\bar{\mathbf{e}}^\pm_i\bar{\mathbf{e}}^\pm_j\bar{\nabla}_l\bar{\nabla}_nY_{\ell m}(\hat{\mathbf{x}}) \nonumber \\
    &=&\frac{1}{2}\sqrt{\frac{\ell+2)!}{(\ell-2)!}}\,\left(\frac{j_\ell(kx)}{x^2}\right)\,{}_{\pm2}Y_{\ell m}(\hat{\mathbf{x}}), \\
     g^{\mu\nu}g^{\lambda\rho}\mathbf{e}^\pm_\mu\mathbf{e}^\mp_\lambda\nabla_\nu\nabla_\rho f_{k,\boldsymbol{\ell}}&=&\frac{j_\ell(kx)}{x^2}\bar{g}^{il}\bar{g}^{jn}\bar{\mathbf{e}}^\pm_i\bar{\mathbf{e}}^\mp_j\bar{\nabla}_l\bar{\nabla}_nY_{\ell m}(\hat{\mathbf{x}})+\frac{1}{x}\frac{\partial j_\ell(kx)}{\partial x}\,Y_{\ell m}(\hat{\mathbf{x}}) \nonumber \\
     &=&\frac{1}{x}\left[\frac{\partial j_\ell(kx)}{\partial x}-\frac{\ell(\ell+1)}{2}\frac{j_\ell(kx)}{x}\right]\,Y_{\ell m}(\hat{\mathbf{x}}).
\end{eqnarray}
We note that the last term can be written as
\ba
    g^{\mu\nu}g^{\lambda\rho}\mathbf{e}^\pm_\mu\mathbf{e}^\mp_\lambda\nabla_\nu\nabla_\rho f_{k,\boldsymbol{\ell}}=\frac{-1}{2}\left[k^2j_\ell(kx)+\frac{\partial^2j_\ell(kx)}{\partial x^2}\right]\,Y_{\ell m}(\hat{\mathbf{x}})
\ea
where we use the differential equation satisfied by Bessel functions 
\ba
\left[\partial_x^2 + \frac{2}{x} \partial_x  - \frac{\ell(\ell+1)}{x^2}\right] j_\ell(kx)= -k^2 j_\ell(kx).
\ea
\subsection{Spin-weighted Gaunt integral}
\label{sapp:defgaunt}
The Gaunt integral is defined as
\be
G_{\boldsymbol{\ell}_1;\boldsymbol{\ell}_2;\boldsymbol{\ell}_3} = \int \dd^2\hn \ Y_{\ell_1 m_1}(\hn) \ Y_{\ell_2 m_2}(\hn) \ Y_{\ell_3 m_3}(\hn).
\ee
It can be extended to the case of spin-weighted spherical harmonics. Suppose that $s_1+s_2+s_3=0$, then the spin-weighted Gaunt integral is
\ba \label{Eq:def-Gaunt-spinweighted}
		G^{s_1;s_2;s_3}_{\boldsymbol{\ell}_1;\boldsymbol{\ell}_2;\boldsymbol{\ell}_3}=\!\int\dd^2\hn \ {}_{s_1}Y_{\ell_1 m_1}(\hn) \ {}_{s_2}Y_{\ell_2 m_2}(\hn) \ {}_{s_3}Y_{\ell_3 m_3}(\hn),
\ea
which is invariant by permuting $(\boldsymbol{\ell}_i,s_i)$ with $(\boldsymbol{\ell}_j,s_j)$ for all $i$ and $j$ (for instance $G^{s_1;s_2;s_3}_{\boldsymbol{\ell}_1;\boldsymbol{\ell}_2;\boldsymbol{\ell}_3}=G^{s_2;s_1;s_3}_{\boldsymbol{\ell}_2;\boldsymbol{\ell}_1;\boldsymbol{\ell}_3}$).
It is expressed using the Wigner-$3j$ symbols:
\ba
	G^{s_1;s_2;s_3}_{\boldsymbol{\ell}_1;\boldsymbol{\ell}_2;\boldsymbol{\ell}_3}=\sqrt{\frac{(2\ell_1+1)(2\ell_2+1)(2\ell_3+1)}{4\pi}}\left(\begin{array}{ccc}
		\ell_1 & \ell_2 & \ell_3 \\
		s_1 & s_2 & s_3
	\end{array}\right)\left(\begin{array}{ccc}
		\ell_1 & \ell_2 & \ell_3 \\
		m_1 &m_2 & m_3
	\end{array}\right),
\ea
under the condition $s_1+s_2+s_3=0$ \cite{varshalovich1988quantum}.

\section{Relating spin-weighted and standard Gaunt coefficients}
\label{apps:spinto0}
We show in this appendix how to express spin-weighted Gaunt integrals as functions of "standard" (spin-0) Gaunt integrals. We first note that the spin-weighted Gaunt integrals we are dealing with in our calculations are
\ba
	G^{s;-s;0}_{\boldsymbol{\ell}_1;\boldsymbol{\ell}_2;\boldsymbol{\ell}_3} = \!\int \dd^2\hat\xx \ {}_sY_{\ell_1m_1}(\hat\xx) \, {}_{-s}Y_{\ell_2m_2}(\hat\xx) \, Y_{\ell_3m_3}(\hat\xx),
\ea
which indeed satisfies that $s_1+s_2+s_3=0$. 
To relate it to the standard Gaunt integral, we first express the spin-weighted Gaunt integrals using spin-raising and spin-lowering operators
\ba
	G^{s;-s;0}_{\boldsymbol{\ell}_1;\boldsymbol{\ell}_2;\boldsymbol{\ell}_3}=(-2)^s\sqrt{\frac{(\ell_1-s)!}{(\ell_1+s)!}}\sqrt{\frac{(\ell_2-s)!}{(\ell_2+s)!}}\underbrace{\!\int \dd^2\hat\xx \left(\nabla_+^sY_{\ell_1m_1}\right)\left(\nabla_-^sY_{\ell_2m_2}\right)Y_{\ell_3m_3}}_{\mathcal{G}^{s;-s;0}_{\boldsymbol{\ell}_1;\boldsymbol{\ell}_2;\boldsymbol{\ell}_3}},
\ea
where in our study we restrict to the cases $s=1$ and $s=2$. We note that $\mathcal{G}^{s;-s;0}_{\boldsymbol{\ell}_1;\boldsymbol{\ell}_2;\boldsymbol{\ell}_3}=\mathcal{G}^{-s;s;0}_{\boldsymbol{\ell}_2;\boldsymbol{\ell}_1;\boldsymbol{\ell}_3}$.

We stress that in our study we are only interested in the sum $\left(\mathcal{G}^{s;-s;0}_{\boldsymbol{\ell}_1;\boldsymbol{\ell}_2;\boldsymbol{\ell}_3}+\mathcal{G}^{-s;s;0}_{\boldsymbol{\ell}_1;\boldsymbol{\ell}_2;\boldsymbol{\ell}_3}\right)$. Hence in the following we will show how to express this sum as a function of the standard Gaunt integral ${G}^{0;0;0}_{\boldsymbol{\ell}_1;\boldsymbol{\ell}_2;\boldsymbol{\ell}_3}$.

\subsection{$s=1$ case}
This case closely follows the one developed in \cite{Goldberg:1999xm,Hu:2000ee}. We start by a first integration by parts to get\footnote{Since the total spin of the integrand in the spin-weighted Gaunt integral is zero, we can use Eq. (\ref{eq:ippspin}) to perform integration by parts.}
\ba
	\mathcal{G}^{1;-1;0}_{\boldsymbol{\ell}_1;\boldsymbol{\ell}_2;\boldsymbol{\ell}_3}=-\!\int \dd^2\hat\xx \ Y_{\ell_1m_1} \left(\nabla_-Y_{\ell_2m_2}\right) \left(\nabla_+Y_{\ell_3m_3}\right) + \frac{1}{2}\ell_2(\ell_2+1) \, {G}^{0;0;0}_{\boldsymbol{\ell}_1;\boldsymbol{\ell}_2;\boldsymbol{\ell}_3},
\ea
Let us now apply a second integration by parts in the first term to get
\ba
	\mathcal{G}^{1;-1;0}_{\boldsymbol{\ell}_1;\boldsymbol{\ell}_2;\boldsymbol{\ell}_3}= & \!\int \dd^2\hat\xx\left(\nabla_-Y_{\ell_1m_1}\right)Y_{\ell_2m_2}\left(\nabla_+Y_{\ell_3m_3}\right) \nonumber \\
	&+\frac{1}{2}\ell_2(\ell_2+1){G}^{0;0;0}_{\boldsymbol{\ell}_1;\boldsymbol{\ell}_2;\boldsymbol{\ell}_3}-\frac{1}{2}\ell_3(\ell_3+1){G}^{0;0;0}_{\boldsymbol{\ell}_1;\boldsymbol{\ell}_2;\boldsymbol{\ell}_3}.
\ea
Performing a last integration by parts on the first term shows that
\ba
	\mathcal{G}^{1;-1;0}_{\boldsymbol{\ell}_1;\boldsymbol{\ell}_2;\boldsymbol{\ell}_3}= & -\!\int \dd^2\hat\xx\left(\nabla_-Y_{\ell_1m_1}\right)\left(\nabla_+Y_{\ell_2m_2}\right)Y_{\ell_3m_3} \nonumber \\
	& +\frac{1}{2}\left[{\ell_1(\ell_1+1)}+{\ell_2(\ell_2+1)}-{\ell_3(\ell_3+1)}\right]{G}^{0;0;0}_{\boldsymbol{\ell}_1;\boldsymbol{\ell}_2;\boldsymbol{\ell}_3},
\ea
where in the first line we recognize $-\mathcal{G}^{-1;1;0}_{\boldsymbol{\ell}_1;\boldsymbol{\ell}_2;\boldsymbol{\ell}_3}$. It gives
\ba
	\mathcal{G}^{1;-1;0}_{\boldsymbol{\ell}_1;\boldsymbol{\ell}_2;\boldsymbol{\ell}_3}=\frac{1}{4}\left[{\ell_1(\ell_1+1)}+{\ell_2(\ell_2+1)}-{\ell_3(\ell_3+1)}\right]{G}^{0;0;0}_{\boldsymbol{\ell}_1;\boldsymbol{\ell}_2;\boldsymbol{\ell}_3}. \label{eq:g1tog0}
\ea
By using $\mathcal{G}^{-1;1;0}_{\boldsymbol{\ell}_1;\boldsymbol{\ell}_2;\boldsymbol{\ell}_3}=\mathcal{G}^{1;-1;0}_{\boldsymbol{\ell}_2;\boldsymbol{\ell}_1;\boldsymbol{\ell}_3}$, an expression for $\mathcal{G}^{-1;1;0}_{\boldsymbol{\ell}_1;\boldsymbol{\ell}_2;\boldsymbol{\ell}_3}$  is easily obtained from Eq. (\ref{eq:g1tog0}) by permuting $\ell_1$ with $\ell_2$. Finally, one makes use of ${G}^{0;0;0}_{\boldsymbol{\ell}_2;\boldsymbol{\ell}_1;\boldsymbol{\ell}_3}={G}^{0;0;0}_{\boldsymbol{\ell}_1;\boldsymbol{\ell}_2;\boldsymbol{\ell}_3}$ to get
\ba
	\mathcal{G}^{-1;1;0}_{\boldsymbol{\ell}_1;\boldsymbol{\ell}_2;\boldsymbol{\ell}_3}=\frac{1}{4}\left[{\ell_1(\ell_1+1)}+{\ell_2(\ell_2+1)}-{\ell_3(\ell_3+1)}\right]{G}^{0;0;0}_{\boldsymbol{\ell}_1;\boldsymbol{\ell}_2;\boldsymbol{\ell}_3}.
\ea
Combining the two we deduce the following expression
\ba
 	\mathcal{G}^{1;-1;0}_{\boldsymbol{\ell}_1;\boldsymbol{\ell}_2;\boldsymbol{\ell}_3}+\mathcal{G}^{-1;1;0}_{\boldsymbol{\ell}_1;\boldsymbol{\ell}_2;\boldsymbol{\ell}_3}=\frac{1}{2}\left[{\ell_1(\ell_1+1)}+{\ell_2(\ell_2+1)}-{\ell_3(\ell_3+1)}\right]{G}^{0;0;0}_{\boldsymbol{\ell}_1;\boldsymbol{\ell}_2;\boldsymbol{\ell}_3}. \label{eq:spin1to0}
\ea

\subsection{$s=2$ case}
In this case our starting point is 
\ba
	\mathcal{G}^{2;-2;0}_{\boldsymbol{\ell}_1;\boldsymbol{\ell}_2;\boldsymbol{\ell}_3}=\!\int \dd^2\hat\xx\left(\nabla_+\nabla_+Y_{\ell_1m_1}\right)\left(\nabla_-\nabla_-Y_{\ell_2m_2}\right)Y_{\ell_3m_3}.
\ea
The strategy here is to express $\mathcal{G}^{2;-2;0}_{\boldsymbol{\ell}_1;\boldsymbol{\ell}_2;\boldsymbol{\ell}_3}$ as a function of $\mathcal{G}^{1;-1;0}_{\boldsymbol{\ell}_1;\boldsymbol{\ell}_2;\boldsymbol{\ell}_3}$. 

We start by a first integration by parts to get
\ba
	\mathcal{G}^{2;-2;0}_{\boldsymbol{\ell}_1;\boldsymbol{\ell}_2;\boldsymbol{\ell}_3} = & -\!\int \dd^2\hat\xx\left(\nabla_+Y_{\ell_1m_1}\right)\left(\nabla_-\nabla_-Y_{\ell_2m_2}\right)\left(\nabla_+Y_{\ell_3m_3}\right) \nonumber \\
	&-\!\int \dd^2\hat\xx\left(\nabla_+Y_{\ell_1m_1}\right)\left(\nabla_+\nabla_-\nabla_-Y_{\ell_2m_2}\right)Y_{\ell_3m_3}.
\ea
The second line can be expressed as a function of $\mathcal{G}^{1;-1;0}_{\boldsymbol{\ell}_1;\boldsymbol{\ell}_2;\boldsymbol{\ell}_3}$. We first note that the spin-raising and spin-lowering operators commute only if they are applied to a scalar function. Since $\nabla Y_{\ell_2m_2}$ is a spin-$(-1)$ function, we cannot make use of commutation in $\left(\nabla_+\nabla_-\nabla_-Y_{\ell_2m_2}\right)$. This is however a spin-$(-1)$ function, hence related to $\nabla_- Y_{\ell_2 m_2}$. To show this, we use Eqs. (\ref{eq:spinrais}) \& (\ref{eq:spinlow}) to get
\ba
	\nabla_-Y_{\ell_2m_2}= &-\sqrt{\frac{\ell_2(\ell_2+1)}{2}}~{}_{-1}Y_{\ell_2m_2}, \\
	\nabla_+\nabla_-\nabla_-Y_{\ell_2m_2} = & (\ell_2-1)(\ell_2+2)\sqrt{\frac{\ell_2(\ell_2+1)}{8}}~{}_{-1}Y_{\ell_2m_2}.
\ea
This allows us to derive the following identity
\ba
	\!\int \dd^2\hat\xx \left(\nabla_+Y_{\ell_1m_1}\right) \left(\nabla_+\nabla_-\nabla_-Y_{\ell_2m_2}\right) Y_{\ell_3m_3} = \frac{-1}{2}(\ell_2-1)(\ell_2+2) \, \mathcal{G}^{1;-1;0}_{\boldsymbol{\ell}_1;\boldsymbol{\ell}_2;\boldsymbol{\ell}_3}.
\ea
 For the first term then, we perform a second integration by parts moving a spin-lowering operator applying on $Y_{\ell_2 m_2}$. This gives
\ba
 	\mathcal{G}^{2;-2;0}_{\boldsymbol{\ell}_1;\boldsymbol{\ell}_2;\boldsymbol{\ell}_3} = & \frac{-1}{2}{\ell_3(\ell_3+1)} \!\int \dd^2\hat\xx \ \left(\nabla_+Y_{\ell_1m_1}\right) \left(\nabla_-Y_{\ell_2m_2}\right) Y_{\ell_3m_3} \\
	& -\frac{1}{2}{\ell_1(\ell_1+1)} \!\int\dd^2\hat\xx \ Y_{\ell_1m_1} \left(\nabla_-Y_{\ell_2m_2}\right) \left(\nabla_+Y_{\ell_3m_3}\right) + \frac{1}{2}(\ell_2-1)(\ell_2+2)\, \mathcal{G}^{1;-1;0}_{\boldsymbol{\ell}_1;\boldsymbol{\ell}_2;\boldsymbol{\ell}_3}. \nonumber
 \ea 
 We can recognize the appearance of $\mathcal{G}^{1;-1;0}_{\boldsymbol{\ell}_1;\boldsymbol{\ell}_2;\boldsymbol{\ell}_3}$ in the first line and of $\mathcal{G}^{0;-1;1}_{\boldsymbol{\ell}_1;\boldsymbol{\ell}_2;\boldsymbol{\ell}_3}$ in the second line. We can directly make use of Eq. (\ref{eq:g1tog0}) to introduce the standard Gaunt integral.\footnote{We note that the strategy developed  for $s=1$ can be directly used to get 
\ba
	\mathcal{G}^{0;-1;1}_{\boldsymbol{\ell}_1;\boldsymbol{\ell}_2;\boldsymbol{\ell}_3}=\frac{1}{4}\left[-{\ell_1(\ell_1+1)}+{\ell_2(\ell_2+1)}+{\ell_3(\ell_3+1)}\right]{G}^{0;0;0}_{\boldsymbol{\ell}_1;\boldsymbol{\ell}_2;\boldsymbol{\ell}_3}.
\ea
 }
 It boils down to
 \ba   
    \mathcal{G}^{2;-2;0}_{\boldsymbol{\ell}_1;\boldsymbol{\ell}_2;\boldsymbol{\ell}_3} = & \frac{1}{8}\Big\{\left[\left(\ell_2-1\right)\left(\ell_2+2\right)-\ell_3(\ell_3+1)\right]\left[\ell_1(\ell_1+1)+\ell_2(\ell_2+1)-\ell_3(\ell_3+1)\right] \nonumber \\
    &-\ell_1(\ell_1+1)\left[-\ell_1(\ell_1+1)+\ell_2(\ell_2+1)+\ell_3(\ell_3+1)\right]\Big\} \ {G}^{0;0;0}_{\boldsymbol{\ell}_1;\boldsymbol{\ell}_2;\boldsymbol{\ell}_3}. \label{eq:g2tog0}
 \ea

The second integral, $\mathcal{G}^{-2;2;0}_{\boldsymbol{\ell}_1;\boldsymbol{\ell}_2;\boldsymbol{\ell}_3}$, is obtained by using $\mathcal{G}^{-2;2;0}_{\boldsymbol{\ell}_1;\boldsymbol{\ell}_2;\boldsymbol{\ell}_3}=\mathcal{G}^{2;-2;0}_{\boldsymbol{\ell}_2;\boldsymbol{\ell}_1;\boldsymbol{\ell}_3}$. Thus one just performs the permutation of $\ell_1$ with $\ell_2$ in Eq. (\ref{eq:g2tog0}) followed by the identity ${G}^{0;0;0}_{\boldsymbol{\ell}_2;\boldsymbol{\ell}_1;\boldsymbol{\ell}_3}={G}^{0;0;0}_{\boldsymbol{\ell}_1;\boldsymbol{\ell}_2;\boldsymbol{\ell}_3}$ to get
 \ba   
    \mathcal{G}^{-2;2;0}_{\boldsymbol{\ell}_1;\boldsymbol{\ell}_2;\boldsymbol{\ell}_3} = & \frac{1}{8}\Big\{\left[\left(\ell_1-1\right)\left(\ell_1+2\right)-\ell_3(\ell_3+1)\right]\left[\ell_1(\ell_1+1)+\ell_2(\ell_2+1)-\ell_3(\ell_3+1)\right] \nonumber \\
    &-\ell_2(\ell_2+1)\left[\ell_1(\ell_1+1)-\ell_2(\ell_2+1)+\ell_3(\ell_3+1)\right]\Big\} \ {G}^{0;0;0}_{\boldsymbol{\ell}_1;\boldsymbol{\ell}_2;\boldsymbol{\ell}_3}.
 \ea
The end result is obtained by combining the two expression, i.e.
\ba
    \mathcal{G}^{2;-2;0}_{\boldsymbol{\ell}_1;\boldsymbol{\ell}_2;\boldsymbol{\ell}_3}+\mathcal{G}^{-2;2;0}_{\boldsymbol{\ell}_1;\boldsymbol{\ell}_2;\boldsymbol{\ell}_3}=\frac{1}{8}\left(\frac{\Delta^{(2)}_{\ell_1}}{\Delta^{(1)}_{\ell_1}}+\frac{\Delta^{(2)}_{\ell_2}}{\Delta^{(1)}_{\ell_2}}+\Delta^{(1)}_{\ell_1}+\Delta^{(1)}_{\ell_2}-2\Delta^{(1)}_{\ell_3}\right)\left(\Delta^{(1)}_{\ell_1}+\Delta^{(1)}_{\ell_2}-\Delta^{(1)}_{\ell_3}\right){G}^{0;0;0}_{\boldsymbol{\ell}_1;\boldsymbol{\ell}_2;\boldsymbol{\ell}_3}, \nonumber \\
\ea
where we introduce $\Delta^{(s)}_\ell\equiv(\ell+s)!/(\ell-s)!$ from which we easily note that $\Delta^{(1)}_\ell=\ell(\ell+1)$ and $\Delta^{(2)}_\ell/\Delta^{(1)}_\ell=(\ell-1)(\ell+2)$.

\subsection{Relation in terms of Wigner-$3j$ symbols}
\label{apps:recwig}
The expressions derived in App. \ref{apps:spinto0}  relate spin-weighted Gaunt integrals to the standard one. By expressing these in terms of Wigner-$3j$ symbols, we can summarize the above results in terms of relations between Wigner-$3j$ symbols. 

By expressing Gaunt integrals with Wigner-$3j$ symbols, we obtain the following relations
\ba
	&\frac{-1}{2}\sqrt{\Delta^{(1)}_{\ell_1}\Delta^{(1)}_{\ell_2}}\left[\left(\begin{array}{ccc}
		\ell_1 & \ell_2 & \ell_3 \\
		1 & -1& 0
	\end{array}\right)
+\left(\begin{array}{ccc}
		\ell_1 & \ell_2 & \ell_3 \\
		-1 & 1 & 0
	\end{array}\right)\right]=\frac{1}{2}\left(\Delta^{(1)}_{\ell_1}+\Delta^{(1)}_{\ell_2}-\Delta^{(1)}_{\ell_3}\right)\left(\begin{array}{ccc}
		\ell_1 & \ell_2 & \ell_3 \\
		0 & 0& 0
	\end{array}\right),
\ea
 and
 \ba
 	\frac{1}{4}\sqrt{\Delta^{(2)}_{\ell_1}\Delta^{(2)}_{\ell_2}}\left[\left(\begin{array}{ccc}
		\ell_1 & \ell_2 & \ell_3 \\
		2 & -2& 0
	\end{array}\right)
+\left(\begin{array}{ccc}
		\ell_1 & \ell_2 & \ell_3 \\
		-2 & 2 & 0
	\end{array}\right)\right]=&\frac{1}{8}\left(\frac{\Delta^{(2)}_{\ell_1}}{\Delta^{(1)}_{\ell_1}}+\frac{\Delta^{(2)}_{\ell_2}}{\Delta^{(1)}_{\ell_2}}+\Delta^{(1)}_{\ell_1}+\Delta^{(1)}_{\ell_2}-2\Delta^{(1)}_{\ell_3}\right)\nonumber  \\
	& \times\left(\Delta^{(1)}_{\ell_1}+\Delta^{(1)}_{\ell_2}-\Delta^{(1)}_{\ell_3}\right)\left(\begin{array}{ccc}
		\ell_1 & \ell_2 & \ell_3 \\
		0 & 0& 0
	\end{array}\right). 
\ea
We note that the above are non zero for even values of $\ell_1+\ell_2+\ell_3$ only.

\section{Computation of \texorpdfstring{$f^{(n)}$}{f\^(n)} }
\label{app:Fn}
Many factors depending on $\ell$ appears in the following calculations. Hence to lighten the expression, we introduce the notations
\ba
	\Gamma_{\ell_1;\ell_2;\ell_3} = \sqrt{\frac{(2\ell_1+1)(2\ell_2+1)(2\ell_3+1)}{4\pi}} ~~~\mathrm{and}~~~ \Delta_{\ell}^{(s)} = \frac{(\ell+s)!}{(\ell-s)!},
\ea
where $s$ is an integer smaller than $\ell$. The first factor naturally appears from Gaunt integrals. The second appears by differentiating (spin-weighted) spherical harmonics.


\subsection{$n=0$ case}
This case is rather simple since it is readily written using a Gaunt integral using $Y^\star_{\ell m}(\hat\xx)=(-1)^mY_{\ell (-m)}(\hat\xx)$. Hence one obtains
\ba
	\!\int\dd^3\xx \ F^{(0)}_{\boldsymbol{\ell}_{123}}\left(k_{123},\xx\right) = & \Gamma_{\ell_1;\ell_2;\ell_3}  \!\int x^2\dd x \, \left[j_{\ell_1}(k_1x) \ j_{\ell_2}(k_2x) \ j_{\ell_3}(k_3x) \right] \nonumber \\
	& \times(-1)^{m_1+m_2+m_3}\left(\begin{array}{ccc}
		\ell_1 & \ell_2 & \ell_3 \\
		0 & 0 & 0
	\end{array}\right)\left(\begin{array}{ccc}
		\ell_1 & \ell_2 & \ell_3 \\
		-m_1 & -m_2 & -m_3
	\end{array}\right).
\ea
Let us now make use of
\ba
	\left(\begin{array}{ccc}
		\ell_1 & \ell_2 & \ell_3 \\
		-m_1 & -m_2 & -m_3
	\end{array}\right)=(-1)^{\ell_1+\ell_3+\ell_3}\left(\begin{array}{ccc}
		\ell_1 & \ell_2 & \ell_3 \\
		m_1 & m_2 & m_3
	\end{array}\right), \nonumber
\ea
which also leads to 
\ba
	\left(\begin{array}{ccc}
		\ell_1 & \ell_2 & \ell_3 \\
		0 & 0 & 0
	\end{array}\right)=0,
\ea
if $\ell_1+\ell_2+\ell_3$ is an odd number. Finally, the triangular conditions imposes $m_1+m_2+m_3=0$. This finally gives 
\ba
	\!\int\dd^3\xx \ F^{(0)}_{\boldsymbol{\ell}_{123}}\left(k_{123},\xx\right) = & \Gamma_{\ell_1;\ell_2;\ell_3}  \left(\begin{array}{ccc}
		\ell_1 & \ell_2 & \ell_3 \\
		0 & 0 & 0
	\end{array}\right)\left(\begin{array}{ccc}
		\ell_1 & \ell_2 & \ell_3 \\
		m_1 &m_2 & m_3
	\end{array}\right)\!\int x^2 \dd x \ I^{(0)}_{\ell_{123}}(k_{123},x),
\ea
where we introduce the radial integrand $I^{(0)}_{\ell_{123}}(k_{123},x)=\left[j_{\ell_1}(k_1x) \ j_{\ell_2}(k_2x) \ j_{\ell_1}(k_2x) \right] $. 

\subsection{$n=1$ case}
For this case, we start by computing the quantity 
\ba
	g^{\mu\nu}\nabla_\mu\left[j_{L_1}(k_1x)Y_{\boldsymbol{L}_1}^\star(\hat{x})\right]\nabla_\nu\left[j_{L_2}(k_2x)Y_{\boldsymbol{L}_2}^\star(\hat{x})\right],
\ea
by splitting it into a radial part and an angular part. Using the $2+1$ decomposition of the metric and the covariant derivative given in App. \ref{app:radangdec}, this gives
\ba
	\nabla_\mu\left[j_{L_1}(k_1x)Y_{\boldsymbol{L}_1}^\star(\hat{x})\right]\nabla^\mu\left[j_{L_2}(k_2x)Y_{\boldsymbol{L}_2}^\star(\hat{x})\right] =& \left[\frac{\partial j_{\ell_1}(k_1x)}{\partial x}\frac{\partial j_{\ell_2}(k_2x)}{\partial x}\right]\left[Y^\star_{\boldsymbol{\ell}_1}(\hat\xx)Y^\star_{\boldsymbol{\ell}_2}(\hat\xx)\right]  \\
	&+\left[\frac{j_{\ell_1}(k_1x)j_{\ell_2}(k_2x)}{x^2}\right]\bar{g}^{ij}\left[\bar{\nabla}_{i}Y^\star_{\boldsymbol{\ell}_1}(\hat\xx)\right]\left[\bar{\nabla}_{j}Y^\star_{\boldsymbol{\ell}_2}(\hat\xx)\right], \nonumber
\ea
where we write the radial part in the first line and the angular part in the second. In the angular part, $\bar{g}^{ij}$ is the inverse of the metric on the sphere and $\bar{\nabla}$ is the covariant derivative on the sphere.

\paragraph*{Radial part--} We first consider the contribution of the radial part to $\!\int\dd^3\xx \, F^{(1)}_{\boldsymbol{\ell}_{123}}\left(k_{123},\xx\right)$. This one is rather straightforward since it is derived in an identical way as the case $n=0$. This boils down for the radial part only to
\ba
	\!\int\dd^3\xx \, F^{(1)\mathrm{S}}_{\boldsymbol{\ell}_{123}}\left(k_{123},\xx\right) = & \Gamma_{\ell_1;\ell_2;\ell_3} \!\int x^2\dd x\left[\frac{\partial j_{\ell_1}(k_1x)}{\partial x}\frac{\partial j_{\ell_2}(k_2x)}{\partial x}j_{\ell_3}(k_3x)\right]\nonumber \\
	& \times\left(\begin{array}{ccc}
		\ell_1 & \ell_2 & \ell_3 \\
		0 & 0 & 0
	\end{array}\right)\left(\begin{array}{ccc}
		\ell_1 & \ell_2 & \ell_3 \\
		m_1 &m_2 & m_3
	\end{array}\right).
\ea
This term will be qualified hereafter as being scalar (hence the "S" superscript) in the sense that it only involves scalar quantities on the hyper-surfaces $\Sigma_x$, i.e. on the sphere. 

 \paragraph*{Angular part--} We now consider the contribution of the angular part. To this end one first plug the decomposition of $\bar\nabla_iY_{\ell m}$ as a function of the spin-$(\pm1)$ spherical harmonics as given in Eq. (\ref{eq:spin1dec}) in the quantity $\bar{g}^{ij}\left[\bar{\nabla}_{i}Y^\star_{\boldsymbol{\ell}_1}(\hat\xx)\right]\left[\bar{\nabla}_{j}Y^\star_{\boldsymbol{\ell}_2}(\hat\xx)\right]$. This gives rise to the appearance of inner-dot product of the form $\bar{g}^{ij}~\bar{\mathbf{e}}^\pm_i~\bar{\mathbf{e}}^\pm_j=0$ and $\bar{g}^{ij}~\bar{\mathbf{e}}^\pm_i~\bar{\mathbf{e}}^\mp_j=1$, where $\bar{\mathbf{e}}^\pm=(\bar{\mathbf{e}}^\theta\pm i\bar{\mathbf{e}}^\varphi)/\sqrt{2}$ is the helicity basis (see App. \ref{sapp:sphere}). This boils down to
\ba
	\bar{g}^{ij}\left[\bar{\nabla}_{i}Y^\star_{\boldsymbol{\ell}_1}(\hat\xx)\right]\left[\bar{\nabla}_{j}Y^\star_{\boldsymbol{\ell}_2}(\hat\xx)\right]= & \frac{-1}{2}\sqrt{\Delta_{\ell_1}^{(1)}\Delta_{\ell_2}^{(1)}}\Big[{}_{1}Y^\star_{\ell_1m_1}(\hat\xx){}_{-1}Y^\star_{\ell_2m_2}(\hat\xx) +{}_{-1}Y^\star_{\ell_1m_1}(\hat\xx){}_{1}Y^\star_{\ell_2m_2}(\hat\xx)\Big],\label{eq:vectpart1}
\ea
where ${}_{\pm1}Y_{\ell m}$ are the spin-$(\pm1)$ spherical harmonics. We note that we also make use of ${}_sY^\star_{\ell m}= (-1)^{s+m}{}_{-s}Y_{\ell (-m)}$. The above is then plugged in the angular part of $\!\int\dd^3\xx \, F^{(1)}_{\boldsymbol{\ell}_{123}}\left(k_{123},\xx\right)$ yielding
\ba
	\!\int\dd^3\xx \, F^{(1)\mathrm{ang}}_{\boldsymbol{\ell}_{123}}\left(k_{123},\xx\right) = & \frac{-1}{2}\sqrt{\Delta_{\ell_1}^{(1)}\Delta_{\ell_2}^{(1)}}\!\int \dd x\left[ j_{\ell_1}(k_1x)j_{\ell_2}(k_2x)j_{\ell_3}(k_3x)\right] \\ 
	& \!\int\dd^2\hat\xx\left[{}_{1}Y^\star_{\ell_1m_1}(\hat\xx){}_{-1}Y^\star_{\ell_2m_2}(\hat\xx)+{}_{-1}Y^\star_{\ell_1m_1}(\hat\xx){}_{1}Y^\star_{\ell_2m_2}(\hat\xx)\right]Y^\star_{\ell_3m_3}(\hat\xx). \nonumber
\ea
The last angular integral can be derived using two different approaches. The first one makes directly used of spinned Gaunt integrals (see App. \ref{sapp:defgaunt}). The second makes use of a reformulation of the spinned Gaunt integral as a function of the spin-0 Gaunt integral (see App. \ref{apps:spinto0}).  For completeness, we provide here both approaches. \\

Using spinned Gaunt integrals, the angular integral gives
\ba
	\rightarrow &(-1)^{m_1+m_2+m_3}\Gamma_{\ell_1;\ell_2;\ell_3}\left(\begin{array}{ccc}
		\ell_1 & \ell_2 & \ell_3 \\
		-m_1 & -m_2 & -m_3
	\end{array}\right)\left[\left(\begin{array}{ccc}
		\ell_1 & \ell_2 & \ell_3 \\
		1 & -1& 0
	\end{array}\right)
+\left(\begin{array}{ccc}
		\ell_1 & \ell_2 & \ell_3 \\
		-1 & 1 & 0
	\end{array}\right)
\right]. \nonumber
\ea
By making use of 
\ba
	\left(\begin{array}{ccc}
		\ell_1 & \ell_2 & \ell_3 \\
		-m_1 & -m_2 & -m_3
	\end{array}\right)=(-1)^{\ell_1+\ell_3+\ell_3}\left(\begin{array}{ccc}
		\ell_1 & \ell_2 & \ell_3 \\
		m_1 & m_2 & m_3
	\end{array}\right), \nonumber
\ea
we easily derive that
\ba
	\left(\begin{array}{ccc}
		\ell_1 & \ell_2 & \ell_3 \\
		1 & -1& 0
	\end{array}\right)
+\left(\begin{array}{ccc}
		\ell_1 & \ell_2 & \ell_3 \\
		-1 & 1 & 0
	\end{array}\right)
=\left[1+(-1)^{\ell_1+\ell_2+\ell_3}\right]\left(\begin{array}{ccc}
		\ell_1 & \ell_2 & \ell_3 \\
		1 & -1& 0
	\end{array}\right).\nonumber
\ea
Hence the above is non vanishing for even values of $\ell_1+\ell_2+\ell_3$ only. Using the same symmetries of the Wigner-$3j$ symbols as for the case $n=0$, the angular contribution boils down to
\ba
	\!\int\dd^3\xx \, F^{(1)\mathrm{V}}_{\boldsymbol{\ell}_{123}}\left(k_{123},\xx\right) = & \frac{-1}{2}\Gamma_{\ell_1;\ell_2;\ell_3}\!\int \dd x\left[ j_{\ell_1}(k_1x)j_{\ell_2}(k_2x)j_{\ell_3}(k_3x)\right] \\
	&\sqrt{\Delta_{\ell_1}^{(1)}\Delta_{\ell_2}^{(1)}}\left[1+(-1)^{\ell_1+\ell_2+\ell_3}\right]\left(\begin{array}{ccc}
		\ell_1 & \ell_2 & \ell_3 \\
		1 & -1& 0
	\end{array}\right)\left(\begin{array}{ccc}
		\ell_1 & \ell_2 & \ell_3 \\
		m_1 & m_2 & m_3
	\end{array}\right). \nonumber
\ea
In the following, we qualify a term of that form as being of vector type (hence the "V" superscript) in the sense that it involves spin-$(\pm1)$ functions on the sphere, that is components of a vector field.

Using instead the standard Gaunt integral (i.e. without spin), one first note that 
\ba
	 \frac{-1}{2}\sqrt{\Delta_{\ell_1}^{(1)}\Delta_{\ell_2}^{(1)}}\!\int\dd^2\hat\xx \ {}_{\pm1}Y^\star_{\ell_1m_1}\,{}_{\mp1}Y^\star_{\ell_2m_2}\,Y^\star_{\ell_3m_3}=\!\int \dd^2\hat\xx \left(\nabla_\pm Y_{\ell_1m_1}\right)\left(\nabla_\mp Y_{\ell_2m_2}\right)Y_{\ell_3m_3},
\ea
with $\nabla_\pm$ the spin-raising and spin-lowering operators (see App. \ref{sapp:spin}). We show in App. \ref{apps:spinto0} how to express the above spinned Gaunt integral as a function of the standard (i.e. spin-0) Gaunt integral. The end result of this is summarized in App. \ref{apps:recwig} in the form of a relation between the Wigner-$3j$ with spin-$(\pm1)$ and the one with spins zero. This finally gives for the angular part 
\ba
	\!\int\dd^3\xx \, F^{(1)\mathrm{V}}_{\boldsymbol{\ell}_{123}}\left(k_{123},\xx\right) = & \frac{1}{2}\Gamma_{\ell_1;\ell_2;\ell_3}\!\int \dd x\left[ j_{\ell_1}(k_1x)j_{\ell_2}(k_2x)j_{\ell_3}(k_3x)\right] \\
	&\left(\Delta_{\ell_1}^{(1)}+\Delta_{\ell_2}^{(1)}-\Delta_{\ell_3}^{(1)}\right)\left(\begin{array}{ccc}
		\ell_1 & \ell_2 & \ell_3 \\
		0 & 0& 0
	\end{array}\right)\left(\begin{array}{ccc}
		\ell_1 & \ell_2 & \ell_3 \\
		m_1 & m_2 & m_3
	\end{array}\right). \nonumber
\ea

\paragraph*{Expression of $F^{(1)}_{\boldsymbol{\ell}_{123}}$--} The full expression of $F^{(1)}_{\boldsymbol{\ell}_{123}}$ is obtained by summing the radial and the angular part. It is written as 
\ba
	\!\int\dd^3\xx \, F^{(1)}_{\boldsymbol{\ell}_{123}}\left(k_{123},\xx\right) = &\Gamma_{\ell_1;\ell_2;\ell_3}\left(\begin{array}{ccc}
		\ell_1 & \ell_2 & \ell_3 \\
		0 & 0& 0
	\end{array}\right)\left(\begin{array}{ccc}
		\ell_1 & \ell_2 & \ell_3 \\
		m_1 & m_2 & m_3
	\end{array}\right) \nonumber \\
	&\times\!\int x^2\dd x \, I^{(1)}_{\ell_{123}}(k_{123},x),
\ea
where the radial integrand is
\ba
	I^{(1)}_{\ell_{123}}(k_{123},x)= & \left[\frac{\partial j_{\ell_1}(k_1x)}{\partial x}\frac{\partial j_{\ell_2}(k_2x)}{\partial x}+\Lambda^{\mathrm{V}}_{\ell_{123}}\frac{j_{\ell_1}(k_1x)}{x}\frac{j_{\ell_2}(k_2x)}{x}\right]j_{\ell_3}(k_3x),
\ea
and where we introduce
\ba
	\Lambda^{\mathrm{V}}_{\ell_{123}} = & \frac{1}{2}\left(\Delta_{\ell_1}^{(1)}+\Delta_{\ell_2}^{(1)}-\Delta_{\ell_3}^{(1)}\right).
\ea

\subsection{$n=2$ case}
Computing this term proceeds similarly as the case $n=1$. We first plug the $2+1$ decomposition of the metric tensor to write it as the sum of four terms, each given by
\ba
	(\alpha) = & \nabla_x\nabla_x \left[j_{\ell_1}(k_1x)Y^\star_{\boldsymbol{\ell}_1}(\hat\xx)\right]\nabla_x\nabla_x \left[j_{\ell_2}(k_2x)Y^\star_{\boldsymbol{\ell}_2}(\hat\xx)\right], \\
	(\beta) = & \frac{\bar{g}^{ij}}{x^2}\nabla_x\nabla_i \left[j_{\ell_1}(k_1x)Y^\star_{\boldsymbol{\ell}_1}(\hat\xx)\right]\nabla_x\nabla_i \left[j_{\ell_2}(k_2x)Y^\star_{\boldsymbol{\ell}_2}(\hat\xx)\right], \\
	(\gamma) = & \frac{\bar{g}^{ij}}{x^2}\nabla_i\nabla_x \left[j_{\ell_1}(k_1x)Y^\star_{\boldsymbol{\ell}_1}(\hat\xx)\right]\nabla_i\nabla_x \left[j_{\ell_2}(k_2x)Y^\star_{\boldsymbol{\ell}_2}(\hat\xx)\right], \\
	(\delta) = & \frac{1}{x^4}\bar{g}^{ij}\bar{g}^{l n} \nabla_i\nabla_l \left[j_{\ell_1}(k_1x)Y^\star_{\boldsymbol{\ell}_1}(\hat\xx)\right]\nabla_j\nabla_n \left[j_{\ell_2}(k_2x)Y^\star_{\boldsymbol{\ell}_2}(\hat\xx)\right].
\ea
We stress that in the above we have not yet introduced the $2+1$ decomposition of the covariant derivative. We now deals with each term successively.

\paragraph*{The $(\alpha)$ term--} This term is the easiest since there is no angular derivative. Using Eq. (\ref{eq:covrad}) it is given by
\ba
	(\alpha) = \left[\partial^2_xj_{\ell_1}(k_1x)\right]\left[\partial^2_xj_{\ell_2}(k_2x)\right]\left[Y^\star_{\boldsymbol{\ell}_1}(\hat\xx)Y^\star_{\boldsymbol{\ell}_2}(\hat\xx)\right].
\ea

\paragraph*{The $(\beta)$ and $(\gamma)$ terms--} These two terms are equal since $\nabla_x\nabla_i=\nabla_i\nabla_x$ as explained in App. \ref{app:radangdec}. Using Eq. (\ref{eq:covmix}) to get the covariant derivative on the sphere, $\bar{\nabla}_i$, they can be expressed as 
\ba
	(\beta)+(\gamma) = & \frac{2}{x^2}\left[\partial_x j_{\ell_1}(k_1x)-\frac{j_{\ell_1}(k_1x)}{x}\right]\left[\partial_x j_{\ell_2}(k_2x)-\frac{j_{\ell_2}(k_2x)}{x}\right]\nonumber \\
	&\times\bar{g}^{ij}\left[\bar{\nabla}_{i}Y^\star_{\boldsymbol{\ell}_1}(\hat\xx)\right]\left[\bar{\nabla}_{j}Y^\star_{\boldsymbol{\ell}_2}(\hat\xx)\right].
\ea
The second line in the above is finally expressed using spin-$(\pm1)$ spherical harmonics in the very same way as for the angular contribution in the case $n=1$ [see Eq. (\ref{eq:vectpart1})], i.e.
\ba
	(\beta)+(\gamma) = & \frac{-1}{x^2}\left[\partial_x j_{\ell_1}(k_1x)-\frac{j_{\ell_1}(k_1x)}{x}\right]\left[\partial_x j_{\ell_2}(k_2x)-\frac{j_{\ell_2}(k_2x)}{x}\right]\nonumber \\
	& \sqrt{\Delta^{(1)}_{\ell_1}\Delta^{(1)}_{\ell_2}}\Big[{}_{1}Y^\star_{\ell_1m_1}(\hat\xx){}_{-1}Y^\star_{\ell_2m_2}(\hat\xx) +{}_{-1}Y^\star_{\ell_1m_1}(\hat\xx){}_{1}Y^\star_{\ell_2m_2}(\hat\xx)\Big].
\ea

\paragraph*{The $\delta$ term--} This term is clearly the most complicated regarding the $2+1$ decomposition of $\nabla_i\nabla_j$ as given in Eq. (\ref{eq:covang}). Plugging this into the expression of $(\delta)$ we split it into four new terms given by
\ba
	(\delta_1) = & \frac{1}{x^2}\bar{g}^{ij}\bar{g}^{l n}\bar{g}_{il}\bar{g}_{j n}\left[\partial_x j_{\ell_1}(k_1x)\right]\left[\partial_x j_{\ell_2}(k_2x)\right]\left[Y^\star_{\boldsymbol{\ell}_1}(\hat\xx)Y^\star_{\boldsymbol{\ell}_2}(\hat\xx)\right], \\
	(\delta_2) = & \frac{1}{x^3}\bar{g}^{ij}\bar{g}^{l n}\bar{g}_{il}\left[\partial_x j_{\ell_1}(k_1x)\right] j_{\ell_2}(k_2x)Y^\star_{\boldsymbol{\ell}_1}(\hat\xx)\left[\bar{\nabla}_j\bar{\nabla}_nY^\star_{\boldsymbol{\ell}_2}(\hat\xx)\right], \\
	(\delta_3) = & \frac{1}{x^3}\bar{g}^{ij}\bar{g}^{l n}\bar{g}_{jn} j_{\ell_1}(k_1x) \left[\partial_xj_{\ell_2}(k_2x)\right]\left[\bar{\nabla}_i\bar{\nabla}_lY^\star_{\boldsymbol{\ell}_1}(\hat\xx)\right]Y^\star_{\boldsymbol{\ell}_2}(\hat\xx), \\
	(\delta_4) = & \frac{1}{x^4}\bar{g}^{ij}\bar{g}^{l n}\left[j_{\ell_1}(k_1x)j_{\ell_2}(k_2x)\right]\left[\bar{\nabla}_i\bar{\nabla}_lY^\star_{\boldsymbol{\ell}_1}(\hat\xx)\right]\left[\bar{\nabla}_j\bar{\nabla}_nY^\star_{\boldsymbol{\ell}_2}(\hat\xx)\right].
\ea
For the first term, we have $\bar{g}^{ij}\bar{g}^{l n}\bar{g}_{il}\bar{g}_{j n}=\delta^j_l\delta^l_j=2$, hence yielding
\ba
	(\delta_1) = & \frac{2}{x^2}\left[\partial_x j_{\ell_1}(k_1x)\right]\left[\partial_x j_{\ell_2}(k_2x)\right]\left[Y^\star_{\boldsymbol{\ell}_1}(\hat\xx)Y^\star_{\boldsymbol{\ell}_2}(\hat\xx)\right].
\ea
For the term $(\delta_2)$, we first have $\bar{g}^{ij}\bar{g}_{il}=\delta^j_l$. This leads to
\ba
	\bar{g}^{ij}\bar{g}^{l n}\bar{g}_{il}\left[\bar{\nabla}_j\bar{\nabla}_nY^\star_{\boldsymbol{\ell}_2}(\hat\xx)\right]=\bar{g}^{l n}\left[\bar{\nabla}_l\bar{\nabla}_nY^\star_{\boldsymbol{\ell}_2}(\hat\xx)\right].
\ea
We are left with the Laplacian of the spherical harmonics given in Eq. (\ref{eq:lapylm}). We finally obtain for this second term
\ba
	(\delta_2) =\frac{-1}{x^3}\Delta^{(1)}_{\ell_2}\left[\partial_x j_{\ell_1}(k_1x)\right] j_{\ell_2}(k_2x)\left[Y^\star_{\boldsymbol{\ell}_1}(\hat\xx)Y^\star_{\boldsymbol{\ell}_2}(\hat\xx)\right].
\ea
The third term is similarly derived by just permuting $(k_1,\boldsymbol{\ell}_1)$ with $(k_2,\boldsymbol{\ell}_2)$ in $(\delta_2)$. This gives
\ba
	(\delta_3) =\frac{-1}{x^3}\Delta^{(1)}_{\ell_1}j_{\ell_1}(k_1x) \left[\partial_xj_{\ell_2}(k_2x)\right]\left[Y^\star_{\boldsymbol{\ell}_1}(\hat\xx)Y^\star_{\boldsymbol{\ell}_2}(\hat\xx)\right].
\ea
For $(\delta_4)$ we use the decomposition of $\bar{\nabla}_i\bar{\nabla}_lY^\star_{\boldsymbol{\ell}_1}$ given in Eq. (\ref{eq:spin2dec}). This decomposition is made of two terms, one being a pure trace, i.e. $\propto \bar{g}_{ij}$, and the second being traceless, i.e. a linear combination of $(\bar{\mathbf{e}}^{+}_i\bar{\mathbf{e}}^{+}_j)$ and $(\bar{\mathbf{e}}^{-}_i\bar{\mathbf{e}}^{-}_j)$. By plugging this in the expression of $(\delta_4)$ the terms mixing a pure trace with a traceless tensor are proportional to 
\ba
	\bar{g}^{ij}\bar{g}^{l n}\bar{g}_{il}(\bar{\mathbf{e}}^{\pm}_j\bar{\mathbf{e}}^{\pm}_n)=\bar{g}^{jn}(\bar{\mathbf{e}}^{\pm}_j\bar{\mathbf{e}}^{\pm}_n),\nonumber 
\ea
hence vanishing for $(\bar{\mathbf{e}}^{\pm}_j\bar{\mathbf{e}}^{\pm}_n)$ is traceless. The only remaining terms finally lead to
\ba
	(\delta_4) = & \frac{1}{2x^4}\Delta^{(1)}_{\ell_1}\Delta^{(1)}_{\ell_2}\left[j_{\ell_1}(k_1x) j_{\ell_2}(k_2x)\right]\left[Y^\star_{\boldsymbol{\ell}_1}(\hat\xx)Y^\star_{\boldsymbol{\ell}_2}(\hat\xx)\right] \\
	& + \frac{1}{4x^4}\sqrt{\Delta^{(2)}_{\ell_1}\Delta^{(2)}_{\ell_2}}\left[j_{\ell_1}(k_1x) j_{\ell_2}(k_2x)\right]\left[{}_2Y^\star_{\boldsymbol{\ell}_1}(\hat\xx){}_{-2}Y^\star_{\boldsymbol{\ell}_2}(\hat\xx)+{}_{-2}Y^\star_{\boldsymbol{\ell}_1}(\hat\xx){}_2Y^\star_{\boldsymbol{\ell}_2}(\hat\xx)\right]. \nonumber
\ea
The first line is the pure trace contribution and the second line is the traceless contribution.

\paragraph*{Expression of $F^{(2)}_{\boldsymbol{\ell}_{123}}$--} All the above are now plugged into the expression of $F^{(2)}_{\boldsymbol{\ell}_{123}}$. We group them with respect to their properties on the sphere. Following the scalar and vector classification we introduced for the case $n=1$, we note that $(\alpha)$, $(\delta_{1,2,3})$ and the pure trace part of $(\delta_4)$ are of scalar type. The term coming from $(\beta)+(\gamma)$ is of vector type. Finally, the term coming from the pure traceless part of $(\delta_4)$ involves spin-$(\pm2)$ and we classify it as tensor type.

The final expression of $F^{(2)}_{\boldsymbol{\ell}_{123}}$ is obtained using Gaunt integrals. Its scalar contribution is
\ba
	\!\int\dd^3\xx \, F^{(2)\mathrm{S}}_{\boldsymbol{\ell}_{123}}\left(k_{123},\xx\right) = & \Gamma_{\ell_1;\ell_2;\ell_3}\left(\begin{array}{ccc}
		\ell_1 & \ell_2 & \ell_3 \\
		0 & 0 & 0
	\end{array}\right)\left(\begin{array}{ccc}
		\ell_1 & \ell_2 & \ell_3 \\
		m_1 &m_2 & m_3
	\end{array}\right)\!\int x^2\dd x\, I^{(2)\mathrm{S}}_{\ell_{123}}(k_{123},x) , 
\ea
where the scalar radial integrand is
\ba
	I^{(2)\mathrm{S}}_{\ell_{123}}(k_{123},x) = &\Big\{\left[\partial^2_xj_{\ell_1}(k_1x)\right]\left[\partial^2_xj_{\ell_2}(k_2x)\right]+\frac{2}{x^2}\left[\partial_x j_{\ell_1}(k_1x)\right]\left[\partial_x j_{\ell_2}(k_2x)\right] \nonumber \\
	&-\frac{1}{x^3}\ell_2(\ell_2+1)\left[\partial_x j_{\ell_1}(k_1x)\right] j_{\ell_2}(k_2x)-\frac{1}{x^3}\ell_1(\ell_1+1)j_{\ell_1}(k_1x) \left[\partial_xj_{\ell_2}(k_2x)\right] \nonumber \\
	&\frac{1}{2x^4}\ell_1(\ell_1+1)\ell_2(\ell_2+1)\left[j_{\ell_1}(k_1x) j_{\ell_2}(k_2x)\right]\Big\}j_{\ell_3}(k_3x).
\ea
The first term involving second derivatives of the Bessel functions comes for $(\alpha)$ while all the others come from $(\delta)$. These second set of contributions can be more compactly written to finally get
\ba
	I^{(2)\mathrm{S}}_{\ell_{123}}(k_{123},x)= & \left[\partial^2_xj_{\ell_1}(k_1x)\right]\left[\partial^2_xj_{\ell_2}(k_2x)\right]\,j_{\ell_3}(k_3x) \\
	 &+\frac{2}{x^2}\left[\partial_x j_{\ell_1}(k_1x)-\frac{\ell_1(\ell_1+1)}{2}\frac{j_{\ell_1}(k_1x)}{x}\right]\left[\partial_x j_{\ell_2}(k_2x)-\frac{\ell_2(\ell_2+1)}{2}\frac{j_{\ell_2}(k_2x)}{x}\right]\,j_{\ell_3}(k_3x). \nonumber
\ea

The vector contribution is given by
\ba
	\!\int\dd^3\xx \, F^{(2)\mathrm{V}}_{\boldsymbol{\ell}_{123}}\left(k_{123},\xx\right) = & \frac{-1}{2}\Gamma_{\ell_1;\ell_2;\ell_3}\left[1+(-1)^{\ell_1+\ell_2+\ell_3}\right]\left(\begin{array}{ccc}
		\ell_1 & \ell_2 & \ell_3 \\
		1 & -1& 0
	\end{array}\right)\left(\begin{array}{ccc}
		\ell_1 & \ell_2 & \ell_3 \\
		m_1 &m_2 & m_3
	\end{array}\right) \nonumber \\
	& \times\sqrt{\Delta^{(1)}_{\ell_1}\Delta^{(1)}_{\ell_2}}\!\int x^2 {\dd x}\, I^{(2)\mathrm{V}}_{\ell_{123}}(k_{123},x),
\ea
where the vector radial integrand is given by
\ba
	I^{(2)\mathrm{V}}_{\ell_{123}}(k_{123},x) =\frac{2}{x^2}\left[\partial_x j_{\ell_1}(k_1x)-\frac{j_{\ell_1}(k_1x)}{x}\right]\left[\partial_x j_{\ell_2}(k_2x)-\frac{j_{\ell_2}(k_2x)}{x}\right]\,j_{\ell_3}(k_3x).
\ea

	Finally, the tensor contribution is obtained using Gaunt integrals with spin-$(\pm2)$. It boils down to
\ba
	\!\int\dd^3\xx \, F^{(2)\mathrm{T}}_{\boldsymbol{\ell}_{123}}\left(k_{123},\xx\right) = & \Gamma_{\ell_1;\ell_2;\ell_3}\left[1+(-1)^{\ell_1+\ell_2+\ell_3}\right]\left(\begin{array}{ccc}
		\ell_1 & \ell_2 & \ell_3 \\
		2 & -2& 0
	\end{array}\right)\left(\begin{array}{ccc}
		\ell_1 & \ell_2 & \ell_3 \\
		m_1 &m_2 & m_3
	\end{array}\right) \nonumber \\
	& \times\frac{1}{4}\sqrt{\Delta^{(2)}_{\ell_1}\Delta^{(2)}_{\ell_2}}\!\int x^2{\dd x}\,I^{(2)\mathrm{T}}_{\ell_{123}}(k_{123},x) .
\ea
The tensor radial integrand is
\ba
	I^{(2)\mathrm{T}}_{\ell_{123}}(k_{123},x) =\frac{1}{x^4}\left[j_{\ell_1}(k_1x) j_{\ell_2}(k_2x)j_{\ell_3}(k_3x)\right].
\ea

We finally combine all the above terms to derive an expression for $F^{(2)}_{\boldsymbol{\ell}_{123}}$. To this end, we first replace the Wigner-$3j$ symbols  with non-zero spins in the vector and tensor contribution as functions of the Wigner-$3j$ symbols with zero spin (see App. \ref{apps:spinto0} for the the full proof and App. \ref{apps:recwig} for the end result). This finally gives
\ba
	\!\int\dd^3\xx \, F^{(2)}_{\boldsymbol{\ell}_{123}}\left(k_{123},\xx\right)= \Gamma_{\ell_1;\ell_2;\ell_3}\left(\begin{array}{ccc}
		\ell_1 & \ell_2 & \ell_3 \\
		0 & 0 & 0
	\end{array}\right)\left(\begin{array}{ccc}
		\ell_1 & \ell_2 & \ell_3 \\
		m_1 &m_2 & m_3
	\end{array}\right)\!\int x^2\dd x\, I^{(2)}_{\ell_{123}}(k_{123},x),
\ea
where the full radial integrand is 
\ba
	I^{(2)}_{\ell_{123}}(k_{123},x)=&I^{(2)\mathrm{S}}_{\ell_{123}}(k_{123},x)+\Lambda^{\mathrm{V}}_{\ell_{123}}I^{(2)\mathrm{V}}_{\ell_{123}}(k_{123},x) +\Lambda^{\mathrm{T}}_{\ell_{123}} I^{(2)\mathrm{T}}_{\ell_{123}}(k_{123},x),
\ea
and where we further introduce
\ba
	\Lambda^{\mathrm{T}}_{\ell_{123}} = \frac{1}{8}\left(\frac{\Delta^{(2)}_{\ell_1}}{\Delta^{(1)}_{\ell_1}}+\frac{\Delta^{(2)}_{\ell_2}}{\Delta^{(1)}_{\ell_2}}+\Delta^{(1)}_{\ell_1}+\Delta^{(1)}_{\ell_2}-2\Delta^{(1)}_{\ell_3}\right)\left(\Delta^{(1)}_{\ell_1}+\Delta^{(1)}_{\ell_2}-\Delta^{(1)}_{\ell_3}\right).
\ea

\section{Removing Bessel derivatives with integration by part}\label{App:remove-bessel-derivs-IPP}

We want to simplify a bispectrum term of the form
\ba
b_{\ell_1,\ell_2,\ell_3} = & \left(\frac{2}{\pi}\right)^3\!\int x^2\dd x \ r^2_{123} \, \dd r_{123}  \ k_{123}^2 \, \dd k_{123} \ F(k_{123},r_{123},x) \ j_{\lu}(k_1 r_1) \, j_{\ld}(k_2 r_2) \, j_{\lt}(k_3 r_3) \nonumber \\
& \times \partial_x^{n_1} j_{\lu}(k_1 x) \, \partial_x^{n_2} j_{\ld}(k_2 x) \, \partial_x^{n_3} j_{\lt}(k_3 x)
\ea
to make it amenable to Limber's approximation or integration with FFTlog methods, i.e. our goal is to get rid of the derivatives of Bessel function and end up only with normal Bessel functions.

We start by noting that
\ba
\partial_x^{n} j_{\ell}(k x) = k^n j^{(n)}_{\ell}(k x) = \left(\frac{k}{x}\right)^n \partial_k^{n} j_{\ell}(k x).
\ea
Then we perform integrals by part in all $k_i$ integrals
\ba
b_{\ell_1,\ell_2,\ell_3} = & \left(\frac{2}{\pi}\right)^3\!\int x^2\dd x \ r^2_{123} \, \dd r_{123}  \int F(k_{123},r_{123},x) \prod_{i=1}^3 k_{i}^2 \, \dd k_{i} \ j_{\ell_i}(k_i r_i) \, \left(\frac{k_i}{x}\right)^{n_i} \partial_{k_i}^{n_i} j_{\ell_i}(k_i x) \\
& = (-1)^{n_1+n_2+n_3} \left(\frac{2}{\pi}\right)^3\!\int \frac{x^2 \, \dd x}{x^{n_1+n_2+n_3}} \ r^2_{123} \, \dd r_{123} \int \left(\prod_{i=1}^3 \dd k_{i} \ j_{\ell_i}(k_i x)\right) \nonumber \\
& \times \frac{\partial^{n_1+n_2+n_3}}{\partial k_1^{n_1} \, \partial k_2^{n_2} \, \partial k_3^{n_3}}\left[F(k_{123},r_{123},x) \prod_{i=1}^3 k_i^{n_i+2} \, j_{\ell_i}(k_i r_i)\right] \label{Eq:bispectrum-IPPs-intermediate-result}
\ea
where boundary terms vanish through careful inspection\footnote{Boundary terms for a $k_i$ integral are of the form $\left[ \partial_{k_i}^{n_i-\alpha} j_{\ell_i}(k_i x) \times \frac{\partial^{\alpha}}{\partial k_i^{\alpha}} \left(F(k_{123},r_{123},x)  k_i^{n_i+2} j_{\ell_i}(k_i r_i) \right) \right]_{k_i=0}^\infty$. This vanishes at $k_i=0$ as long as $F(k_{123},r_{123},x)$ (and its derivatives) is finite there, which is the case for the LSS bispectrum. It also vanishes when $k_i\rightarrow\infty$ as long as $F(k_{123},r_{123},x) \, k_i^{n_i} \rightarrow 0$. The latter holds true for the LSS bispectrum $B\propto P(k_1) P(k_2)$ given that $n_i\leq 2$ and $P(k) \sim k^{-3}$ when $k\rightarrow\infty$ ; the exception is $k_3$ (there is no $P(k_3)$), where inspection of Table \ref{Table:radkernels} shows that we have $n_3=0$ in all cases, so there is no integration by part needed and Eq.\ref{Eq:bispectrum-IPPs-intermediate-result} holds true.}.\\
Then we use the Leibniz rule for the derivative of products to get the Bessel functions out
\ba
b_{\ell_1,\ell_2,\ell_3} = & (-1)^{n_1+n_2+n_3} \left(\frac{2}{\pi}\right)^3\!\int \frac{x^2 \, \dd x}{x^{n_1+n_2+n_3}} \ r^2_{123} \, \dd r_{123} \int \left(\prod_{i=1}^3 \dd k_{i} \ j_{\ell_i}(k_i x)\right) \nonumber \\
& \times \sum_{m_{123}=0}^{n_{123}} \frac{\partial^{n_1+n_2+n_3-(m_1+m_2+m_3)}}{\partial k_1^{n_1-m_1} \, \partial k_2^{n_2-m_2} \, \partial k_3^{n_3-m_3}}\left[F(k_{123},r_{123},x) \prod_{i=1}^3 k_i^{n_i+2} \right] \nonumber \\
& \times \prod_{i=1}^3 \begin{pmatrix}n_i \\ m_i\end{pmatrix} \frac{\partial^{m_i}}{\partial k_i^{m_i}} j_{\ell_i}(k_i r_i)
\ea
and redo the trick on derivatives of $j_\ell$
\ba
\partial_{k_i}^{m_i} j_{\ell_i}(k_i r_i) = r_i^{m_i} j^{(n)}_{\ell_i}(k_i r_i) = \left(\frac{r_i}{k_i}\right)^{m_i} \partial_{r_i}^{m_i} j_{\ell_i}(k_i r_i).
\ea
Finally we redo integrals by parts now in all $r_i$ integrals to obtain the wanted result:
\ba\label{Eq:bispectrum-IPPs-result}
b_{\ell_1,\ell_2,\ell_3} = & \left(\frac{2}{\pi}\right)^3\!\int \frac{x^2 \, \dd x}{x^{n_1+n_2+n_3}} \ \dd r_{123} \int \sum_{m_{123}=0}^{n_{123}} \left(\prod_{i=1}^3 \frac{\dd k_{i}}{k_i^{m_i}} \ j_{\ell_i}(k_i x) j_{\ell_i}(k_i r_i) (-1)^{n_i+m_i} \begin{pmatrix}n_i \\ m_i\end{pmatrix} \right) \nonumber \\
& \times \left(\prod_{i=1}^3 \frac{\partial^{n_i}}{\partial r_i^{m_i} \partial k_i^{n_i-m_i}}\right) \left[F(k_{123},r_{123},x) \prod_{i=1}^3 k_i^{n_i+2} \, r_i^{m_i+2}\right]
\ea
where again boundary terms vanish through careful inspection\footnote{The argument is analog to the earlier case of the $k_i$ integral. Here the condition is that $F(k_{123},r_{123},x) \, r_i^2 \rightarrow 0$ when $r_i\rightarrow 0$ and $r_i\rightarrow \infty$. This holds true for the LSS bispectrum because $F(k_{123},r_{123},x)$ is proportional to the tracer kernel (number density of galaxies, lensing efficiency...), and the condition means that this kernel must be integrable, which is indeed the case.}.

\section{Naive Limber approximation}\label{App:naive-Limber}

\subsection{Power spectrum case}\label{App:naive-Limber-power-spectrum}
A naive way to see the Limber approximation is to replace the Bessel function by a Dirac \cite{Tanidis2019}:
\be\label{App:Eq:bessel=dirac}
j_\ell(kr) \rightarrow \sqrt{\frac{\pi}{2\ell+1}} \delta_D(\ell+\sfrac{1}{2} - kr)
\ee
If we take as example the classical projection of the power spectrum
\be
C_\ell = \frac{2}{\pi} \int x^2 \dd x \, r^2\dd r \, k^2 \dd k \ P(k|x,r) \ j_\ell(kx) \, j_\ell(kr).
\ee
after a bit of algebra this replacement gives 
\be
C_\ell \approx \int x^2 \dd x \ P(k_\ell|x)
\ee
which indeed coincides with the correct result of the Limber approximation obtained through the method of Sect.~\ref{Sect:proj-Limber}.

The situation is a bit more involved if we consider a similar integral containing Bessel derivatives:
\ba
\tilde{C}_\ell &= \frac{2}{\pi} \int x^2 \dd x \, r^2\dd r \, k^2 \dd k \ P(k|x,r) \ \frac{\partial^2 j_\ell(kx)}{\partial k^2} \, j_\ell(kr).
\ea
This type of integral appears for instance when considering the effect of redshift-space distortions \cite[see e.g.][]{Tanidis2019}.\\
One can tackle this integral with the classical Limber approach after integrations by part:
\ba
\tilde{C}_\ell &= \frac{2}{\pi} \int x^2 \dd x \, r^2\dd r \, k^2 \dd k \ P(k|x,r) \ \frac{\partial^2 j_\ell(kx)}{\partial k^2} \,  j_\ell(kr)  \\
&= \frac{2}{\pi} \int x^2 \dd x \, r^2\dd r \, \dd k \ j_\ell(kx) \frac{\partial^2 \left[k^2 \, P(k|x,r) \, j_\ell(kr)\right]}{\partial k^2} \\
&= \frac{2}{\pi} \int x^2 \dd x \, r^2\dd r \, k^2\dd k \ j_\ell(kx) \frac{1}{k^2} \sum_{m=0}^{2}  \begin{pmatrix}2 \\ m\end{pmatrix} \frac{\partial^{2-m} \left[k^2 \, P(k|x,r) \right]}{\partial k^{2-m}} \frac{r^m}{k^m} \frac{\partial^{m} j_\ell(kr)}{\partial r^{m}}\\
&= \frac{2}{\pi} \int x^2 \dd x \, r^2\dd r \, k^2\dd k \ j_\ell(kx) \, j_\ell(kr)  \sum_{m=0}^{2}  \begin{pmatrix}2 \\ m\end{pmatrix} \frac{(-1)^m}{k^{m+2} \, r^2} \frac{\partial^{2} \left[k^2 \, r^{m+2} \, P(k|x,r) \right]}{\partial k^{2-m} \partial r^{m}} \\
&\approx \int x^2 \dd x \left(\sum_{m=0}^{2}  \begin{pmatrix}2 \\ m\end{pmatrix} \frac{(-1)^m}{k^{m+2} \, r^2} \frac{\partial^{2} \left[k^2 \, r^{m+2} \, P(k|x,r) \right]}{\partial k^{2-m} \partial r^{m}} \right)_{k=k_\ell,r=x} \label{App:Eq:Limber-prologue-d2Bessel-method1}
\ea
Alternatively through the Dirac approach one finds
\ba
\tilde{C}_\ell 
& \approx \frac{1}{\ell+\sfrac{1}{2}} \int x^2 \dd x \ \frac{1}{x} \left.\frac{\partial^2 \left[ k \, r_\ell^2 \, P(k|x,r_\ell) \right]}{\partial k^2}\right|_{k=k_\ell,r=x} \label{App:Eq:Limber-prologue-d2Bessel-method2}
\ea
where one must take care that $r_\ell=(\ell+\sfrac{1}{2})/k$ is also hit by the derivative.\\
The two results Eq.~\ref{App:Eq:Limber-prologue-d2Bessel-method1} \& \ref{App:Eq:Limber-prologue-d2Bessel-method2} look dissimilar at first, but after careful algebra one can show that they are equal. We are thus provided with two equivalent analytical forms, where the form Eq.~\ref{App:Eq:Limber-prologue-d2Bessel-method2} is more compact and more useful for physical interpretation, while the form Eq.~\ref{App:Eq:Limber-prologue-d2Bessel-method1} separates the effects of spatial and temporal evolutions of cosmological quantities and is more adequate for numerical implementation.

\subsection{Application to the bispectrum}\label{App:naive-Limber-bispectrum-allell}

We now apply the same process to simplify the more complex case of a bispectrum term of the form
\ba
\nonumber b_{\ell_1,\ell_2,\ell_3} = & \left(\frac{2}{\pi}\right)^3\!\int x^2\dd x \ r^2_{123} \, \dd r_{123}  \ k_{123}^2 \, \dd k_{123} \ F(k_{123},r_{123},x) \ j_{\lu}(k_1 r_1) \, j_{\ld}(k_2 r_2) \, j_{\lt}(k_3 r_3) \\
& \times \partial_x^{n_1} j_{\lu}(k_1 x) \, \partial_x^{n_2} j_{\ld}(k_2 x) \, \partial_x^{n_3} j_{\lt}(k_3 x)
\ea
We first transform the simple Bessel function, leaving the derivatives, and use the Diracs to perform the integrals over $r_{123}$, giving
\ba
b_{\ell_1,\ell_2,\ell_3} = & \left(\frac{2}{\pi}\right)^{3/2} \frac{1}{\sqrt{(\ell+\sfrac{1}{2})_{123}}} \!\int x^2\dd x \ k_{123}^2 \, \dd k_{123} \frac{F(k_{123},r_{123},x) \, r^2_{123}}{k_{123}} \prod_{i=1}^3{\partial_{x}^{n_i} j_{\ell_i}(k_i x)}
\ea
where now $r_i=\frac{\ell_i+\sfrac{1}{2}}{k_i}$. Then we use the trick of Appendix~\ref{App:remove-bessel-derivs-IPP}, namely $\partial_x^{n} j_{\ell}(k x) = \left(\frac{k}{x}\right)^n \partial_k^{n} j_{\ell}(k x)$
\ba
b_{\ell_1,\ell_2,\ell_3} = & \left(\frac{2}{\pi}\right)^{3/2} \frac{1}{\sqrt{(\ell+\sfrac{1}{2})_{123}}} \!\int \frac{x^2\dd x}{x^{n_1+n_2+n_3}} \dd k_{123} \ F(k_{123},r_{123},x) \, r^2_{123} \prod_{i=1}^3{k_i^{n_i+1} \, \partial_{k_i}^{n_i} j_{\ell_i}(k_i x)}
\ea
And perform integration by part on all integrals over $k_{123}$, giving
\ba
\nonumber b_{\ell_1,\ell_2,\ell_3} = &  \left(\frac{2}{\pi}\right)^{3/2} \frac{(-1)^{n_1+n_2+n_3}}{\sqrt{(\ell+\sfrac{1}{2})_{123}}} \!\int \frac{x^2\dd x}{x^{n_1+n_2+n_3}} \ \dd k_{123} \ j_{\lu}(k_1 x) \,  j_{\ld}(k_2 x) \, j_{\lt}(k_3 x) \\
& \times \frac{\partial^{n_1+n_2+n_3}}{\partial k_1^{n_1} \partial k_2^{n_2} \partial k_3^{n_3}} \left[ F(k_{123},r_{123},x) \ r^2_{123} \prod_{i=1}^3{k_i^{n_i+1}}\right]
\ea
We transform again the Bessel functions into Diracs and use them to perform the integrals over $k_{123}$, giving
\ba
\nonumber b_{\ell_1,\ell_2,\ell_3} = &  \frac{(-1)^{n_1+n_2+n_3}}{(\ell+\sfrac{1}{2})_{123}} \!\int \frac{x^2\dd x}{x^{n_1+n_2+n_3}} \frac{1}{x^3}\\
& \times \frac{\partial^{n_1+n_2+n_3}}{\partial k_1^{n_1} \partial k_2^{n_2} \partial k_3^{n_3}} \left[ F\left(k_{123},r_i=\frac{\ell_i+\sfrac{1}{2}}{k_i},x\right) \prod_{i=1}^3{k_i^{n_i+1} \left(\frac{\ell_i+\sfrac{1}{2}}{k_i}\right)^2}\right]_{k_i=k_{\ell_i}, r_i=x}
\ea
By comparison, the other way to Limber's approximation (section~\ref{Sect:proj-Limber}) gives
\ba
b_{\ell_1,\ell_2,\ell_3} = & (-1)^{n_1+n_2+n_3}\int \frac{x^2 \, \dd x}{x^{n_1+n_2+n_3}} \sum_{m_{123}=0}^{n_{123}} \left(\prod_{i=1}^3 (-1)^{m_i} \begin{pmatrix}n_i \\ m_i\end{pmatrix} \right) \nonumber \\
& \times \left[ \left(\prod_{i=1}^3 \frac{1}{r_i^2 k_i^{m_i+2}} \frac{\partial^{n_i}}{\partial r_i^{m_i} \partial k_i^{n_i-m_i}}\right) \left[F(k_{123},r_{123},x) \prod_{i=1}^3 k_i^{n_i+2} \, r_i^{m_i+2}\right] \right]_{k_i=k_{\ell_i}, r_i=x}
\ea
As argued in App.~\ref{App:naive-Limber-power-spectrum}, the two equations are equivalent, but have different advantage in terms of compactness and numerical implementation.

\section{Projecting LSS terms}\label{App:proj-LSS}
In this appendix we apply the general methods of appendices \ref{App:remove-bessel-derivs-IPP} and \ref{App:naive-Limber} to the LSS bispectrum which is the subject of the article.
From Table~\ref{Table:radkernels}, and ignoring permutations, it can be seen that all bispectrum terms that need to be projected can be decomposed into sub-terms where the needed elements are of the form
\ba
\tilde{B}_n(k_{123},r_{123}) &= n_\mr{t_1} \, n_\mr{t_2} \, n_\mr{t_3}\ P(k_1|r_{13}) \, P(k_2|r_{23}) \ b^\mr{t_1}_1 \, b^\mr{t_2}_1 \, b^\mr{t_3}_{a} \ k_1^\mu \, k_2^\nu \\
\mathcal{I}^{(n),A}_{\ell_{123}} &= \frac{B}{x^b}\frac{\partial^c_x j_{\ell_1}}{k_1^\xi} \ \frac{\partial^d_x j_{\ell_2}}{k_2^\rho} \ j_{\ell_3}
\ea
for some integer values of $(\mu,\nu,\xi,\rho,b,c,d)$\footnote{These are not all independent, since $\tilde{B}_n$ is required to have dimension 1/Volume and $\mathcal{I}^{(n)}$ is dimensionless. We however need not enter these considerations here.}, some constant $B$ (that may depend on $\ell_i$) and $a\in\{((0)) ; 1 ; ((2))\}$ so that $b^\mr{t_3}_{a}$ only depends on $k_3$.\\
Such a bispectrum is separable and composed of two types of integrals: first the integral on $k_3$, of the form
\ba
\frac{2}{\pi}\int k^2_3 \, \dd k_3 \ r^2_3 \, \dd r_3 \ n_{t_3} \, b_a^{t_3} \, G(r_3)^2 \ j_{\ell_3}(k_3 x) \, j_{\ell_3}(k_3 r_3)
\ea
and second the integrals on $k_1$ and $k_2$, of the form
\ba
P &= \frac{2}{\pi}\int k^2_i \, \dd k_i \ r^2_i \, \dd r_i \ n_{t_i} \, b_1^{t_i} \, G(r_i) \, P(k_i) / k_i^\alpha \ \partial^c_x j_{\ell_i}(k_i x) \, j_{\ell_i}(k_i r_i)
\ea
We now tackle these integrals and find the total bispectrum in three cases: a first case with Limber's approximation (section~\ref{App:proj-LSS-Limber}), a beyond-Limber (exact) case (section~\ref{App:proj-LSS-noLimber}) and a mixed case with Limber's approximation applied on $k_2$ and $k_3$ but not on $k_1$ (section~\ref{App:proj-LSS-mixedLimber}).

\subsection{With Limber's approximation}\label{App:proj-LSS-Limber}
Limber's approximation allows to trivialise the integral on $k_3$. Remembering $k_{\ell_i} = \frac{\ell_i+\sfrac{1}{2}}{x}$, it gives
\ba
\frac{2}{\pi}\int k^2_3 \, \dd k_3 \ r^2_3 \, \dd r_3 \ n_{t_3}(r_3) \, b_a^{t_3}(k_3,r_3) \, G(r_3)^2 \ j_{\ell_3}(k_3 x) \, j_{\ell_3}(k_3 r_3) = n_{t_3}(x) \, b_a^{t_3}(k_{\ell_3},x) \, G(x)^2
\ea
For the more complex integrals on $k_1$ and $k_2$, we use the approach of App.~\ref{App:naive-Limber-power-spectrum}. Remembering $r_{\ell_i} = \frac{\ell_i+\sfrac{1}{2}}{k_i}$, we find
\ba
P &= \frac{(-1)^c}{(\ell_i+\sfrac{1}{2}) \ x^{c+1}} \left[\frac{\dd^c}{\dd k_i^c} \left[ n_{t_i} \, b_1^{t_i} \, G(r_i) \, \frac{P(k_i)}{k_i^\alpha} k_i^{c+1} r_{\ell_i}^2 \right]  \right]_{k_i=k_{\ell_i},r_{\ell_i}=x} \\
&= \frac{(-1)^c \, (\ell_i+\sfrac{1}{2})}{x^{c+1}} \left[\frac{\dd^c}{\dd k_i^c} \left[ n_{t_i} \, b_1^{t_i} \, G(r_i) \, \frac{P(k_i)}{k_i^\alpha} k_i^{c-1} \right]  \right]_{k_i=k_{\ell_i},r_{\ell_i}=x} \\
\nonumber &= \frac{(-1)^c \, (\ell_i+\sfrac{1}{2})}{x^{c+1}} n_{t_i} \, b_1^{t_i} \, G(x) \, \frac{P(k_{\ell_i})}{k_{\ell_i}^{\alpha+1}} \\
& \quad \times\left[\frac{k_i^c}{n_{t_i} \, b_1^{t_i} \, G(r_i) \, \frac{P(k_i)}{k_i^\alpha} k_i^{c-1}}\frac{\dd^c}{\dd k_i^c} \left[ n_{t_i} \, b_1^{t_i} \, G(r_i) \, \frac{P(k_i)}{k_i^\alpha} k_i^{c-1} \right]  \right]_{k_i=k_{\ell_i},r_{\ell_i}=x} \\
&= \frac{(-1)^c \, (\ell_i+\sfrac{1}{2})}{x^{c+1}} n_{t_i} \, b_1^{t_i} \, G(x) \, \frac{P(k_{\ell_i})}{k_{\ell_i}^{\alpha+1}} \ \mathcal{L}^{t_i,(c)}_{\alpha}(\ell_i,x) \\
&= \frac{(-1)^c}{(\ell_i+\sfrac{1}{2})^\alpha} n_{t_i} \, b_1^{t_i} \, G(x) \, P(k_{\ell_i}) \ \mathcal{L}^{t_i,(c)}_{\alpha}(\ell_i,x) \times \frac{1}{x^\mr{c-\alpha}} \label{App:Eq:proj-LSS-Limber-integral}
\ea
In the equations above, we have defined
\ba\label{App:Eq:def-calL^t,c_alpha}
\mathcal{L}^{t,(c)}_{\alpha}(\ell,x) = \left[\frac{k^c}{n_{t}(r_\ell) \, b_1^{t}(k,r_\ell) \, G(r_\ell) \, \frac{P(k)}{k^\alpha} k^{c-1}}\frac{\dd^c}{\dd k^c} \left[ n_{t}(r_\ell) \, b_1^{t}(k,r_\ell) \, G(r_\ell) \, \frac{P(k)}{k^\alpha} k^{c-1}\right]  \right]_{k=(\ell+\sfrac{1}{2})/x}
\ea
which is a dimensionless quantity with properties detailed in Appendix~\ref{App:props-L-symbols}.\\
Grouping the projections of $k_1, k_2$ and $k_3$, the bispectrum takes the form
\ba
\nonumber b_{\lu,\ld,\lt}^{n,A} = 
\frac{(-1)^{c+d} \, B}{(\lu+\sfrac{1}{2})^\alpha \, (\ld+\sfrac{1}{2})^\beta} \int & x^2 \, \dd x \ n_\mr{t_1} \, n_\mr{t_2} \, n_\mr{t_3} \ b^\mr{t_1}_1 \, b^\mr{t_2}_1 \, b^\mr{t_3}_{a} \ P(k_1|x) \, P(k_2|x) \\
& \times \mathcal{L}^{t_1,(c)}_{\alpha}(\lu,x) \, \mathcal{L}^{t_2,(d)}_{\beta}(\ld,x) \label{App:Eq:proj-LSS-Limber-bispectrum}
\ea
where $\alpha = \xi -\mu$ and $\beta = \rho-\nu$ are the respective exponents of $k_1$ and $k_2$.

\subsection{Exact / beyond Limber}\label{App:proj-LSS-noLimber}

The integral on $k_3$ can be simplified no further analytically, it must be performed numerically, e.g. with the FFTlog algorithm. Instead, we define a tautology whose interest will become apparent later:
\ba
\frac{2}{\pi}\int k^2_3 \, \dd k_3 \ r^2_3 \, \dd r_3 \ n_{t_3}(r_3) \, & b_a^{t_3}(k_3,r_3) \, G(r_3)^2 \ j_{\ell_3}(k_3 x) \, j_{\ell_3}(k_3 r_3) \nonumber \\
&= n_{t_3}(x) \, b_a^{t_3}(k_{\ell_3},x) \, G(x)^2 \times \mathscr{G}^{t_3}_{a}(\ell_3,x)
\ea
with
\ba\label{App:Eq:def-scrG^t_a}
\mathscr{G}^{t_3}_{a}(\ell_3,x) \equiv \frac{2/\pi}{n_{t_3}(x) \, b_a^{t_3}(k_{\ell_3},x) \, G(x)^2}\int k^2_3 \, \dd k_3 \ r^2_3 \, \dd r_3 \ n_{t_3}(r_3) \, b_a^{t_3}(k_3,r_3) \, G(r_3)^2 \ j_{\ell_3}(k_3 x) \, j_{\ell_3}(k_3 r_3)
\ea

For the more complex integrals on $k_1$ and $k_2$, we use the approach of Appendix~\ref{App:remove-bessel-derivs-IPP} in order to transform the derivatives of Bessel functions into normal Bessel functions. We find
\ba
P &= \frac{2}{\pi}\int k^2_i \, \dd k_i \ r^2_i \, \dd r_i \ n_{t_i} \, b_1^{t_i} \, G(r_i) \, P(k_i) / k_i^\alpha \ \partial^c_x j_{\ell_i}(k_i x) \, j_{\ell_i}(k_i r_i) \\
&= \frac{2}{\pi}\int k^2_i \, \dd k_i \ r^2_i \, \dd r_i \ n_{t_i} \, b_1^{t_i} \, G(r_i) \, P(k_i) / k_i^\alpha \ (k_i/x)^c \ \partial^c_{k_i} j_{\ell_i}(k_i x) \, j_{\ell_i}(k_i r_i) \\
&= \frac{(-1)^c}{x^c} \frac{2}{\pi}\int k^2_i \, \dd k_i \ r^2_i \, \dd r_i \ n_{t_i} \, G(r_i) \, \frac{1}{k_i^2}\frac{\partial^c}{\partial k_i^c} \left[b_1^{t_i} \, P(k_i)/ k_i^\alpha \ k_i^{c+2} \  j_{\ell_i}(k_i r_i)\right] \ j_{\ell_i}(k_i x) \\
&= \frac{(-1)^c}{x^c} \!\sum_{m_c=0}^c \!\begin{pmatrix}c \\ m_c\end{pmatrix}\! \frac{2}{\pi}\int k^2_i \, \dd k_i \ r^2_i \, \dd r_i \ n_{t_i} \, G(r_i) \, j_{\ell_i}(k_i x) \frac{1}{k_i^2} \frac{\partial^{c-m_c}}{\partial k_i^{c-m_c}} \!\left[b_1^{t_i} \, P(k_i)/ k_i^\alpha \ k_i^{c+2}\right] \nonumber \\
& \quad \times \partial^{m_c}_{k_i} j_{\ell_i}(k_i r_i) \\
&= \frac{(-1)^c}{x^c} \!\sum_{m_c=0}^c \!\begin{pmatrix}c \\ m_c\end{pmatrix}\! \frac{2}{\pi}\int k^2_i \, \dd k_i \ r^2_i \, \dd r_i \ n_{t_i} \, G(r_i) \, j_{\ell_i}(k_i x) \frac{1}{k_i^2} \frac{\partial^{c-m_c}}{\partial k_i^{c-m_c}} \!\left[b_1^{t_i} \, P(k_i)/ k_i^\alpha \ k_i^{c+2} \right] \nonumber \\
& \quad \times (r_i/k_i)^{m_c} \ \partial^{m_c}_{r_i} j_{\ell_i}(k_i r_i) \\
&= \frac{(-1)^c}{x^c} \!\sum_{m_c=0}^c \!\begin{pmatrix}c \\ m_c\end{pmatrix}\! \frac{2}{\pi}\int k^2_i \, \dd k_i \ r^2_i \, \dd r_i \ j_{\ell_i}(k_i x) \, j_{\ell_i}(k_i r_i) \frac{(-1)^{m_c}}{r_i^2 \, k_i^{2+m_c}} \nonumber \\
& \quad \times \frac{\partial^{c}}{\partial k_i^{c-m_c} \, \partial r_i^{m_c}} \left[ n_{t_i} \, b_1^{t_i} \, G(r_i) \, P(k_i)/ k_i^\alpha \ k_i^{c+2} \, r_i^{m_c+2}\right] \\
&= \frac{(-1)^c}{x^c} \frac{2}{\pi}\int k^2_i \, \dd k_i \ r^2_i \, \dd r_i \ n_{t_i} \, b_1^{t_i} \, G(r_i) \, P(k_i) / k_i^\alpha \ L^{t_i,(c)}_{\alpha}(k_i,r_i) \ j_{\ell_i}(k_i x) \, j_{\ell_i}(k_i r_i)
\ea
where
\ba\label{App:Eq:def-L^t,c_alpha}
L^{t_i,(c)}_{\alpha}(k_i,r_i) &= \frac{1}{n_{t_i} \, b_1^{t_i} \, G(r_i) \, P(k_i) / k_i^\alpha} \sum_{m_c=0}^c \!\begin{pmatrix}c \\ m_c\end{pmatrix}\! \frac{(-1)^{m_c}}{r_i^2 \, k_i^{2+m_c}} \nonumber \\
& \quad \times \frac{\partial^{c}}{\partial k_i^{c-m_c} \, \partial r_i^{m_c}} \left[ n_{t_i} \, b_1^{t_i} \, G(r_i) \, P(k_i)/ k_i^\alpha \ k_i^{c+2} \, r_i^{m_c+2}\right] \\
&= \frac{k_i^\alpha}{p^{t_i}(k_i,r_i)} \sum_{m_c=0}^c \!\begin{pmatrix}c \\ m_c\end{pmatrix}\! \frac{(-1)^{m_c}}{r_i^2 \, k_i^{2+m_c}} \frac{\partial^{c}}{\partial k_i^{c-m_c} \, \partial r_i^{m_c}} \left[ \frac{p^{t_i}(k_i,r_i)}{k_i^\alpha} \ k_i^{c+2} \, r_i^{m_c+2}\right]
\ea
with
\ba
p^t(k,r) &= n_{t}(r) \, b_1^{t}(k,r) \, G(r) \, P(k)
\ea
This can be put in a form similar to Eq.~\ref{App:Eq:proj-LSS-Limber-integral}
\ba
P &= \frac{(-1)^c}{(\ell_i+\sfrac{1}{2})^\alpha} \ p^t(k_{\ell_i},x) \ \mathscr{L}^{t_i,(c)}_{\alpha}(\ell_i,x) \times \frac{1}{x^{sth}}
\ea
with
\ba\label{App:Eq:def-scrL^t,c_alpha}
\mathscr{L}^{t_i,(c)}_{\alpha}(\ell_i,x) &= \frac{2/\pi}{p^t(k_{\ell_i},x) / k_{\ell_i}^\alpha} \int k^2_i \, \dd k_i \ r^2_i \, \dd r_i \ \frac{p^t(k_{i},r_i)}{k_i^\alpha} \ L^{t_i,(c)}_{\alpha}(k_i,r_i) \ j_{\ell_i}(k_i x) \, j_{\ell_i}(k_i r_i)
\ea
Grouping the projections of $k_1, k_2$ and $k_3$, the bispectrum takes a form similar to Eq.~\ref{App:Eq:proj-LSS-Limber-bispectrum}
\ba
\nonumber b_{\lu,\ld,\lt}^{n,A} = 
\frac{(-1)^{c+d} \, B}{(\lu+\sfrac{1}{2})^\alpha \, (\ld+\sfrac{1}{2})^\beta} \int & x^2 \, \dd x \ n_\mr{t_1} \, n_\mr{t_2} \, n_\mr{t_3} \ b^\mr{t_1}_1 \, b^\mr{t_2}_1 \, b^\mr{t_3}_{a} \ P(k_1|x) \, P(k_2|x) \\
& \times \mathscr{L}^{t_1,(c)}_{\alpha}(\lu,x) \, \mathscr{L}^{t_2,(d)}_{\beta}(\ld,x) \, \mathscr{G}^{t_3}_{a}(\ell_3,x) \label{App:Eq:proj-LSS-noLimber-bispectrum}
\ea
where $\alpha = \xi -\mu$ and $\beta = \rho-\nu$ are the respective exponents of $k_1$ and $k_2$.

\subsection{Correspondence Limber / no Limber}\label{App:proj-LSS-LimberCorrespondence}

The Limber result Eq.~\ref{App:Eq:proj-LSS-Limber-bispectrum} and the no Limber result Eq.~\ref{App:Eq:proj-LSS-noLimber-bispectrum} have the same functional form. In this subsection we show that \ref{App:Eq:proj-LSS-Limber-bispectrum} is indeed an approximation of \ref{App:Eq:proj-LSS-noLimber-bispectrum}. For this, it suffices to show
\ba
\mathscr{G}^{t_3}_{a}(\ell_3,x) &\xrightarrow[\mr{Limber}]{} 1 \label{App:Eq:proj-LSS-compLimber-cond1}, \\
\mathscr{L}^{t_1,(c)}_{\alpha}(\lu,x) &\xrightarrow[\mr{Limber}]{} \mathcal{L}^{t_1,(c)}_{\alpha}(\lu,x) \label{App:Eq:proj-LSS-compLimber-cond2}, \\
\mathscr{L}^{t_2,(d)}_{\beta}(\ld,x) &\xrightarrow[\mr{Limber}]{} \mathcal{L}^{t_2,(d)}_{\beta}(\ld,x) \label{App:Eq:proj-LSS-compLimber-cond3},
\ea
when using the Limber approximation Eq.~\ref{Eq:Limber}
\be\label{App:Eq:Limber}
\frac{2}{\pi}\int k^2 \, \dd k \ r^2 \, \dd r \ f(k,r) \ j_{\ell_i}(k r) \, j_{\ell_i}(k x) \xrightarrow[\mr{Limber}]{} f(k_\ell,x).
\ee
For the integral on $k_3$, from its definition Eq.~\ref{App:Eq:def-scrG^t_a} we easily see that $\mathscr{G}^{t_3}_{a}=1$, i.e. the condition Eq.~\ref{App:Eq:proj-LSS-compLimber-cond1} is fulfilled.\\
For the integrals on $k_1, k_2$, starting from the definition Eq.~\ref{App:Eq:def-scrL^t,c_alpha}, we get
\ba
\mathscr{L}^{t,(c)}_{\alpha}(\ell,x) &\xrightarrow[\mr{Limber}]{} L^{t,(c)}_{\alpha}(k_\ell,x)
\ea
Then, to fulfill the conditions Eqs.~\ref{App:Eq:proj-LSS-compLimber-cond2}--\ref{App:Eq:proj-LSS-compLimber-cond3}, it suffices to show that for $c=1,2$ we have
\ba
L^{t,(c)}_{\alpha}(k_\ell,x) = \mathcal{L}^{t,(c)}_{\alpha}(\ell,x)
\ea
This derivation is a bit involved, particularly for $c=2$, but is a straightforward exercise of partial derivatives. In fact it was already noted that it checks out in a similar situation at the end of Appendix~\ref{App:naive-Limber-power-spectrum}. This concludes the proof and establishes the robustness of Eqs.~\ref{App:Eq:proj-LSS-Limber-bispectrum} and \ref{App:Eq:proj-LSS-noLimber-bispectrum}.

Another remark is that the approximation $\mathscr{G}^{t_3}_{a}(\ell_3,x) \simeq 1$ should be an excellent one at all $\lt$. Indeed, the Limber approximation \ref{App:Eq:Limber} is \emph{exact} when $f(k,r)$ is constant as a function of $k$, due to the closure relation of spherical Bessel functions Eq.~\ref{Eq:closure-Bessel}
\be
\int k^2 \, \dd k \ r^2 \ j_{\ell_i}(k r) \, j_{\ell_i}(k x) = \frac{\pi}{2 r^2} \ \delta_D(r-x).
\ee
In the $k_1$ and $k_2$ integrals the $k$ dependence is dictated by the matter power spectrum $P(k)$ so the situation is analogous to the projection of the angular power spectrum $C_\ell$, for which it is well known that Limber's approximation works well at high multipoles but fails at $\ell = \mathcal{O}(1-10)$. By contrast, in the $k_3$ integral, the only $k$ dependence is that from the bias $b_a(k_3,r_3)$, which is fairly weak at low $k$. For instance in the halo model  $b_a(k,r) \rightarrow b_a(r)$ when $k \ll 1/R_h$ with $R_h$ the typical halo radius. Hence at low $\lt$ Limber's approximation will work excellently for this integral, as the integrand will be constant with $k$, and at medium/high $\lt$ it will also work excellently, as it always does. We conclude that the approximation $\mathscr{G}^{t_3}_{a}(\ell_3,x) \simeq 1$ should be excellent for any $\lt$.

\subsection{Mixed Limber}\label{App:proj-LSS-mixedLimber}
In this subsection we derive the bispectrum in a mixed case where Limber's approximation is applied on the $k_2$ and $k_3$ integrals while the $k_1$ integral is kept exact. This is for interest for squeezed bispectrum configuration where $\lu = \mathcal{O}(0-10)$ while $\lu, \lt \gg 1$. These results are used in the main text in the covariance section (Sect.~\ref{Sect:application-covariance}).

We first need a foreword about the multipole permutations (the ``+2 perm.'' term). In appendices \ref{App:proj-LSS-Limber} and \ref{App:proj-LSS-noLimber} we did not need to explicit this term because we either applied the Limber approximation or we did not apply it on all multipoles simultaneously. In the mixed case we break the invariance under permutation so we need to be more careful.\\
From Eq.~\ref{App:Eq:proj-LSS-noLimber-bispectrum} the three permutations are schematically of the form
\ba
(12) = P(k_1) \, P(k_2) \ \mathscr{L}^{t_1,(c)}_{\alpha}(\ell_1,x) \, \mathscr{L}^{t_2,(d)}_{\beta}(\ell_2,x) \, \mathscr{G}^{t_3}_{a}(\ell_3,x) \\
(13) = P(k_1) \, P(k_3) \ \mathscr{L}^{t_1,(c)}_{\alpha}(\ell_1,x) \, \mathscr{G}^{t_2}_{a}(\ell_2,x) \, \mathscr{L}^{t_3,(d)}_{\beta}(\ell_3,x) \\
(23) = P(k_2) \, P(k_3) \ \mathscr{G}^{t_1}_{a}(\ell_1,x) \, \mathscr{L}^{t_2,(c)}_{\alpha}(\ell_2,x) \, \mathscr{L}^{t_3,(d)}_{\beta}(\ell_3,x)
\ea
In all these terms, as argued at the end of Appendix~\ref{App:proj-LSS-LimberCorrespondence}, we can take $\mathscr{G}^{t}_{a} = 1$.\\
The first two permutations contain the power spectrum $P(k_1)$ where $k_1$ is a long wavelength mode (since it projects into a low multipole). For them, we have to use the exact result at $\lu$, $\mathscr{L}^{t_1,(c)}_{\alpha}$, and use the Limber result at $\ld, \lt$, $\mathcal{L}^{t,(c)}_{\alpha}$. The last permutation only contains power spectra of short wavelength modes. For this permutation, since we argued that $\mathscr{G}^{t}_{a}(\lu) = 1$, the contribution is actually the same as in the full Limber case. In total we get
\ba
\nonumber b_{\lu,\ld,\lt}^{n,(12)} =
\frac{(-1)^{c+d} \, B}{(\lu+\sfrac{1}{2})^\alpha \, (\ld+\sfrac{1}{2})^\beta} \int & x^2 \, \dd x \ n_\mr{t_1} \, n_\mr{t_2} \, n_\mr{t_3} \ b^\mr{t_1}_1 \, b^\mr{t_2}_1 \, b^\mr{t_3}_{a} \ P(k_1|x) \, P(k_2|x) \label{App:Eq:proj-LSS-mixedLimber-(12)-bispectrum} \\
& \times \mathscr{L}^{t_1,(c)}_{\alpha}(\lu,x) \, \mathcal{L}^{t_2,(d)}_{\beta}(\ld,x) \\
\nonumber b_{\lu,\ld,\lt}^{n,(13)} =
\frac{(-1)^{c+d} \, B}{(\lu+\sfrac{1}{2})^\alpha \, (\lt+\sfrac{1}{2})^\beta} \int & x^2 \, \dd x \ n_\mr{t_1} \, n_\mr{t_2} \, n_\mr{t_3} \ b^\mr{t_1}_1 \, b^\mr{t_2}_a \, b^\mr{t_3}_{1} \ P(k_1|x) \, P(k_3|x) \\
& \times \mathscr{L}^{t_1,(c)}_{\alpha}(\lu,x) \, \mathcal{L}^{t_3,(d)}_{\beta}(\lt,x) \label{App:Eq:proj-LSS-mixedLimber-(13)-bispectrum} \\
\nonumber b_{\lu,\ld,\lt}^{n,(23)} = \frac{(-1)^{c+d} \, B}{(\ld+\sfrac{1}{2})^\alpha \, (\lt+\sfrac{1}{2})^\beta} \int & x^2 \, \dd x \ n_\mr{t_1} \, n_\mr{t_2} \, n_\mr{t_3} \ b^\mr{t_1}_a \, b^\mr{t_2}_1 \, b^\mr{t_3}_1 \ P(k_2|x) \, P(k_3|x) \nonumber \\
& \times \mathcal{L}^{t_2,(c)}_{\alpha}(\ld,x) \mathcal{L}^{t_3,(d)}_{\beta}(\ld,x) \label{App:Eq:proj-LSS-mixedLimber-(23)-bispectrum}
\ea

\section{Properties of L symbols}\label{App:props-L-symbols}
Here we list useful properties of $L^{t,(c)}_{\alpha}$, $\mathcal{L}^{t,(c)}_{\alpha}$ and $\mathscr{L}^{t,(c)}_{\alpha}$.

\subsection{$L^{t,(c)}_{\alpha}$ and $\mathcal{L}^{t,(c)}_{\alpha}$}\label{App:props-L-and-calL}
$L^{t,(c)}_{\alpha}$ and $\mathcal{L}^{t,(c)}_{\alpha}$ are related by
\ba
L^{t,(c)}_{\alpha}(k_\ell,x) = \mathcal{L}^{t,(c)}_{\alpha}(\ell,x)
\ea
All the properties below are stated for $\mathcal{L}$, since it is the main symbol of interest in the main text, but they equally apply to $L$ due to the relation above.

The quantity $\mathcal{L}^{t,(c)}_{\alpha}$ is defined by
\be
\mathcal{L}^{t,(c)}_{\alpha}(\ell,x) = \left[\frac{k^c}{\frac{p^t(k,r_\ell)}{k^\alpha} k^{c-1}}\frac{\dd^c}{\dd k^c} \left[ \frac{p^t(k,r_\ell)}{k^\alpha} k^{c-1}\right]  \right]_{k=(\ell+\sfrac{1}{2})/x}
\ee
with
\ba
p^t(k,r) &= n_{t}(r) \, b_1^{t}(k,r) \, G(r) \, P(k)
\ea
It has the following algebraic properties:
\ba
\mathcal{L}^{t,(0)}_{\alpha} &= 1 \label{App:Eq:prop-L^t,c_alpha-0} \\
\mathcal{L}^{t,(1)}_{\alpha} &= \mathcal{L}^{t,(1)}_{0} - \alpha \label{App:Eq:prop-L^t,c_alpha-1} \\
\mathcal{L}^{t,(2)}_{\alpha} &= \mathcal{L}^{t,(2)}_{0} - 2 \;\! \alpha \;\! \mathcal{L}^{t,(1)}_{0} + \alpha(\alpha+1) \label{App:Eq:prop-L^t,c_alpha-2}
\ea
meaning that one can restrict the numerical computation to $\alpha=0$.\\
The first order ($c=1$) can be formally written as a log derivative:
\ba
\mathcal{L}^{t,(1)}_{\alpha}(\ell,x) &= \left.\frac{\dd \ln \left[\frac{p^t(k,r_\ell)}{k^\alpha} \right]}{\dd \ln k}\right|_{k=(\ell+\sfrac{1}{2})/x} \\
&= \left.\frac{\dd \ln \left[n_{t}(r_\ell) \, b_1^{t}(k,r_\ell) \, G(r_\ell) \, \frac{P(k)}{k^\alpha} \right]}{\dd \ln k}\right|_{k=(\ell+\sfrac{1}{2})/x}
\ea
For numerical implementation and physical interpretation however, explicit expressions are needed in terms of $k$ or $r$ derivatives.\\
At first order ($c=1$) we have
\ba
\mathcal{L}^{t,(1)}_{\alpha}(\ell,x) &= \left.\frac{\dd \ln \left[\frac{p^t(k,x)}{k^\alpha} \right]}{\dd \ln k}\right|_{k=k_\ell} - \left.\frac{\dd \ln \left[\frac{p^t(k,r)}{k^\alpha}\right]}{\dd \ln r}\right|_{r=x} \\
&= \left.\frac{\dd \ln \left[b_1^{t}(k,x) \, \frac{P(k)}{k^\alpha} \right]}{\dd \ln k}\right|_{k=k_\ell} - \left.\frac{\dd \ln \left[n_{t}(r) \, b_1^{t}(k_\ell,r) \, G(r)\right]}{\dd \ln r}\right|_{r=x} \\
& \equiv g_\mr{dil,\alpha}^{t,(1)}(k_\ell,x) - g_\mr{evo}^{t,(1)}(k_\ell,x)
\ea
with
\ba
g_\mr{dil,\alpha}^{t,(c)}(k,r) &= \frac{k^c}{\frac{p^t(k,r)}{k^\alpha} k^{c-1}}\frac{\partial^c \left[\frac{p^t(k,r)}{k^\alpha} k^{c-1}\right]}{\partial k^c} =  \frac{k^c}{b_1^t(k,r) \, \frac{P(k)}{k^\alpha} k^{c-1}}\frac{\partial^c \left[b_1^t(k,r) \, \frac{P(k)}{k^\alpha} k^{c-1}\right]}{\partial k^c} \\
g_\mr{evo}^{t,(c)}(k,r) &= \frac{r^c}{\frac{p^t(k,r)}{k^\alpha}} \frac{\partial^c \left[\frac{p^t(k,r)}{k^\alpha}\right]}{\partial r^c} = \frac{r^c}{n_t(r) \, b_1^t(k,r) \, G(r)} \frac{\partial^c \left[n_t(r) \, b_1^t(k,r) \, G(r)\right]}{\partial r^c}
\ea
The dilation and evolution terms respectively.\\
At second order ($c=2$), some algebra gives:
\ba
\mathcal{L}^{t,(2)}_{\alpha}(\ell,x) & = g_\mr{dil,\alpha}^{t,(2)}(k_\ell,x) - 2 g_\mr{dil-evo,\alpha}^{t,(2)}(k_\ell,x) + g_\mr{evo}^{t,(2)}(k_\ell,x)
\ea
where we introduced the mixed term
\ba
g_\mr{dil-evo,\alpha}^{t,(2)}(k,r) &= \frac{k r}{\frac{p^t(k,r)}{k^\alpha}} \frac{\partial^2 \left[\frac{p^t(k,r)}{k^\alpha} \right]}{\partial k \ \partial r} = \frac{k r}{n_t(r) \, b_1^t(k,r) \, G(r) \frac{P(k)}{k^\alpha}} \frac{\partial^2 \left[n_t(r) \, b_1^t(k,r) \, G(r) \frac{P(k)}{k^\alpha} \right]}{\partial k \ \partial r}
\ea
which simplifies if $b_1^t(k,r)$ depends only on $k$ or only on $r$ (the latter being a good approximation on large scales):
\be
g_\mr{dil-evo,\alpha}^{t,(2)}(k,r) = g_\mr{dil,\alpha}^{t,(1)}(k,r) \times g_\mr{evo}^{t,(1)}(k,r)
\ee

\subsection{$\mathscr{L}^{t,(c)}_{\alpha}$}\label{App:props-scrL}
We have the definition
\ba
\mathscr{L}^{t,(c)}_{\alpha}(\ell,x) &= \frac{2/\pi}{p^{t}(k_\ell,x) / k_{\ell}^\alpha} \int k^2 \, \dd k \ r^2 \, \dd r \ \frac{p^t(k,r)}{k^\alpha} \ L^{t,(c)}_{\alpha}(k,r) \ j_{\ell}(k x) \, j_{\ell}(k r)
\ea
with
\ba
p^t(k,r) &= n_{t}(r) \, b_1^{t}(k,r) \, G(r) \, P(k)
\ea
We cannot transpose the first properties of $L$ and $\mathcal{L}$ --Eqs~\ref{App:Eq:prop-L^t,c_alpha-0}, \ref{App:Eq:prop-L^t,c_alpha-1} and \ref{App:Eq:prop-L^t,c_alpha-2}-- to $\mathscr{L}$ because the integral does not simplify with the denominator. In particular
\ba
\mathscr{L}^{t,(0)}_{\alpha} &\neq 1 \\
\mathscr{L}^{t,(1)}_{\alpha} &\neq \mathscr{L}^{t,(1)}_{0} - \alpha \\
\mathscr{L}^{t,(2)}_{\alpha} &\neq \mathscr{L}^{t,(2)}_{0} - 2 \;\! \alpha \;\! \mathscr{L}^{t,(1)}_{0} + \alpha(\alpha+1).
\ea
Although these equations asymptotically become (approximate) equalities at high multipole, when Limber's approximation performs best and $\mathscr{L}^{t,(c)}_{\alpha} \approx \mathcal{L}^{t,(c)}_{\alpha}$.

There are however rewritings of $\mathscr{L}^{t,(c)}_{\alpha}$ which are useful for physical interpretation. Indeed for $c=0$ we have 
\ba
\mathscr{L}^{t,(0)}_{\alpha} = \frac{4\pi}{p^{t}(k_\ell,x) / k_{\ell}^\alpha} \int r^2 \, \dd r \ n_{t} \, b_1^{t} \, G \ \sigma^2_{\ell,\alpha}(x,r)
\ea
where
\ba
\sigma^2_{\ell,\alpha}(x,r) = \frac{1}{2\pi^2} \int k^2 \, \dd k \ P(k)/k^\alpha \ j_\ell(kx) \, j_\ell(kr)
\ea
and we neglected the scale dependence of the bias.\\
For $c=1$, a calculation with integration by part gives
\ba
\mathscr{L}^{t,(1)}_{\alpha} = \frac{(-1) \times 4\pi}{p^{t}(k_\ell,x) / k_{\ell}^\alpha} \int r^2 \, \dd r \ n_{t} \, b_1^{t}\, G \ \sigma^2_{\ell,\alpha,(1)}(x,r)
\ea
where
\ba
\sigma^2_{\ell,\alpha,(1)}(x,r) &= \frac{1}{2\pi^2} \int k^2 \, \dd k \, P(k)/k^\alpha \ \underbrace{k x \, j'_\ell(kx)}_{=x \, \partial_x \left[j_\ell(kx)\right]} \, j_\ell(kr) = x \ \partial_x \left[\sigma^2_{\ell,\alpha}(x,r)\right]
\ea
Finally for $c=2$ a longer calculation gives
\ba
\mathscr{L}^{t,(2)}_{\alpha} = \frac{4\pi}{p^{t}(k_\ell,x) / k_{\ell}^\alpha} \int r^2 \, \dd r \ n_{t} \, b_1^{t} \, G \ \sigma^2_{\ell,\alpha,(2)}(x,r)
\ea
where
\ba
\sigma^2_{\ell,\alpha,(2)}(x,r) &= \frac{1}{2\pi^2} \int k^2 \, \dd k \, P(k)/k^\alpha \ \underbrace{(k x)^2 \, j''_\ell(kx)}_{=x^2 \partial^2_x \left[j_\ell(kx)\right]} \, j_\ell(kr) = x^2 \ \partial^2_x \left[\sigma^2_{\ell,\alpha}(x,r)\right]
\ea
These three equations ($c=0,1,2$) can be summarised as
\ba
\frac{p^{t}(k_\ell,x)}{k_{\ell}^\alpha} \ \mathscr{L}^{t,(c)}_{\alpha}(\ell,x) = 4\pi \int r^2 \, \dd r \ n_{t}(r) \, b_1^{t}(r) \, G(r) \ (-x)^c \, \partial^c_x \left[\sigma^2_{\ell,\alpha}(x,r)\right]
\ea

\section{Alternative basis for Super-Sample Covariance}\label{App:alt-base-SSC}

For comparison with literature, we develop here the SSC on the basis $(\sigma^2,\kappa,\tau)$ instead of the basis $(\sigma^2,\upsilon,\tau)$ used in the main text.\\
To this end, we start from the total SSC equation Eq.\ref{Eq:Cov_SSC_total}, the responses definitions Eqs~\ref{Eq:dP/ddeltab}, \ref{Eq:dP/dxb} and \ref{Eq:dP/dyb}
We find 
\ba
\Cov_\mr{SSC} &= \int r^2 \, \dd r \ x^2 \, \dd x \ \frac{\partial n_\mr{t_1}(r)}{\partial \delta_b} \ \frac{\partial P_{n_{t_2},n_{t_3}}(k_{\ell}|x)}{\partial \delta_b} \ \sigma^2_{\delta_b,\delta_b}(r,x) \nonumber \\
&+ \int r^2 \, \dd r \ x^2 \, \dd x \ \frac{\partial n_\mr{t_1}(r)}{\partial \delta_b} \ \frac{\partial P_{n_{t_2},n_{t_3}}(k_{\ell}|x)}{\partial \kappa_b} \ \kappa(r,x) \nonumber \\
&+ \int r^2 \, \dd r \ x^2 \, \dd x \ \frac{\partial n_\mr{t_1}(r)}{\partial \delta_b} \ \frac{\partial P_{n_{t_2},n_{t_3}}(k_{\ell}|x)}{\partial y_b} \ \tau(r,x)
\ea
with the responses (for clarity we temporarily drop the $n_{t_2},n_{t_3}$ subscript of $P$):
\ba
\frac{\partial P}{\partial \delta_b} &\equiv \partial_0 P - 2 \, \partial_2^B P - \frac{1}{2} \partial_1^A P + \partial_2^A P \\
\frac{\partial P}{\partial \kappa_b} &\equiv - \frac{1}{2} \partial_1^A P + \partial_2^A P \\
\frac{\partial P}{\partial y_b} &\equiv \partial_1^B P
\ea
where the terms $\partial_n^{A/B} P$ are defined in the main text in Eqs~\ref{Eq:d0P-shortform}, \ref{Eq:d1AP}, \ref{Eq:d1BP}, \ref{Eq:d2AP-shortform} and \ref{Eq:d2BP-shortform}.\\
We find the explicit expressions:
\ba
\frac{\partial P_{n_{t_2},n_{t_3}}}{\partial \delta_b} &= n_\mr{t_2} \, n_\mr{t_3} \ P(k_{\ell}|x) \times \left( 2 b^\mr{t_2}_1 \, b^\mr{t_3}_{(0)}
-\frac{1}{2} b^\mr{t_2}_1 \, b^\mr{t_3}_{1} \, \mathcal{L}^{t_2,(1)}_{0}
+ b^\mr{t_2}_1 \, b^\mr{t_3}_{(2)} \left(1 - \frac{\mathcal{L}^{t_2,(2)}_{2}}{(\ell+\sfrac{1}{2})^2} \right)
+ (2\leftrightarrow 3) \right) \\
&= n_\mr{t_2} \, n_\mr{t_3} \ P(k_{\ell}|x) \times \Bigg[
b^\mr{t_2}_1 \left(\frac{12}{7}b^\mr{t_3}_1 + b^\mr{t_3}_2 + \frac{1}{3} b^\mr{t_3}_{s^2}\right)
+ \left(\frac{12}{7}b^\mr{t_2}_1 + b^\mr{t_2}_2 + \frac{1}{3} b^\mr{t_2}_{s^2}\right) b^\mr{t_3}_1 \nonumber \\
& - \frac{1}{2} b^\mr{t_2}_1 \, b^\mr{t_2}_1 \left( \mathcal{L}^{t_2,(1)}_{0} + \mathcal{L}^{t_3,(1)}_{0} \right)
- \frac{ b^\mr{t_2}_1 \left(\frac{2}{7}b^\mr{t_3}_1 + b^\mr{t_3}_{s^2}\right) \mathcal{L}^{t_2,(2)}_{2} + \left(\frac{2}{7}b^\mr{t_2}_1 + b^\mr{t_2}_{s^2}\right) b^\mr{t_3}_1 \mathcal{L}^{t_3,(2)}_{2} }{(\ell+\sfrac{1}{2})^2} 
\Bigg] \\
\frac{\partial P_{n_{t_2},n_{t_3}}}{\partial \kappa_b} &= 
n_\mr{t_2} \, n_\mr{t_3} \ P(k_{\ell}|x) \times \left( -\frac{1}{2} b^\mr{t_2}_1 \, b^\mr{t_3}_{1} \, \mathcal{L}^{t_2,(1)}_{0} + b^\mr{t_2}_1 \, b^\mr{t_3}_{(2)} \left( 1 + \frac{3\mathcal{L}^{t_2,(2)}_{2}}{(\ell+\sfrac{1}{2})^2}\right)
+ (2\leftrightarrow 3) \right) \\
&= n_\mr{t_2} \, n_\mr{t_3} \ P(k_{\ell}|x) \times \Bigg[ b^\mr{t_2}_1 \, b^\mr{t_3}_{1} \left(\frac{4}{7} -\frac{\mathcal{L}^{t_2,(1)}_{0}+\mathcal{L}^{t_3,(1)}_{0}}{2} \right) + b^\mr{t_2}_1 \, b^\mr{t_3}_{s^2} + b^\mr{t_2}_{s^2} \, b^\mr{t_3}_{1} \nonumber \\
& + \frac{3}{(\ell+\sfrac{1}{2})^2} \left( b^\mr{t_2}_1 \left(\frac{2}{7} b^\mr{t_3}_{1} + b^\mr{t_3}_{s^2} \right) \mathcal{L}^{t_2,(2)}_{2} + \left(\frac{2}{7} b^\mr{t_2}_{1} + b^\mr{t_2}_{s^2} \right) b^\mr{t_3}_1 \mathcal{L}^{t_3,(2)}_{2} \right) \Bigg] \\
\frac{\partial P_{n_{t_2},n_{t_3}}}{\partial y_b} 
=& \ \frac{n_\mr{t_2} \, n_\mr{t_3} \ P(k_{\ell}|x)}{(\ell+\sfrac{1}{2})^2} \times  b^\mr{t_2}_1 \, b^\mr{t_3}_{1} \left(\mathcal{L}^{t_2,(1)}_{2} + \mathcal{L}^{t_3,(1)}_{2} \right).
\ea
Equivalently, the logarithmic responses are
\ba
\frac{\partial \ln P_{n_{t_2},n_{t_3}}}{\partial \delta_b} 
=&  \frac{24}{7} + \frac{b^\mr{t_2}_2 + \frac{1}{3} b^\mr{t_2}_{s^2}}{b^\mr{t_2}_1}
+ \frac{b^\mr{t_3}_2 + \frac{1}{3} b^\mr{t_3}_{s^2}}{b^\mr{t_3}_1} 
- \frac{1}{2} \left( \mathcal{L}^{t_2,(1)}_{0} + \mathcal{L}^{t_3,(1)}_{0} \right) \nonumber \\
& - \frac{1}{(\ell+\sfrac{1}{2})^2} \left( \frac{\frac{2}{7} b^\mr{t_2}_1 + b^\mr{t_2}_{s^2}}{b^\mr{t_2}_1} \mathcal{L}^{t_3,(2)}_{2} + \frac{\frac{2}{7} b^\mr{t_3}_1 + b^\mr{t_3}_{s^2}}{b^\mr{t_3}_1} \mathcal{L}^{t_2,(2)}_{2} \right) \label{App:Eq:dlnP/ddeltab-altbasis} \\
\frac{\partial \ln P_{n_{t_2},n_{t_3}}}{\partial \kappa_b} 
=&  \frac{4}{7} -\frac{\mathcal{L}^{t_2,(1)}_{0}+\mathcal{L}^{t_3,(1)}_{0}}{2} +  \frac{b^\mr{t_2}_{s^2}}{b^\mr{t_2}_{1}} + \frac{b^\mr{t_3}_{s^2}}{b^\mr{t_3}_{1}} \nonumber \\
& + \frac{3}{(\ell+\sfrac{1}{2})^2} \left[ \left(\frac{2}{7} + \frac{b^\mr{t_3}_{s^2}}{b^\mr{t_3}_{1}} \right) \mathcal{L}^{t_2,(2)}_{2} + \left(\frac{2}{7} + \frac{b^\mr{t_2}_{s^2}}{b^\mr{t_2}_{1}} \right) \mathcal{L}^{t_3,(2)}_{2} \right] \\
\frac{\partial \ln P_{n_{t_2},n_{t_3}}}{\partial y_b} 
=& \ \frac{\mathcal{L}^{t_2,(1)}_{2} + \mathcal{L}^{t_3,(1)}_{2}}{(\ell+\sfrac{1}{2})^2}.
\ea
We will use this for comparison with past 3D literature in Section~\ref{Sect:discussion}.

\section{A decomposition in Scalar-Vector-Tensor spectra}\label{App:SVT-decomposition}
In this appendix, we write the reduced bispectrum making explicit the scalar-vector-tensor decomposition on the celestial sphere. It shows that it can expressed as a sum where each term is given by the product of two angular power spectra cross-correlating the density field with its gradients or velocity. In the context of the covariance of the angular power spectrum with number counts, only the scalar part of the bispectrum contributes to the SSC while the full scalar-vector-tensor terms contribute to the ISC.

\subsection{Decomposition definition}\label{App:SVT-decomposition-definition}

The 3D bispectrum which is considered here allows us to write the reduced bispectrum in a separable manner. To this end, we decompose it over its scalar, vector and tensor contributions as follows
\ba 
    b_{\ell_1,\ell_2,\ell_3} = & \ds\int x^2\dd x\sum_{\mr{A=S,V,T}}b^{(\mr{A})}_{\ell_1,\ell_2,\ell_3}(x).
\ea
First, the scalar contribution reads
\begin{eqnarray}
    b^{(\mr{S})}_{\ell_1,\ell_2,\ell_3}(x) &=& \Bigg\{S^{(0),\mr{t_1}}_{\ell_1}(x)\,S^{(0),\mr{t_2}}_{\ell_2}(x)\,O^{(0),\mr{t_3}}_{\ell_3}(x) \nonumber \\
    && + \frac{1}{2} \left[ S^{(1,-2),\mr{t_1}}_{\ell_1}(x) \ S^{(1,0),\mr{t_2}}_{\ell_2}(x) + S^{(1,0),\mr{t_1}}_{\ell_1}(x) \ S^{(1,-2),\mr{t_2}}_{\ell_2}(x) \right] \, O^{(1),\mr{t_3}}_{\ell_3}(x) \nonumber \\
    &&+\left[ S^{(2,0),\mr{t_1}}_{\ell_1}(x)\,S^{(2,0),\mr{t_2}}_{\ell_2}(x) + \frac{1}{2} S^{(2,1),\mr{t_1}}_{\ell_1}(x)\,S^{(2,1),\mr{t_2}}_{\ell_2}(x) \right]\,O^{(2),\mr{t_3}}_{\ell_3}(x)\Bigg\}+2 \ \mr{perm.}
\end{eqnarray}
where, for the first two tracers $\mr{t_1}$ and $\mr{t_2}$, we have
\begin{eqnarray}
    S^{(0),\mr{t_i}}_{\ell_i}(x)&=&\frac{2}{\pi}\ds\int\left(r_i^2\dd r_i\right)\left(k_i^2\dd k_i\right)n_\mr{t_i}b_1^\mr{t_i}G(r_i)P(k_i)\,j_{\ell_i}(k_ir_i)j_{\ell_i}(k_ix), \label{eq:s0} \\
    S^{(1,a),\mr{t_i}}_{\ell_i}(x)&=&\frac{2}{\pi}\ds\int\left(r_i^2\dd r_i\right)\left(k_i^2\dd k_i\right)n_\mr{t_i}b_1^\mr{t_i}G(r_i)\left[P(k_i) \, k_i^{a}\right]\,j_{\ell_i}(k_ir_i)\left[{\partial_x j_{\ell_i}(k_ix)}\right], \label{eq:s1a} \\
    S^{(2,0),\mr{t_i}}_{\ell_i}(x)&=&\frac{2}{\pi}\ds\int\left(r_i^2\dd r_i\right)\left(k_i^2\dd k_i\right)n_\mr{t_i}b_1^\mr{t_i}G(r_i)\left[\frac{P(k_i)}{k_i^2}\right]\,j_{\ell_i}(k_ir_i)\left[\partial^2_x j_{\ell_i}(k_ix)\right] \label{eq:s20}\\
    S^{(2,1),\mr{t_i}}_{\ell_i}(x)&=&\frac{2}{\pi}\ds\int\left(r_i^2\dd r_i\right)\left(k_i^2\dd k_i\right)n_\mr{t_i}b_1^\mr{t_i}G(r_i)P(k_i)\,j_{\ell_i}(k_ir_i)\left[j_{\ell_i}(k_ix)+\frac{1}{k_i^2}\,\partial^2_x j_{\ell_i}(k_ix)\right]. \nonumber \\ \label{eq:s21}
\end{eqnarray}
The extra $k_i^a$ in $S^{(1,a),\mr{t_i}}_{\ell_i}(x)$ results from the factor $\left(\frac{1}{k_1^2}+\frac{1}{k_2^2}\right)$ in the $F_2$ kernel.\\
For the third tracer $\mr{t_3}$, we have 
\ba
    O^{(n),\mathrm{t_3}}_{\ell_3}(x)=\frac{2}{\pi}\ds\int\left(r_3^2\dd r_3\right)\left(k_3^2\dd k_3\right) n_\mr{t_3} \, G^2(r_3) \, b_{(n)}^\mr{t_3} \ j_{\ell_3}(k_3r_3) \, j_{\ell_3}(k_3x), \label{eq:Ot3}
\ea
where we recall $b_{(0)}^\mr{t_3}= \frac{5}{7} b_1^\mr{t_3}+ \frac{1}{2} b_2^\mr{t_3} -\frac{1}{3} b_{s^2}^\mr{t_3}$, $b_{(1)}^\mr{t_3}=b_1^\mr{t_3}$, and $b_{(2)}^\mr{t_3}= \frac{2}{7} b_1^\mr{t_3}+b_{s^2}^\mr{t_3}$.

Second, the vector contribution is given by
\begin{eqnarray}
    b^{(\mr{V})}_{\ell_1,\ell_2,\ell_3}(x) &=& \Lambda^{(\mr{V})}_{\ell_{123}} \Bigg\{ \frac{1}{2}\left[ V^{(1,-2),\mr{t_1}}_{\ell_1}(x) \, V^{(1,0),\mr{t_2}}_{\ell_2}(x) + V^{(1,0),\mr{t_1}}_{\ell_1}(x) \, V^{(1,-2),\mr{t_2}}_{\ell_2}(x) \right] O^{(1),\mr{t_3}}_{\ell_3}(x) \nonumber \\
    &&+2 V^{(2),\mr{t_1}}_{\ell_1}(x)\,V^{(2),\mr{t_2}}_{\ell_2}(x)\,O^{(2),\mr{t_3}}_{\ell_3}(x)\Bigg\}+2 \ \mr{perm.}
\end{eqnarray}
where 
\begin{eqnarray}
    V^{(1,a),\mr{t_i}}_{\ell_i}(x)&=&\frac{2}{\pi}\ds\int\left(r_i^2\dd r_i\right)\left(k_i^2\dd k_i\right) n_\mr{t_i}b_1^\mr{t_i}G(r_i) \left[ P(k_i) \, k_i^{a}\right] \, j_{\ell_i}(k_ir_i)\left[\frac{j_{\ell_i}(k_ix)}{x}\right], \label{eq:v1a}\\
    V^{(2),\mr{t_i}}_{\ell_i}(x)&=&\frac{2}{\pi}\ds\int\left(r_i^2\dd r_i\right)\left(k_i^2\dd k_i\right) n_\mr{t_i}b_1^\mr{t_i}G(r_i)\,\left[\frac{P(k_i)}{k_i^2}\right]\,j_{\ell_i}(k_ir_i)\left\{\partial_x\left[\frac{j_{\ell_i}(k_ix)}{x}\right]\right\}. \nonumber \\ \label{eq:v2}
\end{eqnarray}
Here again, we stress the extra $k_i^a$ in $V^{(1,a),\mr{t_i}}_{\ell_i}(x)$.

Third, the tensor contribution is given by
\begin{eqnarray}
    b^{(\mr{T})}_{\ell_1,\ell_2,\ell_3}(x)&=&\Lambda^{(\mr{T})}_{\ell_{123}}\Bigg\{T^{(2),\mr{t_1}}_{\ell_1}(x)\,T^{(2),\mr{t_2}}_{\ell_2}(x)\,O^{(2),\mr{t_3}}_{\ell_3}(x)\Bigg\}+2 \ \mr{perm.}
\end{eqnarray}
where 
\begin{eqnarray}
    T^{(2),\mr{t_i}}_{\ell_i}(x)&=&\frac{2}{\pi}\ds\int\left(r_i^2\dd r_i\right)\left(k_i^2\dd k_i\right) n_\mr{t_i}b_1^\mr{t_i}G(r_i)\,\left[\frac{P(k_i)}{k_i^2}\right]\,j_{\ell_i}(k_ir_i)\left[\frac{j_{\ell_i}(k_ix)}{x^2}\right]. \label{eq:t2}
\end{eqnarray}

\subsection{Relation with $\mathscr{L}$ coefficients}\label{App:rels-Ls-and-S/V/T/O}

From the definition of the $S$/$V$/$T$ coefficients, we can pull the derivatives $\partial_x^n$ outside of the integrals. This allows to find a first set of relations between the $S$/$V$/$T$ coefficients and $\mathscr{L}^{t,(c)}_{\alpha}(\ell,x)$ coefficients\footnote{In the following we drop the arguments $x$ or $(\ell,x)$ for simplicity}:
\ba
S^{(0),\mr{t}}_{\ell}   &= p^t_{\ell} \ \mathscr{L}^{t,(0)}_{0}
\qquad &
S^{(1,a),\mr{t}}_{\ell} &= \partial_x \left[ \frac{x^{a} \ p^t_{\ell} \ \mathscr{L}^{t,(0)}_{a}}{(\ell+\sfrac{1}{2})^{a}} \right] \\
S^{(2,0),\mr{t}}_{\ell} &= \partial^2_x \left[ \frac{x^2 \ p^t_{\ell} \ \mathscr{L}^{t,(0)}_{2}}{(\ell+\sfrac{1}{2})^2} \right]
\qquad &
S^{(2,1),\mr{t}}_{\ell} &= \left[p^t_{\ell,0} \ \mathscr{L}^{t,(0)}_{0} + \partial^2_x \left[ \frac{x^2 \ p^t_{\ell} \ \mathscr{L}^{t,(0)}_{2}}{(\ell+\sfrac{1}{2})^2} \right] \right] \\
V^{(1,a),\mr{t}}_{\ell} &= \frac{p^t_{\ell} \ \mathscr{L}^{t,(0)}_{a}}{(\ell+\sfrac{1}{2})^{a}} \times x^{a-1}
\qquad &
V^{(2),\mr{t}}_{\ell}   &= \partial_x \left[\frac{x \ p^t_{\ell} \ \mathscr{L}^{t,(0)}_{2}}{(\ell+\sfrac{1}{2})^2} \right] \\ 
T^{(2),\mr{t}}_{\ell}   &= \frac{p^t_{\ell} \ \mathscr{L}^{t,(0)}_{2}}{(\ell+\sfrac{1}{2})^2}. \qquad
\ea
where we defined $p^t_{\ell} \equiv p^t(k_\ell,x) = n_{t}(x) \, b_1^{t}(k_{\ell},x) \, G(x) \, P(k_{\ell})$.\\
In particular this allows to easily establish the units of the scalar/vector/tensor coefficients:
\ba
\left[S^{(0)}\right] &= \left[S^{(2,0)}\right] = \left[S^{(2,1)}\right] = \left[V^{(2)}\right] = \left[T^{(2)}\right] = 1 \\
\left[S^{(1,a)}\right] &= \left[V^{(1,a)}\right] = \mr{Mpc}^{a-1}
\ea

A second set of relations between the $S$/$V$/$T$ coefficients and $\mathscr{L}^{t,(c)}_{\alpha}$ can be derived by applying the method of Appendix~\ref{App:proj-LSS-noLimber}. These relations are useful for numerical implementation of the bispectrum computation. They are 
\ba
S^{(0),\mr{t}}_{\ell} &= p^t_{\ell} \ \mathscr{L}^{t,(0)}_{0}
\qquad &
S^{(1,a),\mr{t}}_{\ell} &= - \frac{p^t_{\ell} \ \mathscr{L}^{t,(1)}_{a}}{(\ell+\sfrac{1}{2})^{a}} \times x^{a-1} \\
S^{(2,0),\mr{t}}_{\ell} &= \frac{p^t_{\ell} \ \mathscr{L}^{t,(2)}_{2}}{(\ell+\sfrac{1}{2})^{2}}
\qquad &
S^{(2,1),\mr{t}}_{\ell} &= p^t_{\ell} \left[ \mathscr{L}^{t,(0)}_{0} + \frac{\mathscr{L}^{t,(2)}_{2}}{(\ell+\sfrac{1}{2})^{2}} \right] \\
V^{(1,a),\mr{t}}_{\ell} &= \frac{p^t_{\ell} \ \mathscr{L}^{t,(0)}_{a}}{(\ell+\sfrac{1}{2})^{a}}  \times x^{a-1}
\qquad &
V^{(2),\mr{t}}_{\ell} &= - \frac{p^t_{\ell} \left(\mathscr{L}^{t,(0)}_{2} + \mathscr{L}^{t,(1)}_{2}\right)}{(\ell+\sfrac{1}{2})^{2}}  \\ 
T^{(2),\mr{t}}_{\ell} &= \frac{p^t_{\ell} \ \mathscr{L}^{t,(0)}_{2}}{(\ell+\sfrac{1}{2})^2}. \qquad
\ea

Lastly we consider the $O_\ell$ coefficient, defined by
\ba
O^{(n),\mathrm{t}}_{\ell}(x) = \frac{2}{\pi} \ds\int r^2\dd r \ k^2\dd k \ n_\mr{t} \, b_{(n)}^\mr{t} \, G^2(r) \ j_{\ell}(k r) \, j_{\ell}(k x).
\ea
It is related to $\mathcal{G}_a$ by
\ba
O^{(n),\mathrm{t}}_{\ell}(x) = n_\mr{t}(x) \, b_{(n)}^\mr{t}(k_\ell,x) \, G^2(x) \ \mathcal{G}^t_{(n)}(\ell,x)
\ea

\section{Numerical details: survey specifications and kernel computations}\label{App:numerical-details}
\subsection{Specifications and modelling for a Euclid-like photometric tracer}\label{App:specs-models-phot}

We assume a standard $\Lambda$CDM model with parameters given by \cite{Planck2018-params}, namely $(\omega_\mr{b},\omega_\mr{cdm},n_S,A_S)=(0.022,0.12,0.96,2.035\cdot10^{-9})$. The proportionality factor in the Lemaître \cite{Lemaitre1927} distance-redshift relation is 67 km$\cdot$s$^{-1} \cdot$Mpc$^{-1}$. The background evolution and the matter power spectrum are computed with \texttt{CLASS} \cite{Lesgourgues2011,Blas2011}, specifically the Python wrapper \texttt{classy}.\\
We consider a Euclid-like sample of photometric galaxies, whose redshift distribution is given by
\ba
n(z) = \left(\frac{z}{z_0}\right)^2 \exp\left[-(z/z_0)^{1.5}\right]
\ea
where $z_0=\frac{0.9}{\sqrt{2}}$ following \cite{Euclid2019}. \\
The first order galaxy bias is given by the polynomial expansion 
\ba
b_1^\mr{gal}(z) = b_\mr{gal,0} + b_\mr{gal,1} \ z + b_\mr{gal,2} \ z^2 + b_\mr{gal,3} \ z^3
\ea
with $(b_\mr{gal,0},b_\mr{gal,1},b_\mr{gal,2},b_\mr{gal,3})=(1.33291, -0.72414, 1.01830, -0.14913)$ following \cite{Euclid2024}. \\
For the quadratic galaxy bias, we use the relation
\ba
b_2 = \alpha_0 + \alpha_1 \ b_1 + \alpha_2 \ b_1^2
\ea
with $(\alpha_0,\alpha_1,\alpha_2)=(0.51, -2.21, 1)$ following \cite{Hoffmann2015}. \\
Finally, the quadratic tidal tensor bias is given by the co-evolution relation \cite{Baldauf2012}
\ba
b_{s^2} = -\frac{2}{7} \left( b_1 - 1\right).
\ea

\subsection{Numerical computation of kernels for Super-Sample Covariance}\label{App:num-kernels-SSC}

In the computation of SSC (Section~\ref{Sect:SSC}), the two following quantities intervene:
\ba
\sigma^2_{\delta_b,\delta_b}(r,x) &= \frac{1}{2\pi^2} \int k^2 \, \dd k \ P(k|r,x) \ j_0(kr) \, j_0(kx) \\
\kappa(r,x) &= \frac{1}{2\pi^2} \int k^2 \, \dd k \ P(k|r,x) \ j_0(kr) \, j''_0(kx) \\
\upsilon(r,x) &= \frac{1}{2\pi^2} \int k^2 \, \dd k \ P(k|r,x) \ j_0(kr) \, \frac{j'_0(kx)}{kx} \\
\tau(r,x) &= \frac{1}{2\pi^2} \int k^2 \, \dd k \ P(k|r,x) \ j_0(kr) \, kx \, j'_0(kx).
\ea
To compute these numerically, a brute force integration is possible but expensive due to the highly oscillatory behaviour of the Bessel functions. Instead, we can express the Bessel functions as sums of sines and cosines:
\ba
j_0(t) = \frac{\sin t}{t} \qquad
\mr{,} \qquad j'_0(t) = -\frac{\sin t}{t^2}  + \frac{\cos t}{t} \qquad
\mr{and} \qquad j''_0(t) = \frac{2 \sin t}{t^3}  - \frac{2\cos t}{t^2} - \frac{\sin t}{t}
\ea
and use laws of trigonometry to express products of sines/cosines. This leads to expressions for the kernels as functions on Fourier transforms of the power spectrum multiplied by power laws. Specifically, if we define 
\ba
I_{c,n}(r) \equiv & \int \dd k \ P(k)/k^n \cos(kr), & \qquad & I_{c,n}^{\pm}(r_1,r_2) \equiv I_{c,n}(r_1\pm r_2), \\
I_{s,n}(r) \equiv & \int \dd k \ P(k)/k^n \ \sin(kr), & \qquad & I_{s,n}^{\pm}(r_1,r_2) \equiv I_{s,n}(r_1\pm r_2).
\ea
then we get
\ba
\sigma^2_{\delta_b,\delta_b}(r,x) &= \frac{G(r) \, G(x)}{4\pi^2} \times \frac{ I_{c,0}^{-}(r,x) - I_{c,0}^{+}(r,x)}{r \, x}\\
\kappa(r,x) &= \frac{G(r) \, G(x)}{4\pi^2} \left[ 
2 \frac{I_{c,2}^{-}(r,x) - I_{c,2}^{+}(r,x)}{r \, x^3}
- 2 \frac{I_{s,1}^{-}(r,x) + I_{s,1}^{+}(r,x)}{r \, x^2}
- \frac{I_{c,0}^{-}(r,x) - I_{c,0}^{+}(r,x)}{r \, x}
\right] \\
\upsilon(r,x) &= \frac{G(r) \, G(x)}{4\pi^2} \left[ 
- \frac{I_{c,2}^{-}(r,x) - I_{c,2}^{+}(r,x)}{r \, x^3}
+ \frac{I_{s,1}^{-}(r,x) + I_{s,1}^{+}(r,x)}{r \, x^2}
\right] \\
\tau(r,x) &= \frac{G(r) \, G(x)}{4\pi^2} \left[
- \frac{ I_{c,0}^{-}(r,x) - I_{c,0}^{+}(r,x)}{r \, x}
+ \frac{I_{s,-1}^{-}(r,x) + I_{s,-1}^{+}(r,x)}{r}
\right].
\ea
The interest of this is that the $I_{c/s,n}$ integrals can be computed cheaply and accurately with a Fast Fourier Transform (FFT) and tabulated. We call this first approach the trigonometric method. We remark that there can be divergent integrals in the IR (when $k\rightarrow 0$), specifically $I_{c,n}$ diverges for $n \geq 2$ and $I_{s,n}$ diverges for $n \geq 3$ (which does not appear here). Formally, this issue is cured by the fact that an $I_{c,n}^{-}$ is always paired with an $I_{c,n}^{+}$ of opposite sign and divergences are cancelled out. Numerically however, this may be a problem because high cancellations are prone to numerical errors. In the case presented in this article where we have a single bessel derivative, the troublesome integral is $I_{c,2}$ which is logarithmically divergent. We wanted to have a point of comparison for this case.

That is why we  have also used a second method to derive $\kappa$, $\upsilon$ and $\tau$. Indeed, one can remark that
\ba
\kappa(r,x) &= G(r) G(x) \frac{\partial^2}{\partial x^2}\left[\frac{1}{2\pi^2} \int k^2 dk \, \frac{P(k)}{k^2} \ j_0(k r) \, j_0(k x)\right] = G(r) G(x)\frac{\partial^2}{\partial x^2}\left[\frac{I_{c,2}^{-}(r,x) - I_{c,2}^{+}(r,x)}{4\pi^2 \ r x}\right] \\
\upsilon(r,x) &= \frac{G(r) G(x)}{x} \frac{\partial}{\partial x}\left[\frac{1}{2\pi^2} \int k^2 dk \, \frac{P(k)}{k^2} \ j_0(k r) \, j_0(k x)\right] = \frac{G(r) G(x)}{x}\frac{\partial}{\partial x}\left[\frac{I_{c,2}^{-}(r,x) - I_{c,2}^{+}(r,x)}{4\pi^2 \ r x}\right] \\
\tau(r,x) &= x G(r) G(x) \frac{\partial}{\partial x}\left[\frac{1}{2\pi^2} \int k^2 dk \, P(k) \ j_0(k r) \, j_0(k x)\right] = xG(r) G(x)\frac{\partial}{\partial x}\left[\frac{I_{c,0}^{-}(r,x) - I_{c,0}^{+}(r,x)}{4\pi^2 \ r x}\right].
\ea
Hence we can compute all kernels by tabulating $I_{c,0}$ and $I_{c,2}$ and applying numerical derivatives. We call this second approach the derivative method. We can expect it to be more robust against the IR divergence because the derivatives involve only differences of such integrals.

Regarding the new kernels $\kappa$, $\upsilon$ and $\tau$ \footnote{$\sigma$ is not new to this article, it has been studied in past literature.}, we implemented the two approaches --the trigonometric and the derivative methods-- and compared them. For $\upsilon$ we found agreement between both methods, except for a 10\% discrepancy at low redshift which we attribute to the issue of IR divergence in the trigonometric method. For $\kappa$ we also found agreement, except at the peak where the derivative method fails due to the peak's steepness\footnote{We found that the derivative method works well for a single derivative, as for $\tau$, but fails increasingly for second and higher order derivatives. Indeed we found higher disagreement in exploratory calculation of integrals involving $j'_0 \times j''_0$ or $j''_0 \times j''_0$. We expect these integrals to appear in the trispectrum, beyond the case of the bispectrum studied in this article.}. For $\tau$ we found that both methods gave the same order of magnitude and similar shapes for the curves, but the level of precision was not satisfactory. This may be due to the extra $k$ factor which makes the integral UV sensitive. More numerical study would be needed to reach satisfactory precision. Fortunately, the order of magnitude is a sufficient result here, since we show in the main text that the SSC term associated to $\tau$ is negligible with respect to the other SSC terms. Detailed plots showing all these comparisons are available at the end of the notebook made available with this article, for the reader avid of numerical details.


\bibliography{bibliography}
\bibliographystyle{JHEP}


\end{document}